\documentclass[12pt]{article}

\usepackage{xr-hyper}
\pdfoutput=1
\usepackage{hyperref}
\hypersetup{
    colorlinks=true,
    linkcolor=black,
    citecolor=black,
    filecolor=black,
    urlcolor=black,
}

\RequirePackage{amssymb}
\RequirePackage{amsmath}

\DeclareMathOperator*{\cov}{cov}
\DeclareMathOperator*{\var}{var}
\newcommand{\R}{\ensuremath{\mathbb{R}}}
\newcommand{\Exp}{\ensuremath{\mathbb{E}}}
\newcommand{\Prob}{\ensuremath{\mathbb{P}}}

\usepackage{bbm}

\def\independenT#1#2{\mathrel{\rlap{$#1#2$}\mkern2mu{#1#2}}}

\makeatletter
\newcommand*{\indep}{
  \mathbin{
    \mathpalette{\@indep}{}
  }
}
\newcommand*{\nindep}{
  \mathbin{
    \mathpalette{\@indep}{\not}
  }
}
\newcommand*{\@indep}[2]{
  \sbox0{$#1\perp\m@th$}
  \sbox2{$#1=$}
  \sbox4{$#1\vcenter{}$}
  \rlap{\copy0}
  \dimen@=\dimexpr\ht2-\ht4-.2pt\relax
  \kern\dimen@
  {#2}
  \kern\dimen@
  \copy0
} 
\makeatother

\usepackage[longnamesfirst]{natbib}
\usepackage{ulem}
\RequirePackage{amsthm}

\theoremstyle{definition}

\newtheorem{definition}{Definition}

\theoremstyle{definition}

\newtheorem{property}{Property}
\newtheorem{theorem}{Theorem}
\newtheorem{proposition}{Proposition}
\newtheorem{corollary}{Corollary}

\newtheorem{lemma}[theorem]{Lemma}

\newtheorem{Aassumption}{Assumption}

\newtheorem{Bassumption}{Assumption}

\newtheorem{Cassumption}{Assumption}

\newcounter{homogSection}
\newcommand{\aAssump}{A\arabic{homogSection}}
\newcounter{het1section}
\newcommand{\bAssump}{B\arabic{het1section}}
\newcounter{het2section}
\newcommand{\cAssump}{C\arabic{het2section}}

\newtheoremstyle{theoremSuppressedNumber}{}{}{}{}{\bfseries}{.}{ }{\thmname{#1}\thmnote{ (\mdseries #3)}}
\theoremstyle{theoremSuppressedNumber}

\newcounter{peq}
\renewcommand{\thepeq}{P\arabic{peq}}

\newenvironment{Pequation}
  {
    \refstepcounter{peq}
    \begin{equation*}
  }
  {
    \tag{\thepeq}
    \end{equation*}\ignorespacesafterend
  }

\usepackage{mathtools}

\newcommand{\Lin}{\ensuremath{\mathbb{L}}}

\usepackage{color}

\usepackage{wrapfig}
\usepackage{tikz-cd}
\usepackage{tikz}
\usepackage{pgfplots}

\usepackage{rotating}
\usepackage{lscape}

\usepackage{array}
\usepackage{chngpage}
\usepackage{multirow}
\usepackage{tabulary}

\usepackage{tabularx}
\usepackage{booktabs}

\usepackage{tabularray}
\UseTblrLibrary{booktabs}

\def\Wob{W_1}

\def\Wu{W_2}

\def\avgk{\frac{1}{K} \sum_{i=1}^K}
\def\pconv{\overset{p}{\rightarrow}}

\def\1{\mathbbm{1}}

\usepackage{subfig}
\usepackage{wrapfig}
\usepackage{float}

\usepackage{epstopdf}

\usepackage{setspace}
\usepackage{relsize}
\newcommand*{\sbullet}{\mathrel{\mathsmaller{\bullet}}}

\title{\textbf{An Axiomatic Approach to \\ Comparing Sensitivity Parameters}\footnote{This paper is a revised, shorter version of our now-superseded previous working paper titled ``Assessing Omitted Variable Bias when the Controls are Endogenous'' (\citealt{DMP2023}), without the identification analysis or empirical results of the former sections 4 and 5. The identification analysis and empirical results can now be found in our companion paper \cite{DMP2023v5}. We thank audiences at various seminars and conferences, as well as Joe Altonji, Isaiah Andrews, Peter Hull, Evan Rose, and Jon Roth for helpful conversations and comments. We thank Gianna Fenaroli, Hongchang Guo, and Muyang Ren for excellent research assistance. Masten thanks the National Science Foundation for research support under Grant 1943138.}}

\author{Paul Diegert\footnote{Toulouse School of Economics,        \texttt{paul.diegert@tse-fr.eu}} \qquad
Matthew A. Masten\footnote{Department of Economics, Duke University,
        \texttt{matt.masten@duke.edu}} \qquad Alexandre Poirier\thanks{
    Department of Economics, Georgetown University,
 \texttt{alexandre.poirier@georgetown.edu}}
}

\date{February 4, 2026}

\begin{document}
\maketitle
\begin{abstract}
Many methods are available for assessing the importance of omitted variables in linear regression. These methods typically make different, non-falsifiable assumptions. Hence the data alone cannot tell us which method is most appropriate. Since it is unreasonable to expect results to be robust against all possible robustness checks, researchers often use methods deemed ``interpretable,'' a subjective criterion with no formal definition. In contrast, we develop the first formal, axiomatic framework for comparing and selecting among these methods. Our framework is analogous to the standard approach for comparing estimators based on their sampling distributions. We propose that sensitivity parameters be selected based on their \emph{covariate sampling distributions}, a design distribution of parameter values induced by an assumption on how covariates are assigned to be observed or unobserved. Using this idea, we define new concepts of parameter consistency and monotonicity, and argue that a reasonable sensitivity parameter should satisfy both properties. We prove that the literature's most popular approach is inconsistent and non-monotonic, while several alternatives satisfy both.
\end{abstract}



\allowdisplaybreaks

\onehalfspacing
\normalsize
\newpage

\section{Introduction}\label{sec:intro}

\citet[page 74]{AngristPischke2015} argue that ``careful reasoning about OVB [omitted variables bias] is an essential part of the 'metrics game.'' Largely for this reason, researchers have eagerly adopted new tools that let them quantitatively assess the impact of omitted variables on their linear regression results. In particular, researchers now widely use the sensitivity analysis methods developed in \cite{AltonjiElderTaber2005} and \cite{Oster2019}. These methods have been extremely influential, with about 4800 and 5600 Google Scholar citations as of January 2026, respectively. Every top journal in economics is now regularly publishing papers using these methods.\footnote{For example, we surveyed papers published in the top five economics journals in 2024 and 2025. We found 19 which discuss the results of \cite{Oster2019}, 12 of which implement them numerically, including \citet[\emph{AER}]{AndrabiBauDasKhwaja2025}, \citet[\emph{QJE}]{AizerEtAl2024}, \citet[\emph{REStud}]{FrakesGruber2025}, \citet[\emph{JPE}]{HowellEtAl2025}, and \citet[\emph{ECMA}]{Rustagi2024}, for example. For data from 2019--2021, see the survey in Appendix A of \cite{MastenPoirier2024}, who note that this method is more commonly used than 2SLS.}

These sensitivity analyses are part of a broader econometric literature on robustness checks. After researchers perform a baseline analysis, using data to draw conclusions about a parameter of interest under a set of baseline assumptions, they typically use various methods to explore what could happen if those baseline assumptions do not hold. Which of these robustness checks should researchers use? There are infinitely many possibilities. In this paper, focusing on the context of assessing robustness to omitted variables, \textbf{we provide a formal, axiomatic framework that applied researchers can use to narrow down the set of robustness checks that they should implement}. 

Specifically, a key innovation of \cite{AltonjiElderTaber2005} was the idea that the magnitude of OVB can be assessed by reasoning about the relative importance of observed and unobserved variables. This idea can be formalized by defining a specific \emph{sensitivity parameter} which measures the impact of omitted variables on treatment and/or outcomes. Assumptions about this measure lead to bounds on OVB, thus providing researchers a way to quantitatively assess the impact of omitted variables. However, there are many different ways to formally define the ``relative importance'' of observed and unobserved variables, each leading to a different sensitivity parameter and thus a different robustness check. Moreover, assumptions about these different sensitivity parameters are usually non-nested and non-falsifiable. Thus the data alone cannot tell us which method or sensitivity parameter is the `correct' one to use. Indeed, the data alone cannot refute the stronger assumption that the omitted variable bias is zero. Consequently, researchers face a practical problem: Which of the many different sensitivity analyses should they implement? 

The current, common answers are: (1) Use a method that is well known and widely used in their field\footnote{This choice can be clearly seen in citation patterns: For example, \cite{AltonjiElderTaber2005} and \cite{Oster2019} are primarily used in economics, \cite{CinelliHazlett2020} is primarily used in statistics and political science, and the marginal sensitivity model of \cite{Tan2006} is primarily used in statistics and medicine.} and (2) Use a method that is ``interpretable,'' a subjective criterion with no formal definition. In contrast, we provide the first formal, mathematical framework that can be used to select which sensitivity analysis to perform.

Since the data itself cannot be used to select among the various methods, we propose an approach that is inspired by classical frequentist theory. That classical theory starts from a frequentist thought experiment where repeated samples are drawn from a data generating process, which induces a frequentist distribution for any given estimator or test statistic. Estimators and tests are viewed as \emph{procedures} which have properties like consistency and efficiency for estimators or size-validity and test consistency for hypothesis tests. Those properties are then used to justify the choice of a specific estimator or a specific hypothesis test. The early statistics literature defined many different formal properties that estimators and tests may have, which have since been used evaluate the desirability of newly proposed estimators and tests.

Analogously, for sensitivity analysis, inspired by the analysis in \cite{AltonjiElderTaber2005}, we imagine repeatedly drawing sets of covariates from a universe of potentially observable covariates, with each draw partitioning that universe into observed and unobserved covariates. The sensitivity parameters measure the relative importance of the observed and unobserved covariates. Hence their exact value will differ across different draws of observed covariates. This repeated sampling of covariates induces a distribution of the sensitivity parameters, which we call their \emph{covariate sampling distribution}. Following classical frequentist theory, we suggest using this distribution to compare and evaluate different sensitivity parameters. Specifically, we use this distribution to define two new properties for sensitivity parameters:
\begin{enumerate}
\item The first property, \emph{consistency}, requires the sensitivity parameter's covariate sampling distribution to collapse to 1 as the number of covariates gets large, whenever covariates have an equal chance of being observed versus unobserved. This property reflects the core idea of \cite{AltonjiElderTaber2005} that the usefulness of using sensitivity parameters based on the \emph{relative} importance of observed and unobserved variables comes from its ability to provide a known benchmark value of 1, typically called ``equal selection.'' 

\item The second property, called \emph{monotonicity in selection}, says that the sensitivity parameter's covariate sampling distribution should shift below 1 when covariates are more likely to be observed than not, and shift above 1 when covariates are less likely to be observed than not. This property is a refinement of consistency that requires the sensitivity parameter to move in the expected direction.
\end{enumerate}
We argue that reasonable sensitivity parameters should satisfy both properties. At first glance, it may appear that any sensitivity parameter based on the equal selection benchmark of \cite{AltonjiElderTaber2005} would easily satisfy both properties. However, we show that a variety of parameters do not satisfy both properties. This includes the parameter studied in \cite{Oster2019}, the most widely used parameter in the literature. This parameter is based on a particular residualization, which uses a version of the omitted variable that has been residualized to project out its correlation with the observed variables. Our results show that this transformation is not innocuous and changes the interpretation of the sensitivity parameter so that it no longer captures the notion of equal selection from \cite{AltonjiElderTaber2005}. In the next two subsections we discuss the implications of our results for both empirical and theoretical research.

\subsubsection*{Practical Implications for Applied Researchers}

As described above, the purpose of our paper is to help researchers narrow down and formally justify the set of robustness checks they choose. Concretely, we focus on the question: How should researchers assess the impact of omitted variables in linear regression? An immediate, practical implication of our results is that applied researchers should carefully consider whether to continue using the very popular sensitivity analysis developed by \cite{Oster2019}. This conclusion follows since that method relies on a sensitivity parameter which we prove to be inconsistent and non-monotonic, implying that researchers who use this parameter will generally draw the wrong conclusions about robustness. Instead, researchers can choose among various alternative methods, including \cite{CinelliHazlett2020} or our companion paper \cite{DMP2023v5}, both of which use sensitivity parameters that we prove are consistent and monotonic in the present paper. Thus our paper provides a formal, theoretical justification that empirical researchers can use to explain why they use one of these robustness checks, but not another.

Finally, note that although our method can help researchers narrow down the set of robustness checks they use, our results do not deliver a single, unique ``best'' approach. This is not surprising, however, since it is unlikely that any theory could argue that researchers should perform one and only one robustness check. Moreover, we conjecture that further distinctions between different sensitivity parameters could be obtained by examining other distributional features of covariate sampling distributions besides the two properties we focus on; we leave these extensions to future work.

\subsubsection*{Implications for Theoretical Researchers}

We develop our framework in the concrete setting of methods for assessing sensitivity to omitted variables in linear regression. Beyond this setting, our framework can be used to help researchers narrow down the set of robustness checks in any other setting where researchers are concerned about omitted variables. For example:
\begin{enumerate}
\item Time-varying covariates are often included in difference-in-differences analyses to adjust for observed differences in trends between treated and untreated units. The conditional parallel trends assumption may still fail, however, if there are further relevant omitted time-varying covariates. Thus one could examine robustness to parallel trend violations by using a sensitivity parameter that compares the relative impact of the observed and unobserved time-varying covariates. 

\item Similarly, covariates are often included in instrumental variable models to ensure that the instrument exogeneity condition holds. Omitted variables can lead to exogeneity failures, however. Again, one could imagine analyzing robustness to such failures by comparing the impact of the observed and unobserved covariates.
\end{enumerate}
In both settings our framework can be used by theoretical researchers who must define and defend a sensitivity parameter used to measure those relative impacts; we leave the full development of these extensions to future work.

\subsubsection*{Overview of Our Results}

In section \ref{sec:model}, we review the identification problem caused by omitted variables, and set up the main challenge: How can the magnitude of the omitted variable bias be parameterized in a useful way? We survey eleven different measures from the literature in section \ref{sec:HowToMeasureSelection}. We then develop our framework for choosing among these various measures in section \ref{sec:comparingratios}. As in \cite{AltonjiElderTaber2005}, our framework asserts that observed covariates are selected probabilistically from a finite universe of potentially observable covariates. For any specific sensitivity parameter, a covariate selection mechanism induces a distribution over possible values of that sensitivity parameter, which we call the \emph{covariate sampling distribution} of the sensitivity parameter. The key idea then is to compare this induced distribution across different sensitivity parameters.

We formally define \emph{equal selection} of covariates as occurring when an equal number of covariates are observed and unobserved, and the specific split of these covariates is chosen uniformly at random.
Under equal selection of covariates, from the ex ante perspective before the covariates have been selected, the covariates we observe are equally important for determining treatment as the covariates we do not observe. 
Put differently, equal selection of covariates implies equal impact of observed and unobserved variables on treatment, ex ante. 
Therefore, under equal selection, any sensitivity parameter that measures the \emph{relative} importance of observed and unobserved variables should be approximately centered around the value 1. 
In this case, the value 1 represents the benchmark of equal selection against which other values of the sensitivity parameter can be compared.

While equal selection may not reflect the actual data-collection process in a given empirical study, we view it as a device that provides a tractable framework and also serves as a useful benchmark for analyzing sensitivity parameters.
We perform these analyses in section \ref{sec:designBasedAsymptotics}.
Formally, we approximate the distribution of sensitivity parameters induced by covariate selection using an asymptotic approximation based on a data-generating process (dgp) with an increasing number of covariates.
This approximation simplifies the analysis substantially, analogous to the simplification that occurs in standard frequentist analysis, whereby the exact finite sample distribution of a statistic is approximated by its asymptotic distribution. 
In our main result, we prove that the most commonly used approach in the literature is \emph{not} necessarily centered at 1 under equal selection. 
Specifically, in Theorem \ref{thm:delta_resid} we prove that Oster's \citeyearpar{Oster2019} delta parameter can converge to any real number, even under equal selection, unless an exogeneity assumption is imposed on control variables.
Consequently, the value 1 is not the correct benchmark for equal selection, for Oster's delta. 
Intuitively, this bias away from 1 arises because this parameter involves an asymmetry in how the importance of observed and unobserved variables is measured.
Empirical researchers widely use the value 1 as a cutoff for assessing robustness. 
This focus on the value 1 is motivated by its interpretation as the benchmark value of the parameter under equal selection. 
Thus, our result shows that when the controls are endogenous, researchers generally use the wrong benchmark for equal selection and therefore draw the wrong conclusions about robustness.

We also formally study the behavior of several other sensitivity parameters used in practice, including those proposed by \cite{AET2019}, \cite{CinelliHazlett2020}, and \cite{DMP2023v5}. We show that all of these sensitivity parameters concentrate around 1 under equal selection. We also study whether these sensitivity parameters satisfy a monotonicity property which, roughly, states that the sensitivity parameter's values increase as the number of unobservable variables increases in our framework. We expect reasonable sensitivity parameters to satisfy this property as they purportedly measure the importance of unobservable covariates relative to observable covariates. We show the parameters of \cite{CinelliHazlett2020} and \cite{DMP2023v5} are monotonic, while the ones proposed by \cite{AltonjiElderTaber2005}, \cite{AET2019}, and \cite{Oster2019} are not. 

These results are shown under high-level conditions on the distribution of covariates; in particular, the structure of the covariances across all observed and unobserved covariates. In section \ref{sec:AR-MA-Ex} we show that these high-level conditions are implied by four different kinds of lower-level conditions on the distribution of covariates. In particular, we show that variance matrices with a moving-average, autoregressive, exchangeable, or factor structure all satisfy our high-level assumptions.

Finally, to complement the asymptotic results of section \ref{sec:designBasedAsymptotics}, we study the exact, non-asymptotic distribution of sensitivity parameters in section \ref{sec:empiricalDesignBased}. We construct an empirically calibrated dgp by treating the data from the paper \cite{BFG2020} as if it were the true population. Hence we treat the twenty-two observed covariates in this dataset as the universe of all potentially observable covariates. We then compute and compare the exact distribution of various sensitivity parameters induced by random selection of a fixed number of these twenty-two covariates. Overall, our non-asymptotic findings agree with our asymptotic results.

\subsubsection*{Related Literature}

\citet[page 170]{AltonjiElderTaber2005} discuss the idea of sampling covariates. They argue that random covariate sampling can justify an assumption that their sensitivity parameter equals 1 (their Condition 1), which can then be used for identification. They do not use covariate sampling to compare and contrast different sensitivity parameters, however, which is our focus. The structure of our asymptotic analysis is similar to \cite{AET2019}, who also use high level assumptions (e.g., their Assumption  2) accompanied by lower level sufficient conditions (e.g., their appendices A.1 and A.5). However, again the goals of the papers are quite different: \cite{AET2019} use assumptions about the covariates for identification. In contrast, we focus on the choice of sensitivity parameter, with the goal of selecting a sensitivity parameter that has desirable properties for a large class of covariate distributions. The selected sensitivity parameter can then be used in an identification analysis which does not explicitly impose any structure on the covariates, like \cite{Krauth2016}, \cite{Oster2019}, \cite{CinelliHazlett2020}, or the one in our companion paper \cite{DMP2023v5}. Another difference is that \cite{AET2019} assume that the inclusion indicators are iid Bernoulli random variables (their Assumption  4), whereas we use uniform selection (our \ref{assn:sampling_of_S}), which allows us to control the ex post proportion of observed covariates.

The contribution of our companion paper \cite{DMP2023v5} is to propose new sensitivity parameters ($r_X$ and $r_Y$, see section \ref{sec:HowToMeasureSelection}) and derive new identification results and sensitivity analyses based on those parameters. In contrast, the present paper does not derive any identification results or propose any new sensitivity analyses. We instead focus on providing a general, axiomatic approach to comparing and selecting sensitivity parameters, as described in section \ref{sec:comparingratios} and applied in section \ref{sec:designBasedAsymptotics}. In particular, we also apply this approach to study the properties of the sensitivity parameters proposed in \cite{AltonjiElderTaber2005} and \cite{Oster2019}. Several previous papers have also studied and critiqued those specific parameters and their uses, including \cite{DeLucaMagnusPeracchi2019}, \citet[section 6.3]{CinelliHazlett2020}, \cite{Basu2022}, and \cite{MastenPoirier2024}. Our paper complements this literature by providing a new and distinct appraisal of those specific parameters. Furthermore, while this prior literature focused on specific parameters, this paper provides a general framework for studying a wide variety of sensitivity parameters, including the eleven parameters described in section \ref{sec:HowToMeasureSelection}. 

Finally, our framework uses a finite population design-based setup, similar to recent work like \cite{AbadieAtheyImbensWooldridge2020}; see \cite{BorusyakHull2024} for a review and further citations. In our analysis, the population consists of covariates rather than units, and the design distribution is a set of indicators denoting which covariates are observed, rather than indicators denoting which units are treated.

\subsubsection*{Notation Remarks}

For random vectors $A$ and $B$, let $\cov(A,B)$ be the $\text{dim}(A) \times \text{dim}(B)$ matrix whose $(i,j)$th element is $\cov(A_i, B_j)$. Define $A^{\perp B} \coloneqq A - \cov(A,B)\var(B)^{-1}B$. This is the sum of the residual from a linear projection of $A$ onto $(1,B)$ and the intercept in that projection. Many of our equations therefore do not include intercepts because they are absorbed into $A^{\perp B}$ by definition. Note also that $\cov(A^{\perp B},B) = 0$ by definition. Let $R_{A \sim B \sbullet C}^2$ denote the R-squared from a regression of $A^{\perp C}$ on $(1,B^{\perp C})$. This is sometimes called the partial R-squared. $\iota_K$ denotes a $K \times 1$ vector of ones and $\textbf{I}_K$ denotes a $K \times K$ identity matrix.

\section{The Regression You Have and the Regression You Want}\label{sec:model}

Let $Y$ be a scalar outcome variable, $X$ a scalar regressor of interest (often a treatment variable), and $\Wob$ be a vector of observed covariates. Consider the OLS estimand of $Y$ on $(1,X,\Wob)$. Let $(\beta_\text{med}, \gamma_{1,\text{med}})$ denote the coefficients on $(X,\Wob)$. Then we can write
\[
	Y = \beta_\text{med} X + \gamma_{1,\text{med}}' \Wob + Y^{\perp X,\Wob}
\]
where $Y^{\perp X,\Wob}$ is defined to be the OLS residual plus the intercept term, and hence is uncorrelated with each component of $(X,\Wob)$ by construction. Researchers often begin their analyses by computing an estimate of coefficients like these. The second step typically asks: How would the coefficient on $X$ change if we include additional unobserved covariates? Let $\Wu$ denote the vector of these unobserved variables. Let $W \coloneqq (\Wob, \Wu)$, where $W_1$ has dimension $d_1$, $W_2$ has dimension $d_2$, and $W$ has dimension $K \coloneqq d_1 + d_2$. Consider the long OLS estimand $Y$ on $(1,X,\Wob, \Wu)$. Let $(\beta_\text{long}, \gamma_1, \gamma_2)$ denote the coefficients on $(X,\Wob,\Wu)$. Then we can write
\begin{equation}\label{eq:outcome}
	Y = \beta_\text{long} X + \gamma_1' \Wob + \gamma_2' \Wu + Y^{\perp X,W}
\end{equation}
where $Y^{\perp X,W}$ is defined to be the OLS residual plus the intercept term. To ensure that these OLS estimands are well defined, we maintain the following assumption throughout the paper.

\begin{Aassumption}\label{assump:posdefVar}
The variance matrix of $(Y,X,\Wob,\Wu)$ is finite and positive definite.
\end{Aassumption}

Suppose our goal is to learn about the parameter $\beta_\text{long}$. Section 4 of our companion paper \cite{DMP2023v5} discusses causal models that lead to this specific OLS estimand as the parameter of interest, using either unconfoundedness, difference-in-differences, or instrumental variables as an identification strategy. Alternatively, it may be that we are simply interested in $\beta_\text{long}$ as a descriptive statistic. The specific motivation for interest in $\beta_\text{long}$ does not affect our technical analysis.

The identification problem is that $W_2$ is not observed, and therefore researchers cannot run the long regression. Hence the omitted variable bias,
\[
	\text{OVB} \coloneqq \beta_\text{med} - \beta_\text{long},
\]
is completely unidentified without further assumptions. Rather than simply assuming that OVB is zero, there is now a large literature on sensitivity analysis that allows researchers to directly reason about and bound the magnitude of OVB. A first pass, naive approach would directly assume $| \text{OVB} | \leq M$ for some known sensitivity parameter $M \geq 0$, which immediately implies that $\beta_\text{long}$ is within $\pm M$ of $\beta_\text{med}$. The problem with this approach is that it is not clear how to select $M$. To avoid this problem, most methods do not reason about OVB directly, but rather parameterize OVB in terms of an ``interpretable'' sensitivity parameter, and then ask practitioners to make assumptions about this sensitivity parameter. These assumptions can then be translated into bounds on OVB, and hence bounds on $\beta_\text{long}$.

The key difference between the various methods in the literature is therefore how they define the sensitivity parameter which practitioners must reason about. In general, these parameters are completely unidentified from the data alone---recall that the data cannot even refute the claim that OVB is zero---and therefore the data cannot help practitioners select which sensitivity parameters to work with. In light of this challenge, we propose a new theoretical framework that can be used to compare and contrast the different sensitivity parameters. We develop this approach in section \ref{sec:comparingratios}. First, however, we review a variety of different sensitivity parameters from the literature in section \ref{sec:HowToMeasureSelection}.

\section{How to Measure Selection on Unobservables?}\label{sec:HowToMeasureSelection}

Many approaches to measuring the importance of omitted variables follow the important and influential work of \cite{AltonjiElderTaber2005}, who proposed that researchers make assumptions on \emph{selection ratios}, measures which compare the \emph{relative} importance of observed and unobserved variables. This kind of sensitivity parameter allows empirical researchers to make statements like ``to attribute the entire OLS estimate to selection effects, selection on unobservables would have to be at least three times greater than selection on observables'' (\citealt{NunnWantchekon2011}, \emph{AER}, page 3238). There are many different ways to formally define a selection ratio, however. In this section we survey eleven different measures from the literature, numbered P1--P11. Some of these are similar and hence we combine them into four groups.

\subsubsection*{1. Ratios of Regression Coefficients}

Following \cite{AltonjiElderTaber2005}, \cite{Oster2019} defines the following sensitivity parameter:
\begin{Pequation}\label{eq:DeltaOrig}
	\delta_\text{orig} \coloneqq \frac{ \cov(X, \gamma_2' \Wu) }{ \var(\gamma_2' \Wu) } \Bigg/ \frac{\cov(X,\gamma_1' \Wob) }{\var(\gamma_1' \Wob)}.
\end{Pequation}
The numerator is a measure of selection on unobservables while the denominator is a measure of selection on observables. \cite{Oster2019} provides identification results for $\beta_\text{long}$ under three assumptions: (i) $\delta_\text{orig}$ is known, (ii) $R_{Y \sim X,W_1,W_2}^2$, the R-squared in the long regression of equation \eqref{eq:outcome}, is known, and (iii) exogenous controls, $\cov(W_1,W_2) = 0$. However, her identification results do not hold under knowledge of $\delta_\text{orig}$ if the controls are endogenous ($\cov(W_1, W_2) \neq 0$). To allow for endogenous controls, she suggests replacing $\delta_\text{orig}$ with a different sensitivity parameter. Specifically, consider the linear projection of $\gamma_2'W_2$ onto $(1,W_1)$:
\[
	\gamma_2'W_2 = \phi' W_1 + (\gamma_2'W_2)^{\perp W_1}.
\]
Here $\cov(W_1, (\gamma_2'W_2)^{\perp W_1}) = 0$ by construction and $\phi \coloneqq  \var(W_1)^{-1}\cov(W_1,\gamma_2'W_2)$. Now define
\begin{Pequation}\label{eq:deltaResid}
	\delta_\text{resid} \coloneqq \frac{ \cov(X, (\gamma_2'W_2)^{\perp W_1}) }{ \var( (\gamma_2'W_2)^{\perp W_1}) } \Bigg/ \frac{\cov(X,(\gamma_1 + \phi)' W_1) }{\var((\gamma_1 + \phi)' W_1)}.
\end{Pequation}
Oster's \citeyearpar{Oster2019} results now hold so long as the assumptions are stated in terms of $\delta_\text{resid}$, even if exogenous controls fails. We further discuss the technical details behind this residualization in Appendix \ref{sec:osterredef}.

\cite{AET2019} propose the following third variation on this type of parameter:
\begin{Pequation}
	\delta_\text{ACET} \coloneqq \frac{\cov(X,(\gamma_2'W_2)^{\perp W_1})}{\var((\gamma_2'W_2)^{\perp W_1})} \Bigg/ \frac{\cov(X,(\gamma_1'W_1)^{\perp W_2})}{\var((\gamma_1'W_1)^{\perp W_2})},
\end{Pequation}
which uses the same numerator but a different denominator. They then provide identification results that involve assumptions on this parameter.

\subsubsection*{2. The Relative Variation in Selection Indices}

Our next parameter depends on the OLS estimand of $X$ on $(1,\Wob,\Wu)$. Let $(\pi_1,\pi_2)$ denote the coefficients on $(\Wob,\Wu)$. Then we can write
\begin{equation}\label{eq:XprojectionW1W2}
	X = \pi_1' \Wob + \pi_2' \Wu + X^{\perp W}
\end{equation}
where $X^{\perp W}$ is defined to be the OLS residual plus the intercept term, and hence is uncorrelated with each component of $W$ by construction. Our companion paper \cite{DMP2023v5} measures selection on unobservables by the standard deviation of the index of unobservables, $\sqrt{\var(\pi_2' W_2)}$. Likewise, that paper measures selection on observables by the standard deviation of the index of the observables, $\sqrt{\var(\pi_1' W_1)}$. This leads to the following selection ratio:
\begin{Pequation}\label{eq:rX}
	r_X \coloneqq \frac{\sqrt{\var(\pi_2'\Wu)}}{\sqrt{\var(\pi_1'\Wob)}}.
\end{Pequation}
That paper also considers a similar measure, but based on the outcome equation \eqref{eq:outcome},
\begin{Pequation}
	r_Y \coloneqq \frac{\sqrt{\var(\gamma_2'\Wu)}}{\sqrt{\var(\gamma_1'\Wob)}}.
\end{Pequation}

\subsubsection*{3. Ratios of R-Squared's}

The parameters in the next group are defined using R-squared's. \cite{CinelliHazlett2020} use the sensitivity parameter
\begin{Pequation}
	k_X 
	\coloneqq \frac{R_{X \sim W_1,W_2}^2 - R_{X \sim W_1}^2}{R_{X \sim W_1}^2 - 0} 
	= \frac{R_{X \sim W_2^{\perp W_1}}^2}{R_{X \sim W_1}^2}.
\end{Pequation}
They also consider the sensitivity parameter
\begin{Pequation}
	k_Y  
	\coloneqq
	\frac{R_{Y \sim W_2^{\perp W_1} \sbullet X}^2}{R_{Y \sim W_1 \sbullet X}^2}.
\end{Pequation}
The following variations can also be considered, although we are not aware of any identification results based on these:
\begin{align*}
	k_{X,\text{alt}} &\coloneqq \frac{R_{X \sim W_2}^2}{R_{X \sim W_1}^2} \tag{P8} \\
	k_{Y,\text{alt}} &\coloneqq \frac{R_{Y \sim X,W_2}^2}{R_{Y \sim X,W_1}^2} \tag{P9} \\
	k_{Y,\text{alt,2}} &\coloneqq \frac{R^2_{Y \sim X,W_1,W_2} - R^2_{Y \sim X,W_1}}{R^2_{Y \sim X,W_1} - R^2_{Y \sim X}}. \tag{P10}
\end{align*}
\setcounter{peq}{10}

\subsubsection*{4. Ratios of Correlations}

Finally, \cite{Krauth2016} derived identification results under assumptions on the following parameter:
\begin{Pequation}
	\lambda \coloneqq \frac{\text{corr}(X, \gamma_2 W_2^{\perp W_1})}{\text{corr}(X, (\gamma_1 + \gamma_2 \phi) W_1)}
\end{Pequation}
where $W_2 = \phi W_1 + W_2^{\perp W_1}$, assuming $W_1$ and $W_2$ are scalars for simplicity here.

\section{Comparing Selection Ratios}\label{sec:comparingratios}

We now have \textbf{eleven} different ways of measuring selection on unobservables. And more measures are likely to be proposed in future research. Which should practitioners use to perform sensitivity analyses? All of the measures in section \ref{sec:HowToMeasureSelection} are functions of the unobserved variables and hence are not identified from the data. Therefore the data alone cannot tell researchers which parameter they should use to measure relative selection. To address this challenge, in this section we develop a formal, axiomatic approach for choosing between different measures of selection. This approach is based on comparing the properties of each selection ratio under a model of covariate selection that determines which covariates are observed.

We start by recalling the baseline model from section \ref{sec:model}, but now from the \emph{ex ante} perspective where it is not yet known which covariates will actually be observed. Thus we let $W \in \R^K$ denote the random vector of all potentially observable covariates. $Y$ and $X$ are random scalars as before. Given the random vector $(Y,X,W)$, we can consider the linear projections
\begin{align*}
	Y &= \beta_\text{long} X + \gamma'W + Y^{\perp X,W}\\
	X &= \pi'W + X^{\perp W}.
\end{align*}
These equations are identical to equations \eqref{eq:outcome} and \eqref{eq:XprojectionW1W2}, except that we have not specified which covariates are observed. 

Inspired by the analysis in \cite{AltonjiElderTaber2005}, we consider a model by which a subset of the components of $W$ will be observed. For each component $W_k$ of $W = (W_1,\ldots,W_K)$, let $S_k$ be a binary random variable denoting whether $W_k$ is observed or not. Let $S \coloneqq (S_1,\ldots,S_K)$ be a random vector with support $\{0,1\}^K$. The distribution of $S$ is called the \emph{design distribution}. For any given realization $s$ of $S$, define
\begin{align*}
	W_1(s) &\coloneqq \{ W_k: s_k = 1\} \quad &W_2(s) &\coloneqq \{ W_k: s_k = 0\}\\
	\gamma_1(s) &\coloneqq \{ \gamma_k: s_k = 1\} \quad &\gamma_2(s) &\coloneqq \{ \gamma_k: s_k = 0\}\\
	\pi_1(s) &\coloneqq \{ \pi_k: s_k = 1\} \quad &\pi_2(s) &\coloneqq \{ \pi_k: s_k = 0\}.
\end{align*}
This notation emphasizes that the identity of the observed and unobserved covariates, along with their corresponding coefficients in equations \eqref{eq:outcome} and \eqref{eq:XprojectionW1W2}, is determined by a random draw of $S$. For each sensitivity parameter in section \ref{sec:HowToMeasureSelection}, we can use this notation to denote its value for each realization $s$. For example, the sensitivity parameter \ref{eq:deltaResid} is
\[
	\delta_\text{resid}(s) \coloneqq \frac{\cov(X, \gamma_2(s)'W_2(s)^{\perp W_1(s)})}{\var( \gamma_2(s)'W_2(s)^{\perp W_1(s)} )} 
	\hspace{-1mm} \Bigg/ \hspace{-1mm}
	\frac{\cov(X, (\gamma_1(s) + \phi(s))' W_1(s))}{\var( (\gamma_1(s) + \phi(s))' W_1(s))}
	\tag{\ref{eq:deltaResid}$^\prime$}
\]
where $\phi(s) \coloneqq \var(W_1(s))^{-1}\cov(W_1(s),\gamma_2(s)'W_2(s))$, and \ref{eq:rX} is
\[
	r_X(s) \coloneqq \frac{\sqrt{ \var(\pi_2(s)' W_2(s))} }{ \sqrt{\var(\pi_1(s)'W_1(s))} }.
	\tag{\ref{eq:rX}$^\prime$}
\]
This notational dependence on $s$ highlights the dependence of the sensitivity parameter values on the identity of the observed and unobserved covariates. Let $\theta(s)$ denote a generic sensitivity parameter as a function of the covariate selection realization, such as any of the parameters defined in section \ref{sec:HowToMeasureSelection}.

\begin{definition}
Call the probability distribution of $\theta(S)$ induced by the design distribution of $S$ the \emph{covariate sampling distribution} of the sensitivity parameter.
\end{definition}

We can now state the main idea of this paper:
\begin{itemize}
\medskip
\item[] Different sensitivity parameters $\theta(\cdot)$ can be systematically and formally compared by comparing their covariate sampling distributions.
\medskip
\end{itemize}
This idea is directly analogous to the long-established practice in frequentist statistics of comparing the performance of estimators, hypothesis tests, etc., in terms of their sampling distributions. The main difference is the kind of frequentist thought experiment we are using. Here we consider repeated sampling of covariates, whereas traditional statistics considers repeated sampling of units. Despite this difference, this analogy is useful for framing our subsequent analysis, because many of the challenges that arise in implementing classical frequentist analysis arise here as well.

Specifically, the covariate sampling distributions of the sensitivity parameters depend on two things:
\begin{enumerate}
\item The design distribution of $S$; that is, how covariates are selected.

\item The matrix $\var(Y,X,W)$.
\end{enumerate}
For the second aspect, our analysis will allow for lower-level assumptions on the joint covariance matrix of $(Y,X,W)$; we discuss this further in section \ref{sec:designBasedAsymptotics} below. 
For the first aspect, we in principle could carry out all of the analysis below under any posited design distribution of $S$, in the same way that properties of different estimators are studied under different sampling distributions of the observations. As we will discuss, the following design distribution provides a tractable and useful starting point.

\begin{Bassumption}[Uniform random covariate selection]\label{assn:sampling_of_S}
For a given $K$ and $d_1 \in \{1,\ldots,K\}$,
\[
	\Prob(S = s) = \binom{K}{d_1}^{-1} \mathbbm{1} \left( \sum_{k=1}^K s_k = d_1 \right).
\]
That is, exactly $d_1$ covariates are observed, $d_2 = K - d_1$ are unobserved, and the observed covariates are selected according to a uniform distribution over all covariate subsets of size $d_1$.
\end{Bassumption}

Importantly, for most actual empirical studies, we do not view \ref{assn:sampling_of_S} as an accurate description of the true process by which covariates are observed. Instead, we view it as a simplified and tractable model of covariate selection which is useful for comparing sensitivity parameters. \textbf{If a sensitivity parameter has undesirable properties under \ref{assn:sampling_of_S}, it is unlikely to have better properties under more realistic and complicated models of covariate selection.} 

The main purpose of \ref{assn:sampling_of_S} is to provide a benchmark for formally comparing sensitivity parameters. Indeed, a major motivation for using selection ratios to measure the importance of omitted variables is that they allow us to think about the benchmark case where the observed and unobserved variables are equally important, typically called ``equal selection.'' As we discuss in section \ref{sec:frameworkImplications}, ``equal selection'' is by far the most commonly used benchmark in empirical work, and is commonly considered to be the cutoff between a robust and non-robust result. It is therefore a particularly important case to study. One way to formalize this notion of ``equal selection'' is based on the hypothetical covariate selection distribution in \ref{assn:sampling_of_S}, as follows.

\begin{definition}\label{def:equalSelection}
Suppose $S$ satisfies Assumption  \ref{assn:sampling_of_S}. Say there is...
\begin{enumerate}
\item \emph{equal selection} when $d_1 = d_2$, so that exactly half of all covariates are observed.

\item \emph{more} selection on unobservables than on observables when $d_2 > d_1$, so that fewer than half of all covariates are observed.

\item \emph{less} selection on unobservables than on observables when $d_2 < d_1$, so that more than half of all covariates are observed.
\end{enumerate}
\end{definition}

The motivation for this definition is similar to the analysis in \cite{AltonjiElderTaber2005}: When we see exactly half of all covariates, and which covariates we see are chosen at random, then \emph{from the ex ante perspective} before the covariates have been selected, the covariates we do observe will be equally important for determining treatment as the covariates we do not observe. So this definition explicitly formalizes the concept of ``equal selection'' in terms of a specific mechanism by which covariates are observed. Likewise, this definition formalizes the concept of more and less selection on unobservables.

The ex ante perspective here is directly analogous to the design-based analysis of randomized experiments (e.g., \citealt{ImbensRubin2015} chapter 5). Randomized treatment assignment does not guarantee that potential outcomes will be perfectly balanced across treatment groups ex post. But it does guarantee ex ante balance, or balance across repeated randomizations. Similarly, equal selection does not guarantee that any given realization $s$ of $S$ will lead to a partition $W_1(s)$ and $W_2(s)$ of $W$ such that the observed variables are equally as important as the unobserved variables. But it does guarantee this ex ante, or across repeated selections of the covariates.

We can now define two properties of selection ratios that formalize the idea that they measure the relative importance of observed and unobserved variables. 

\begin{property}[Consistency]
When \ref{assn:sampling_of_S} holds and $S$ satisfies equal selection, $| \theta(S) | \xrightarrow{p} 1$ as $K \rightarrow \infty$.
\end{property}

\begin{property}[Monotonicity in Selection]
When \ref{assn:sampling_of_S} holds, $| \theta(S) | \xrightarrow{p} C$ as $K \rightarrow \infty$ for some constant $C$ where (1) $C > 1$ if $S$ satisfies more selection, (2) $C = 1$ if $S$ satisfies equal selection, and (3) $C < 1$ if $S$ satisfies less selection.
\end{property}

Both properties are asymptotic, as the number of covariates gets large. Convergence in probability here refers to the covariate sampling distribution. We discuss these asymptotics in detail in section \ref{sec:designBasedAsymptotics} below. It is possible to define non-asymptotic versions of these properties, but just like the usual analysis of frequentist exact sampling distributions, it is difficult to prove exact results for large classes of data generating processes. Property 1 is a consistency requirement. We view this as a weak and natural requirement for any sensitivity parameter $\theta(\cdot)$ which aims to measure the relative importance of unobserved and observed covariates. Specifically, if $\theta(\cdot)$ is defined as the ratio of a measure of the importance of unobservables to a measure of the importance of observables, then Property 1 says that the absolute value of this ratio should be approximately centered around the value 1---nominally meaning that the unobservables and observables are equally important---when the covariates in fact satisfy equal selection. Property 2 is a stronger version of this requirement. This stronger version requires not only that the sensitivity parameter be approximately 1 under equal selection, but that it is approximately larger than 1 when there is more selection and it is approximately smaller than 1 when there is less selection. We also view this as a weak and natural requirement, since it says that the relative measure $\theta(\cdot)$ should say the unobservables are more important than the observables whenever there is more selection on unobservables. Likewise it says the relative measure $\theta(\cdot)$ should say unobservables are less important than the observables whenever there is less selection on unobservables.

Finally, as in classical frequentist statistics, there are other properties of the covariate sampling distributions we could consider. For example, we could study the asymptotic distribution of $\theta(S)$, rather than just its probability limit. However, as we show below, it turns out that even the weak requirements of Properties 1 and 2 above are often sufficient criteria to choose between the sensitivity parameters present in the literature. So we leave a careful study of other properties of this asymptotic distribution to future work.

\section{Which Selection Ratios Satisfy Properties 1 and 2?}\label{sec:designBasedAsymptotics}

In the previous section we described a general approach for comparing selection ratios: Compare their covariate sampling distributions. We then defined two properties that these sampling distributions could have, and argued that these are desirable properties. In this section we study whether these properties hold for several of the specific selection ratios defined in section \ref{sec:HowToMeasureSelection}. For brevity we provide results for the first six parameters only, which include the most widely used approaches by empirical researchers.

As mentioned above, the covariate sampling distributions depend on features of the joint distribution of $(Y,X,W)$. In practice, these are unknown (because not all components of $W$ are observed). This is analogous to the fact that the exact sampling distribution of estimators also typically depends on the exact dgp, which is unknown. In this section we address this challenge by using asymptotics as a tool to approximate their exact distributions. Specifically, we study their probability limits as the number of covariates $K$ gets large. This approach is directly analogous to the literature on design-based inference on treatment effects in finite populations. That literature uses sequences of finite populations of growing size to approximate exact distributions for populations of fixed size. For example, see \cite{LiDing2017} or \cite{AbadieAtheyImbensWooldridge2020}. In section \ref{sec:empiricalDesignBased} we examine the quality of the asymptotic approximation by showing that several features of the asymptotic analysis appear in the exact, non-asymptotic covariate sampling distributions that arise when the population dgp is constructed using data from a published empirical application.

\subsection*{High Level Assumptions}

To analyze the probability limits of the sensitivity parameters, we make several high level assumptions on the dgp. In section \ref{sec:AR-MA-Ex} we give various sets of lower-level sufficient conditions for these high level assumptions. These lower-level conditions suggest that our high level assumptions are compatible with a large range of data generating processes.

First, we consider sequences where the limiting relative proportion of unobserved to observed covariates is nontrivial.

\begin{Bassumption}[Nontrivial relative proportion] \label{assn:limit_d2d1}
As $K \rightarrow \infty$, $d_2/d_1 \rightarrow r$ for some $r \in (0,\infty)$.
\end{Bassumption}

This assumption requires both $d_1$ and $d_2$ to grow to infinity as $K$ grows. When $r = 1$, this corresponds to (asymptotic) equal selection. This is a weaker version of equal selection than that given in definition \ref{def:equalSelection}, which would require $d_1 = d_2$ at every point along the sequence.

Second, we impose some regularity assumptions on the data generating process. Here we use the notation $W^K$ synonymously with $W$ to emphasize that the dimension of the covariates depends on $K$. Likewise for their coefficients, $\pi^K$ and $\gamma^K$. In this assumption, the subscript $S$ in $\var_S$ refers to operators taken with respect to the design distribution of $S$. This is to distinguish it from operators taken with respect to the distribution of the random vector $(Y,X,W)$, which appear without subscripts.

\begin{Bassumption}\label{assn:pi_and_VarW}
The following hold:
\begin{enumerate}
\item (Nondegenerate and finite variance) For a fixed $D > 1$, $\var(\sum_{i=1}^K \pi_i^K W_i^K) \in (D^{-1},D)$ for all $K$.

\item (Conditions for LLN) $\lim_{K\to\infty} \var_S\left(\sum_{i=1}^K \sum_{j=1}^K S_i S_j \cov(\pi_i^K W_i^K,\pi_j^K W_j^K)\right) = 0$ and \\ $\lim_{K\to\infty} \var_S\left(\sum_{i=1}^K \sum_{j=1}^K (1-S_i)(1-S_j) \cov(\pi_i^K W_i^K,\pi_j^K W_j^K)\right) = 0$.

\item (Converging relative contribution to variance) $\lim_{K\to\infty} \frac{\sum_{i=1}^K  \var(\pi_i^K W_i^K)}{\var(\sum_{i=1}^K \pi_i^K W_i^K)} = c_\pi$ where $c_\pi \in [0,\infty)$.
\end{enumerate}
\end{Bassumption}

Assumption \ref{assn:pi_and_VarW}.1 says that the variance of $\pi'W$, the projection $X$ on $(1,W)$, is bounded from above and away from zero as the number of covariates $K$ varies. By properties of projections $\var(\pi'W) \leq \var(X)$. Thus an unbounded variance $\var(\pi'W)$ would be incompatible with a finite variance for $X$. This assumption requires that either the $K$ entries in the coefficient vector $\pi$ shrink with $K$ or that the $K^2$ elements of $\var(W)$ shrink with $K$. 

Assumption \ref{assn:pi_and_VarW}.2 is a tail condition that we use to apply a law of large numbers to replace sample averages with their expectations. We then use Assumption  \ref{assn:pi_and_VarW}.3 to ensure convergence of a scaled version of those expectations. To understand that assumption, note that the variance of any sum of random variables equals the sum of the variances plus other terms that depend on covariances. For a given $K$ and distribution of $(Y,X,W)$, the ratio of these two components is fixed at some number. Assumption \ref{assn:pi_and_VarW}.3 allows this ratio to vary with $K$ so long as it eventually converges. In this respect it is similar to Assumption  \ref{assn:limit_d2d1}. When the covariates are all mutually uncorrelated---which implies that the observed variables are always exogenous---Assumption  \ref{assn:pi_and_VarW}.3 holds with $c_\pi =1$. When the covariates are correlated, however, generally $c_\pi \neq 1$.

We will also use the same regularity conditions, but replacing the selection equation coefficients with the outcome equation coefficients.

\begin{Bassumption}\label{assn:gamma_and_VarW}
Assumption \ref{assn:pi_and_VarW} holds when we replace $(\pi^K, c_\pi)$ with $(\gamma^K, c_\gamma)$.
\end{Bassumption}

The interpretation of \ref{assn:gamma_and_VarW} is similar to \ref{assn:pi_and_VarW}. Our final assumption restricts the impact of a single covariate on (a) treatment and (b) the portion of outcomes accounted for by covariates.

\begin{Bassumption}[No outlier covariates]\label{assn:limited_dep}
$\sum_{i=1}^K \cov(X,\gamma_i^K W_i^K)^2 \rightarrow 0$ and \\ $\sum_{i=1}^K \cov(\gamma'W,\gamma_i^K W_i^K)^2 \rightarrow 0$ as $K \rightarrow \infty$.
\end{Bassumption}

For example, $\sup_{i =1,\ldots,K} | \cov(X, \gamma_i^K W_i^K) | = o(1 / \sqrt{K})$ is a sufficient condition for the first part of \ref{assn:limited_dep}.

\subsection*{Results}

We can now derive the probability limits of the sensitivity parameters under consideration. We start with $r_X$ and $r_Y$, because they are particularly simple.

\begin{theorem}[Convergence of $r_X$]\label{thm:rx_conv}
Suppose assumptions \ref{assn:sampling_of_S}--\ref{assn:pi_and_VarW} hold. Then, as $K \to \infty$,
\[
	r_X(S) \pconv \sqrt{ \frac{r(r+c_\pi)}{1+ r c_\pi}}.
\]
Consequently, under these assumptions, $r_X(S)$ satisfies Properties 1 and 2.
\end{theorem}

The limiting value of this sensitivity parameter depends on two factors: the ratio of unobserved to observed covariates $r$, and the nonnegative constant $c_\pi$ which depends on the correlation in the covariates. Importantly, for any value of $c_\pi \geq 0$, this ratio is strictly increasing in $r$, and it equals 1 when $r = 1$. Hence, under assumptions \ref{assn:sampling_of_S}--\ref{assn:pi_and_VarW}, $r_X(S)$ satisfies Properties 1 and 2 from section \ref{sec:comparingratios}. In this sense, the sensitivity parameter $r_X$ correctly measures the importance of unobservables relative to observables.

If we replace $r_X$ with $r_Y$ and $c_\pi$ with $c_\gamma$ in Theorem  \ref{thm:rx_conv}, the result continues to hold. This follows since these two parameters have similar definitions. See Appendix \ref{sec:additionalAsymptotics} for details.

Next consider
\[
	\delta_\text{orig}(s)
	\coloneqq \frac{\cov(X,\gamma_2(s)'W_2(s))}{\var(\gamma_{2}(s)'W_2(s))} \hspace{-1mm} \Bigg/ \frac{\cov(X,\gamma_1(s)'W_1(s))}{\var(\gamma_1(s)'W_1(s))}.
\]

\begin{theorem}[Convergence of $\delta_\text{orig}$]\label{thm:delta_orig_conv}
Suppose assumptions \ref{assn:sampling_of_S}, \ref{assn:limit_d2d1}, \ref{assn:gamma_and_VarW}, and \ref{assn:limited_dep} hold. Suppose $\cov(X,\gamma^{K\prime}W^K)$ is bounded above and bounded away from 0. Then, as $K \rightarrow \infty$,
\[
	\delta_\text{orig}(S) \pconv \frac{1 + rc_\gamma}{r + c_\gamma}.
\]
Consequently, under these assumptions, $\delta_\text{orig}(S)$ satisfies Property 1 but not Property 2.
\end{theorem}

Like our other high level assumptions, the additional assumption on $\cov(X, \gamma'W)$ in this theorem holds under a variety of lower level conditions, like MA, AR, or factor covariates; see section \ref{sec:AR-MA-Ex}. Similar to $r_X$, the limiting value of the sensitivity parameter $\delta_\text{orig}(S)$ depends on the ratio of unobserved to observed covariates $r$ and the nonnegative constant $c_\gamma$ which depends on the correlation in the covariates. And like $r_X$, for any $c_\gamma \geq 0$, this limiting value equals 1 when $r=1$. Hence $\delta_\text{orig}(S)$ is consistent. However, it does not satisfy the monotonicity Property 2. Specifically, when $c_\gamma \in [0,1)$, such as when the covariates are exchangeable ($c_\gamma = 0$), this limiting value is monotonic, but in the wrong direction---when a higher proportion of variables are not observed ($r > 1$) the limiting value is \emph{smaller} than 1, nominally suggesting that the unobserved variables are less important than the observed variables. Conversely, when a higher proportion of variables are observed ($r < 1$), the limiting value is \emph{larger} than 1, nominally suggesting that the unobserved variables are \emph{more} important than the observed values. Another unusual property is that when the covariates are uncorrelated, $c_\gamma = 1$, which implies that $\delta_\text{orig}(S) \xrightarrow{p} 1$ regardless of the value of $r$; that is, regardless of whether most covariates are observed, or whether most covariates are unobserved.

Next we consider the residualized version of this sensitivity parameter, $\delta_\text{resid}(S)$. Residualization makes both the behavior and analysis of this parameter much more complicated than the previous parameters. The following theorem considers the case where the covariates are exchangeable. Note that \ref{assn:EXvar} is formally defined on page \pageref{assn:EXvar}.

\begin{theorem}[Behavior of $\delta_\text{resid}$]\label{thm:delta_resid}
Suppose assumptions \ref{assn:sampling_of_S} and \ref{assn:limit_d2d1} hold. Let $\rho \in (0,1)$ and suppose \ref{assn:EXvar} holds with $R=1$, $\Lambda^K = \iota_K \sqrt{\rho}$ and $\sigma^2_E = 1-\rho$ (that is, the covariates are exchangeable). Suppose $\sum_{i=1}^K \pi_i$ and $\sum_{i=1}^K \gamma_i$ are bounded away from 0. Then, as $K \rightarrow \infty$,
\begin{equation}\label{eq:deltaresid_limit}
	\delta_\text{resid}(S)
	= \frac{r \left(\sum_{i=1}^K \pi_i\right) \left(\sum_{i=1}^K \gamma_i\right)^2 + K \left(\sum_{i=1}^K \pi_i \gamma_i\right) \left(\sum_{i=1}^K \gamma_i\right)}{r \left(\sum_{i=1}^K \pi_i\right) \left(\sum_{i=1}^K \gamma_i\right)^2 + K \left(\sum_{i=1}^K \pi_i \right) \left(\sum_{i=1}^K \gamma_i^2\right)} + o_p(1).
\end{equation}
\end{theorem}

Theorem \ref{thm:delta_resid} provides an asymptotic representation for $\delta_\text{resid}(S)$. This representation has two main consequences.

\begin{corollary}[(Non)-Convergence of $\delta_\text{resid}$]\label{corr:delta_r_noconv}
Suppose Assumption  \ref{assn:sampling_of_S} holds. Fix any $r \in (0,\infty)$ and suppose Assumption  \ref{assn:limit_d2d1} holds. Then for any $C \in \R$ there exist sequences of vectors $\gamma^K \in \R^K$, $\pi^K \in \R^K$, and variance matrices $\var(W^K) \in \R^{K \times K}$ satisfying assumptions \ref{assn:pi_and_VarW}, \ref{assn:gamma_and_VarW}, and \ref{assn:limited_dep} such that 
\[
	\delta_\text{resid}(S) \pconv C
\]
as $K \to \infty$. Consequently, under assumptions \ref{assn:sampling_of_S}--\ref{assn:limited_dep}, $\delta_\text{resid}(S)$ does \emph{not} satisfy either Property 1 or 2.
\end{corollary}

Thus, for any value of $r$ and any number $C$, we can find a sequence of dgps such that $\delta_\text{resid}$ converges to $C$ as $K\rightarrow \infty$. In particular, this result holds even if $r = 1$. Corollary \ref{corr:delta_r_noconv} therefore shows that, under the same assumptions by which we showed that $r_X(S)$ satisfies Properties 1 and 2, $\delta_\text{resid}(S)$ does not satisfy either of those properties. It is possible that $\delta_\text{resid}(S)$ may satisfy these properties under stronger assumptions on the dgp, and this is one possible direction for future work. In this case, empirical researchers will have to argue that they believe their dgp satisfies these narrower restrictions in order to ensure that $\delta_\text{resid}$ satisfies the properties in their application.

To build intuition for Corollary \ref{corr:delta_r_noconv}, compare equation \eqref{eq:DeltaOrig} for $\delta_\text{orig}$ with equation \eqref{eq:deltaResid} for $\delta_\text{resid}$. The equation for $\delta_\text{orig}$ treats $W_1$ and $W_2$ symmetrically. Thus, under equal selection, the numerator and denominator of $\delta_\text{orig}(S)$ have the same covariate sampling distribution. In contrast, $\delta_\text{resid}$ treats $W_1$ and $W_2$ asymmetrically: $W_2$ is always projected onto $W_1$ because of residualization. This residualization breaks the symmetry between the numerator and denominator of $\delta_\text{orig}$. Consequently, when the controls are endogenous, the numerator and denominator do \emph{not} have the same covariate sampling distribution, even under equal selection; equation \eqref{eq:deltaresid_limit} in Theorem  \ref{thm:delta_resid} illustrates this asymmetry formally.

The following corollary gives an even stronger negative result.

\begin{corollary}[(Non)-Convergence of $\delta_\text{resid}$, part 2]\label{corr:delta_r_noconv_pt2}
Suppose Assumption  \ref{assn:sampling_of_S} holds. Suppose Assumption  \ref{assn:limit_d2d1} holds. Then there exist sequences of vectors $\gamma^K \in \R^K$, $\pi^K \in \R^K$, and variance matrices $\var(W^K) \in \R^{K \times K}$ satisfying assumptions \ref{assn:pi_and_VarW}, \ref{assn:gamma_and_VarW}, and \ref{assn:limited_dep} such that $\text{plim}_{K \rightarrow \infty} \; \delta_\text{resid}(S)$ is a decreasing function of $r$.
\end{corollary}

This result shows that, like $\delta_\text{orig}(S)$, $\delta_\text{resid}(S)$ can exhibit a reverse monotonicity property---it is large when $r$ is small and small when $r$ is large. This is the opposite of Property 2. In section \ref{sec:empiricalDesignBased} we show that this reverse monotonicity is not just a theoretical phenomenon but can arise in real empirical datasets.

Next consider the third variation on this type of sensitivity parameter,
\[
	\delta_\text{ACET}(s)
	\coloneqq \frac{\cov(X,\gamma_2(s)'W_2(s)^{\perp \gamma_1(s)'W_1(s)})}{\var(\gamma_2(s)'W_2(s)^{\perp \gamma_1(s)'W_1(s)})}
	\hspace{-1mm} \Bigg/ \hspace{-1mm}
	\frac{\cov(X,\gamma_1(s)'W_1(s)^{\perp \gamma_2(s)'W_2(s)})}{\var(\gamma_1(s)'W_1(s)^{\perp \gamma_2(s)'W_2(s)})}.
\]

\begin{theorem}[Convergence of $\delta_\text{ACET}$]\label{thm:delta_ACET}
Suppose assumptions \ref{assn:sampling_of_S}, \ref{assn:limit_d2d1}, \ref{assn:gamma_and_VarW}, and \ref{assn:limited_dep} hold. Suppose $c_\gamma > 0$ and that $\cov(X,\gamma^{K\prime}W^K)$ is bounded away from zero and infinity. Then, as $K \rightarrow \infty$,
\[
	\delta_\text{ACET}(S) \pconv 1.
\]
Consequently, $\delta_\text{ACET}(S)$ satisfies Property 1 but not Property 2.
\end{theorem}

Like $\delta_\text{orig}(S)$, $\delta_\text{ACET}(S)$ treats the observed and unobserved variables symmetrically, and hence it generally satisfies Property 1. Unlike $\delta_\text{orig}(S)$ or $\delta_\text{resid}(S)$, however, it does not exhibit a reverse monotonicity property. Instead, it is asymptotically a constant function in $r$. Theorem \ref{thm:delta_ACET} is similar to Theorem  1 and Corollary  1 of \cite{AET2019}, which show that their parameter converges to 1 under certain assumptions.

Finally, consider 
\[
	k_X(s)
	\coloneqq \frac{R^2_{X \sim W} - R^2_{X \sim W_1(s)}}{R^2_{X \sim W_1(s)}}.
\]
Like $\delta_\text{resid}(S)$, this sensitivity parameter is somewhat complicated to analyze. Here we consider the case where the covariates are exchangeable with shrinking variance (formally stated as \ref{assn:EXvar_shrink} in section \ref{sec:AR-MA-Ex}).

\begin{theorem}[Convergence of $k_X$]\label{thm:k_X_conv}
Suppose assumptions \ref{assn:sampling_of_S} and \ref{assn:limit_d2d1} hold. Suppose \ref{assn:EXvar_shrink} holds with parameters $d_{\pi,\text{EX}}$ and $\alpha$. Then,
\begin{align*}
		k_X(S) \pconv \frac{\alpha r^2 + d_{\pi,\text{EX}} r(1 + r + \alpha)}{\alpha ((1+r)(1+\alpha) + r) + d_{\pi,\text{EX}} (1 + r + \alpha)},
\end{align*}
an increasing function of $r$ for any $\alpha > -1$. Consequently, under these assumptions, $k_X(S)$ satisfies Properties 1 and 2.
\end{theorem}

Theorem \ref{thm:k_X_conv} shows that, when the covariates are exchangeable with small covariances, $k_X(S)$ satisfies both Properties 1 and 2. Under these same assumptions, both $r_X(S)$ and $r_Y(S)$ also satisfy Properties 1 and 2. This shows that these two properties alone are not sufficient to fully distinguish between all possible selection ratios. This is analogous to how many---but not all---estimators are consistent, and hence we have to look beyond their probability limits to further distinguish between them; for example, by studying their asymptotic distributions. The same situation arises here, and we conjecture that examining the asymptotic distribution of selection ratios that satisfy Properties 1 and 2 will raise further differences between them. We leave that to future work.

Note that if the covariates are uncorrelated, $k_X(s) = r_X(s)^2$ and hence $k_X(S) \xrightarrow{p} r$ as $K \rightarrow \infty$ in this case. Another open question concerns the behavior of $k_X(S)$ under more general conditions on the dgp than we give in Theorem \ref{thm:k_X_conv}. In section \ref{sec:empiricalDesignBased} we show that $k_X(S)$ generally performs well in an empirical dataset that does not satisfy the exchangeability assumption. This suggests that it likely performs well in wider classes of dgps, although we leave the full theoretical analysis to future work.

\section{Lower Level Conditions for Assumptions \ref{assn:pi_and_VarW}--\ref{assn:limited_dep}}\label{sec:AR-MA-Ex}

Our asymptotic analysis in section \ref{sec:designBasedAsymptotics} used various high level conditions on the dgp, assumptions \ref{assn:pi_and_VarW}, \ref{assn:gamma_and_VarW}, and \ref{assn:limited_dep}. In this section we give simple lower level conditions on the covariance structure of $\var(W^K)$ for this set of assumptions. Specifically, we show that if the components of $W^K$ (a) satisfy a moving average (MA) process, (b) satisfy an autoregressive (AR) process, (c) satisfy a factor model, or (d) are exchangeable with shrinking covariances, then assumptions \ref{assn:pi_and_VarW}, \ref{assn:gamma_and_VarW}, and \ref{assn:limited_dep} hold. 

\subsection{Moving Average Variance Structure}

We begin by considering covariates that have moving average type dependence.

\begin{Cassumption}[MA(1) Covariates] \label{assn:MAvar}
The following hold.
\begin{enumerate}
\item (MA variance matrix) For $i,j =1,\ldots,K$,
\[
	\cov(W_i^K,W_j^K) = \mathbbm{1}(i = j) + \rho \mathbbm{1}(|i-j| = 1)
\]
where $|\rho| < 1/2$.
	
\item (Bounded coefficients) There exists a $C < \infty$ such that $\sup_{i =1,\ldots,K} |\pi_i^K| \leq C/\sqrt{K}$ and $\sup_{i =1,\ldots,K} |\gamma_i^K | \leq C/\sqrt{K}$.
	
\item (Stable relative contribution to variance) $\lim_{K \to \infty} \frac{\sum_{i=1}^{K-1}\pi_i^K \pi_{i+1}^K}{\sum_{i=1}^K (\pi_i^K)^2} = d_{\pi,\text{MA}}$ and \\ $\lim_{K \to \infty} \frac{\sum_{i=1}^{K-1}\gamma_i^K \gamma_{i+1}^K}{\sum_{i=1}^K (\gamma_i^K)^2} = d_{\gamma,\text{MA}}$ where $d_{\pi,\text{MA}}, d_{\gamma,\text{MA}} \in \R$.
	
\item (Boundedness from below) There exists an $\epsilon > 0$ such that $\var(\sum_{i=1}^K \pi_i^K W_i^K) > \epsilon$ and $\var(\sum_{i=1}^K \gamma_i^K W_i^K) > \epsilon$ for all $K$.
\end{enumerate}
\end{Cassumption}

\ref{assn:MAvar}.1 says that covariates are correlated only if they are at most one index apart. The magnitude restriction on $\rho$ ensures that the covariates' variance matrix is positive semi-definite for all $K$. \ref{assn:MAvar}.2 says that none of the covariates is too important, in the sense that their maximal coefficient value is bounded. This bound also shrinks with $K$ to ensure that $\var(\sum_{i=1}^K \pi_i^K W_i^K)$ does not diverge. To understand \ref{assn:MAvar}.3, note that \ref{assn:MAvar}.1 implies
\begin{equation}\label{eq:MAtotalVariance}
	\var \left( \sum_{i=1}^K \pi_i^K W_i^K \right) = \sum_{i=1}^K (\pi_i^K)^2 + 2\rho \sum_{i=1}^{K-1}\pi_i^K \pi_{i+1}^K.
\end{equation}
So \ref{assn:MAvar}.3 says that the $\rho$ parameter and the coefficient sequence are such that the relative contribution of these two terms of the variance converges. This is a more primitive version of \ref{assn:pi_and_VarW}.3. Like that high level assumption, \ref{assn:MAvar}.3 says that we only consider sequences where this relative contribution is stable. Finally, \ref{assn:MAvar}.4 simply says that covariates are not degenerate in the treatment effect equation. 

Most of the selection ratios we consider in section \ref{sec:comparingratios} are invariant to the scale of $\pi$ or $\gamma$. This suggests that \ref{assn:MAvar}.2 can always be satisfied by simply scaling the coefficients down appropriately. However, a naive rescaling will violate \ref{assn:MAvar}.4. So the two assumptions are in fact restrictive.

\begin{proposition}\label{prop:MAvar}
Suppose assumptions \ref{assn:sampling_of_S}, \ref{assn:limit_d2d1}, and \ref{assn:MAvar} hold. Then assumptions \ref{assn:pi_and_VarW}, \ref{assn:gamma_and_VarW}, and \ref{assn:limited_dep} hold with $c_\pi = (1 + 2 \rho d_{\pi,\text{MA}})^{-1}$ and $c_\gamma = (1 + 2 \rho d_{\gamma,\text{MA}})^{-1}$.
\end{proposition}

This result can be extended to the $\text{MA}(q)$ for $q \geq 1$ case at the expense of a longer proof.

\subsection{Autoregressive Variance Structure}

Consider the following assumption.

\begin{Cassumption}[AR(1) Covariates] \label{assn:ARvar} The following hold.
\begin{enumerate}
\item (AR variance matrix) For $i,j =1,\ldots,K$, 
\begin{align*}
	\cov(W_i^K,W_j^K) = \rho^{|i-j|}
\end{align*}
where $|\rho| < 1$.
\item (Bounded coefficients) There exists a $C < \infty$ such that $\sup_{i =1,\ldots,K} |\pi_i^K| \leq C/\sqrt{K}$ and $\sup_{i =1,\ldots,K} |\gamma_i^K| \leq C/\sqrt{K}$.

\item (Stable relative contribution to variance) $\lim_{K \to\infty} \frac{\sum_{i,j:i \neq j} \pi_i^K \pi_j^K \rho^{|i-j|}}{\sum_{i=1}^K (\pi_i^K)^2} = d_{\pi,\text{AR}}$ and \\ $\lim_{K \to\infty} \frac{\sum_{i,j:i \neq j} \gamma_i^K \gamma_j^K \rho^{|i-j|}}{\sum_{i=1}^K (\gamma_i^K)^2} = d_{\gamma,\text{AR}}$ where $d_{\pi,\text{AR}}, d_{\gamma,\text{AR}} \in (-1,\infty]$.

\item (Boundedness from below) There exists an $\epsilon > 0$ such that $\var(\sum_{i=1}^K \pi_i^K W_i^K) > \epsilon$ and $\var(\sum_{i=1}^K \gamma_i^K W_i^K) > \epsilon$ for all $K$. 
\end{enumerate}
\end{Cassumption}

When $\rho = 0$ in \ref{assn:ARvar}.1, we adopt the convention that $0^0 = 1$. \ref{assn:ARvar}.1 says that the magnitude of the covariance between any two covariates decays as their indices become farther apart. The interpretation of the rest of \ref{assn:ARvar} is very similar to \ref{assn:MAvar}, which we discussed above. The main difference is \ref{assn:ARvar}.3, which arises since \ref{assn:ARvar}.1 yields the following structure on the variance of the sum:
\[
	\var \left( \sum_{i=1}^K \pi_i^K W_i^K \right)
	=
	\sum_{i=1}^K (\pi_i^K)^2 + \sum_{i,j:i \neq j} \pi_i^K \pi_j^K \rho^{|i-j|},
\]
reflecting the different covariance structure for the AR dgp compared to the MA dgp.

\begin{proposition}\label{prop:ARvar}
Suppose assumptions \ref{assn:sampling_of_S}, \ref{assn:limit_d2d1}, and \ref{assn:ARvar} hold. Then assumptions \ref{assn:pi_and_VarW}, \ref{assn:gamma_and_VarW}, and \ref{assn:limited_dep} hold with $c_\pi = (1 +  d_{\pi,\text{AR}})^{-1}$ and $c_\gamma = (1 + d_{\gamma,\text{AR}})^{-1}$.
\end{proposition}

This result can be extended to the $\text{AR}(p)$ for $p \geq 1$ case at the expense of a longer proof.

\subsection{Factor Models} 

Consider the following assumption.

\begin{Cassumption}[Factor Model Covariates] \label{assn:EXvar} The following hold.
\begin{enumerate}
\item (Factor structure) There is an integer $R > 0$ such that
\[
	W^K = \Lambda^K F + E^K
\]
where $\Lambda^K \coloneqq \begin{bmatrix} \lambda_1^K & \cdots & \lambda_R^K \end{bmatrix}$ is a $K \times R$ matrix of (constant) factor loadings, $F$ is an $R$-vector of factors, $\var(F) = \textbf{I}_R$, $\var(E^K) = \sigma^2_E \cdot \textbf{I}_K$, and $\cov(E^K,F) = \textbf{0}_{K \times R}$. 

\item (Bounded coefficients) There exists a $C < \infty$ such that $\sup_{i =1,\ldots,K} |\pi_i^K| \leq C/K$ and \\ $\sup_{i =1,\ldots,K} |\gamma_i^K| \leq C/K$. 

\item (Bounded factor loadings) There exists a $\overline{\lambda}$ such that $\|\Lambda^K\|_{\max} \leq \overline{\lambda}$ for all $K$, where $\|\cdot\|_{\max}$ denotes the maximum of the absolute value of all elements.

\item (Boundedness from below) There exists an $\epsilon > 0$ such that $\var(\sum_{i=1}^K \pi_i^K W_i^K) > \epsilon$ and $\var(\sum_{i=1}^K \gamma_i^K W_i^K) > \epsilon$ for all $K$. 
\end{enumerate}
\end{Cassumption}

\ref{assn:EXvar}.1 imposes a factor structure on the covariates' variance matrix. The special case of exchangeable covariates with covariance $\rho \in (0,1)$ obtains by letting $R=1$, $\Lambda^K = \iota_K \sqrt{\rho}$ and $\sigma^2_E = 1-\rho$. (The exchangeable case with covariance $\rho = 0$ is included as a special case of either the earlier MA(1) or AR(1) assumptions.) \ref{assn:EXvar}.2 is similar to the bounded coefficients assumptions from the MA and AR cases, except that the bound is now of order $K^{-1}$ instead of $K^{-1/2}$. In the exchangeable case, for example, $\var(\sum_{i=1}^K \pi_i^K W_i^K)$ is of order $K^2 \cdot O(\sup_{i =1,\ldots,K} | \pi_i^K |^2)$. Hence we need the largest $\pi_i^K$ values to shrink at least at the $K^{-1}$ rate to keep the variance finite. \ref{assn:EXvar}.3 requires the factor loadings to be uniformly bounded. 
  
\begin{proposition}\label{prop:EXvar}
Suppose assumptions \ref{assn:sampling_of_S}, \ref{assn:limit_d2d1}, and \ref{assn:EXvar} hold. Then assumptions \ref{assn:pi_and_VarW}, \ref{assn:gamma_and_VarW}, and \ref{assn:limited_dep} hold with $c_\pi = c_\gamma = 0$.
\end{proposition}

For one of our results in section \ref{sec:designBasedAsymptotics} we use the following low level conditions, which impose that the covariates are exchangeable with a shrinking covariance. We state these assumptions separately from \ref{assn:EXvar}---which includes exchangeable covariates with non-shrinking covariances as a special case---because the shrinking covariance requires a different scaling of the coefficients $\pi$ and $\gamma$.

\begin{Cassumption}[Exchangeable Covariates with Shrinking Covariances] \label{assn:EXvar_shrink} The following hold.
\begin{enumerate}
\item (Exchangeable variance matrix) For $i,j =1,\ldots,K$, 
\begin{align*}
	\cov(W_i^K,W_j^K) = \1(i = j) + \frac{\alpha}{K} \1(i \neq j)
\end{align*}
where $\alpha \in (-1,K)$.

\item (Bounded coefficients) There exists a $C < \infty$ such that $\sup_{i =1,\ldots,K} |\pi_i^K| \leq C/\sqrt{K}$ and $\sup_{i =1,\ldots,K} |\gamma_i^K| \leq C/\sqrt{K}$.

\item (Stable relative contribution to variance) $\lim_{K\to\infty}\frac{K \sum_{i=1}^K (\pi_i^K)^2}{(\sum_{i=1}^K \pi_i^K)^2} = d_{\pi,\text{EX}}$ and \\ $\lim_{K\to\infty}\frac{K \sum_{i=1}^K (\gamma_i^K)^2}{(\sum_{i=1}^K \gamma_i^K)^2} = d_{\gamma,\text{EX}}$ where $d_{\pi,\text{EX}}, d_{\gamma,\text{EX}} \in [1,\infty]$.

\item (Boundedness from below) There exists an $\epsilon > 0$ such that $\var(\sum_{i=1}^K \pi_i^K W_i^K) > \epsilon$ and $\var(\sum_{i=1}^K \gamma_i^K W_i^K) > \epsilon$ for all $K$. 
\end{enumerate}
\end{Cassumption}

\begin{proposition}\label{prop:EXvar_shrink}
Suppose assumptions \ref{assn:sampling_of_S}, \ref{assn:limit_d2d1}, and \ref{assn:EXvar_shrink} hold. Then assumptions \ref{assn:pi_and_VarW}, \ref{assn:gamma_and_VarW}, and \ref{assn:limited_dep} hold with $c_\pi = d_{\pi,\text{EX}}/(d_{\pi,\text{EX}} + \alpha)$ and $c_\gamma = d_{\gamma,\text{EX}}/(d_{\gamma,\text{EX}} + \alpha)$.
\end{proposition}

\section{An Empirical Comparison of Selection Ratios}\label{sec:empiricalDesignBased}

In section \ref{sec:designBasedAsymptotics} we used asymptotics to approximate and compare the covariate sampling distributions of various selection ratios. That analysis raises several questions: (1) How accurate are the asymptotic approximations? (2) How restrictive are the regularity conditions used to obtain these approximations? (3) Is the non-convergence of $\delta_\text{resid}(S)$ to 1 under equal selection that we showed in Theorem \ref{corr:delta_r_noconv} typical in any sense, or is it a rare knife-edge case that can be safely ignored? We next address these concerns by comparing the selection ratios' covariate sampling distributions in an exact, non-asymptotic setting based on an empirically calibrated dgp using data from \citet[\emph{Econometrica}]{BFG2020}. We find similar patterns as in the asymptotic analysis: For example, the distribution of $r_X(S)$ is tightly centered around 1 under equal selection, and satisfies a natural monotonicity property. In contrast, the distribution of $| \delta_\text{resid}(S) |$ is highly variable under equal selection, and follows a \emph{reverse} monotonicity property: It is \emph{larger} when most covariates \emph{are} observed and it is \emph{smaller} when most covariates are \emph{not} observed. This is the exact opposite pattern from its nominal interpretation, where large values of $| \delta_\text{resid}(S) |$ are supposed to represent cases where the unobserved variables are substantially more important than observed variables. Overall, our findings show that the asymptotic results reflect properties of the sensitivity parameters that arise in real data sets with a finite number of covariates.

\subsection*{Setup}

Since our goal is simply to create a realistic joint distribution of $(Y,X,W)$ from which we can sample covariates, we omit a description of the dataset and refer interested readers to \cite{BFG2020} for details. Here we provide the minimum detail needed for replication. We let $Y$ denote Republican vote share, $X$ total frontier experience. For the covariates $W$, we pick the 10 covariates that the authors use in their baseline specifications (their Table 3) as well as 12 of the 13 additional variables that the authors consider in their appendix. We omit one variable (contemporary population density) so that the total number of covariates $K$ is even; similar results obtain if we include it but then equal selection is not exactly satisfied. This makes for a total of 22 covariates in the vector $W$. As in the authors' sensitivity analysis, we also treat state fixed effects as variables that are not used for calibration, and hence we project them out from all other variables. We thus let $(Y,X,W)$ have the empirical distribution of all these residualized variables in the data. Using this joint distribution of $(Y,X,W)$, $(\beta_\text{long},\gamma)$ are defined as the estimated coefficients on $(X,W)$ from OLS of $Y$ on $(1,X,W)$. Likewise, $\pi$ is the vector of estimated coefficients on $W$ from OLS of $X$ on $(1,W)$. For this exercise, we therefore treat these estimates as the population values of these coefficients.

\subsection*{Results}

Having specified the distribution of $(Y,X,W)$, we can now compute the exact covariate sampling distributions of any sensitivity parameter $\theta(S)$ by fixing a number $d_1$ of covariates to observe from the $K=22$ total covariates, computing $\theta(s)$ for all possible values of $s$ with $\sum_{k=1}^K s_k = d_1$, and then putting equal weight on all of these values.

\begin{figure}[th]
    \centering
    \begin{tabular}{
    >{\centering\arraybackslash} m{0.01\textwidth}
    >{\centering\arraybackslash} m{0.29\textwidth}
    >{\centering\arraybackslash} m{0.29\textwidth}
    >{\centering\arraybackslash} m{0.29\textwidth}
    }
    & {\footnotesize More Observed (19 of 22)} & {\footnotesize Equal Selection (11 of 22)} & {\footnotesize More Unobserved (3 of 22)} \\

    \rotatebox{90}{{\footnotesize $r_X (S)$}} &
    \includegraphics[width=0.26\textwidth]{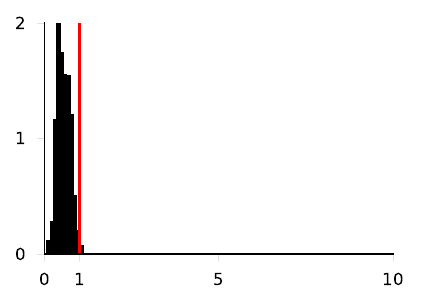} &
    \includegraphics[width=0.26\textwidth]{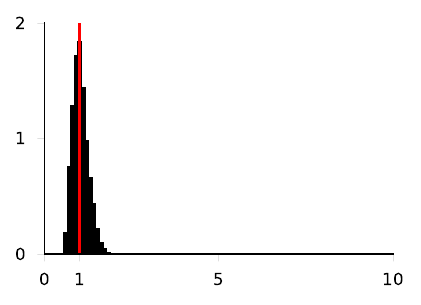} &
    \includegraphics[width=0.26\textwidth]{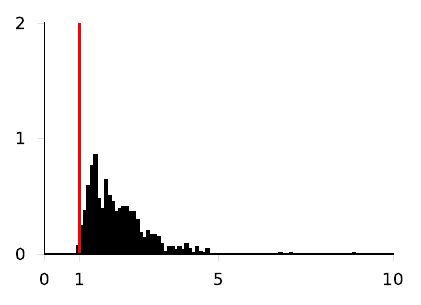} \\

    \rotatebox{90}{{\footnotesize $|\delta_{\text{resid}} (S)|$}}&
    \includegraphics[width=0.26\textwidth]{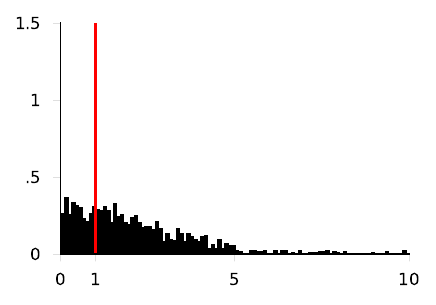} &
    \includegraphics[width=0.26\textwidth]{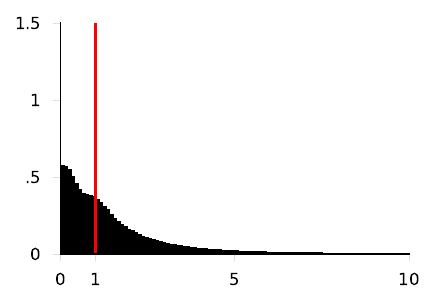} &
    \includegraphics[width=0.26\textwidth]{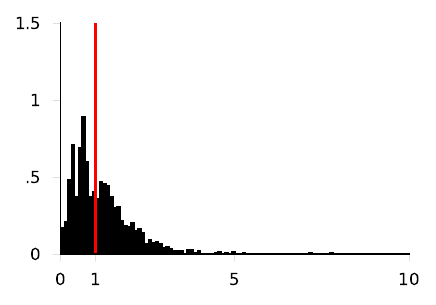} \\

    \end{tabular}
    \\

    \caption{Covariate Sampling Distributions. Top row: Distributions of $r_X(S)$. Bottom row: Distributions of $| \delta_\text{resid}(S) |$. Left column: $d_1 = 19$ of 22 covariates observed. Middle: $d_1 = 11$ of 22 covariates observed. Right: $d_1 = 3$ of 22 covariates observed.\label{fig:designBasedEmpirical}}
\end{figure}

We compute these covariate sampling distributions for the six sensitivity parameters which we analyzed asymptotically in section \ref{sec:designBasedAsymptotics}. For brevity, we focus on $r_X(S)$ and $| \delta_\text{resid}(S) |$ here; Appendix \ref{sec:empiricalAppendix} shows the results for the other four parameters. Figure \ref{fig:designBasedEmpirical} plots histograms of the covariate sampling distributions for three choices of $d_1$, from left to right: More covariates observed ($d_1 = 19$), Equal selection ($d_1 = 11$), and More covariates unobserved ($d_1 = 3$). First consider the top row, which shows the distributions of $r_X(S)$. From the middle plot, we see an empirical analog of Property 1: The distribution of $r_X(S)$ is tightly centered at 1 under equal selection. From the left and right plots we see an empirical analog of Property 2: When most covariates are observed, $r_X(S)$ is mostly below 1, indicating that the unobserved covariates are not as important as the observed covariates. Likewise, when most covariates are not observed, $r_X(S)$ is mostly above 1, indicating that the unobserved covariates are more important than the observed covariates.

Next consider the bottom row, which shows the distributions of $| \delta_\text{resid}(S) |$. From the middle plot, which shows the equal selection case, we see that the distribution is very spread out, and does not appear to have any discernible concentration near one. From the left and right plots, we see that $| \delta_\text{resid}(S) |$ also does \emph{not} satisfy any kind of empirical analog of the monotonicity Property 2. In fact, the distribution shifts closer to zero as we increase the number of unobserved covariates.

\begin{table}[t]
\caption{How Often is the Sensitivity Parameter Below the Equal Selection Benchmark?\label{table:designTable1}}
\centering
\resizebox{0.70\textwidth}{!}{
\begin{tabular}{l c c c}
\toprule
& \multicolumn{3}{c}{Number of Observed Covariates} \\
& More Observed & Equal Selection & More Unobserved \\
& 19 of 22& 11 of 22& 3 of 22\\
\midrule
$\mathbb{P}_S(r_X(S) \le 1)$
&  99.2\%&  50.0\%&  0.8\% \\
$\mathbb{P}_S(r_Y(S) \le 1)$
&  99.4\%&  50.0\%&  0.6\%\\
$\mathbb{P}_S(k_X (S) \le 1)$
&  99.9\%
&  51.8\%
&  0.0\%
 \\
$\mathbb{P}_S(|\delta_\text{orig}(S)| \le 1)$
&  47.1\%
&  50.0\%
&  52.9\%
 \\
$\mathbb{P}_S(|\delta_\text{ACET}(S)| \le 1)$
&  42.2\%
&  50.0\%
&  57.8\%
 \\
$\mathbb{P}_S(|\delta_\text{resid}(S)| \le 1)$
&  28.7\%
&  45.8\%
&  49.5\%
 \\
\bottomrule
\end{tabular}
}
\end{table}

This reverse monotonicity property can also be seen by considering various summary statistics for these distributions. Table \ref{table:designTable1} shows the proportion of realizations of the sensitivity parameters that are below the equal selection benchmark of 1. For $r_X(S)$ this happens 99.2\% of the time when most covariates are observed, exactly 50\% of the time under equal selection, and only 0.8\% of the time when most covariates are not observed. In contrast, $| \delta_\text{resid}(S) |$ is smaller than 1 only 28.7\% of the time when most covariates are observed, and this \emph{increases} up to 49.5\% of the time when most covariates are unobserved. This is the opposite pattern we would expect from the nominal interpretation of $\delta_\text{resid}(S)$, whereby values smaller than 1 are supposed to indicate that the unobserved variables are less important than the observed variables. Table \ref{table:designTable1} also shows the results for the other four sensitivity parameters we consider. Again we see that these finite sample properties mirror their asymptotic properties.

\def\mystrut{\rule{0pt}{1.25\normalbaselineskip}}
\begin{table}[t]
\centering
\SetTblrInner[talltblr]{rowsep=0pt}
\resizebox{.95\textwidth}{!}{
\begin{talltblr}[
    caption = {Covariate Sampling Distribution Summary Statistics.\label{table:designTable2}},
    remark{Note} = {These summary statistics correspond to the full distributions displayed in Figure \ref{fig:designBasedEmpirical}.},
  ]{p{0.35\textwidth} | *{3}{>{\centering \arraybackslash}p{0.17\textwidth}}}
\toprule
\rule{0pt}{1.25\normalbaselineskip}
\rule{0pt}{\normalbaselineskip}
& \SetCell[c=3]{c} Number of Observed Covariates \\
& More Observed & Equal Selection & More Unobserved \\
& 19 of 22& 11 of 22& 3 of 22\\
Number of Support Points, ${K \choose d_1}$ & 1,540 & 705,432 & 1,540 \\
\hline
\SetCell[c=4]{l} Panel A. Distribution of $r_X(S)$ \\
\hline
\ \ \ \ Minimum &  0.055 &  0.438 &  0.904 \\[2pt]
\ \ \ \ 25th Percentile &  0.396 &  0.861 &  1.447 \\[2pt]
\ \ \ \ Median &  0.525 &  1.000 &  1.906 \\[2pt]
\ \ \ \ 25th Percentile &  0.691 &  1.161 &  2.522 \\[2pt]
\ \ \ \ Maximum &  1.107 &  2.285 &  18.316 \\[2pt]
\ \ \ \ Mean &  0.542 &  1.024 &  2.158 \\[2pt]
\ \ \ \ Standard Deviation &  0.190 &  0.226 &  1.147 \\[5pt]
\hline
\SetCell[c=4]{l} Panel B. Distribution of $|\delta_{\text{resid}}(S)|$ \\
\hline
\ \ \ \ Minimum &  0.010 &  0.000 &  0.002 \\[2pt]
\ \ \ \ 25th Percentile &  0.879 &  0.469 &  0.580 \\[2pt]
\ \ \ \ Median &  1.774 &  1.118 &  1.017 \\[2pt]
\ \ \ \ 25th Percentile &  3.102 &  2.283 &  1.650 \\[2pt]
\ \ \ \ Maximum & 79.732& 10,346,687& 176.641\\[2pt]
\ \ \ \ Mean &  2.402 &  20.365 &  1.583 \\[2pt]
\ \ \ \ Standard Deviation & 3.065& 12,335& 5.435\\[5pt]
\hline
\end{talltblr}
}
\end{table}

Similar findings can be seen in Table \ref{table:designTable2}, which shows additional summary statistics for two of these distributions (Appendix Table \ref{table:designAppendixTable1} shows results for the other four distributions). When most covariates are observed, the maximum value of $r_X(S)$ is 1.107, with a median value of 0.525. Thus the $r_X(S)$ parameter almost always correctly reports that the unobserved covariates are less important than the observed covariates. Likewise, when most covariates are not observed, the minimum value is 0.904, with a median value of 1.906. Again, the $r_X(S)$ parameter almost always correctly reports that the unobserved covariates are more important than the observed covariates. In contrast, $| \delta_\text{resid}(S) |$ is mostly \emph{large} when most covariates are observed, with a median of 1.774, a 25th percentile of 0.879, and a standard deviation of 3.06. When most covariates are not observed, this distribution shifts down, with the 25th, 50th, and 75th percentiles all decreasing---the opposite direction we would expect from the nominal interpretation of $| \delta_\text{resid}(S) |$. Under equal selection, $| \delta_\text{resid}(S) |$ is somewhat biased upwards, with a median of 1.118, but it is also very spread out, with a very large upper tail and a standard deviation of about 12,300. In contrast, $r_X(S)$ is tightly distributed exactly around 1 under equal selection.

\section{Conclusion}\label{sec:frameworkImplications}

Since the original work of \cite{AltonjiElderTaber2005}, empirical researchers now regularly discuss the robustness of their results to omitted variables in relative terms, comparing the magnitudes of selection on observables with unobservables. In particular, they use the value 1 as an important benchmark for their sensitivity parameter, nominally interpreted as meaning ``equal selection'' (e.g., this is true for all of the empirical papers published in top 5 journals from 2019--2021 in the survey in Appendix A of \citealt{MastenPoirier2024}). From our results in section \ref{sec:designBasedAsymptotics}, we see that the value 1 is usually \emph{not} the correct benchmark of equal selection for the most popularly used sensitivity parameter, $\delta_\text{resid}$. This implies that researchers who wish to compare values of their sensitivity parameter to the benchmark of equal selection will generally draw the wrong conclusions about robustness if they use 1 as the benchmark. 

One possible response to this result is to continue to use $\delta_\text{resid}$, but to change the benchmark value. The problem with this approach is that the correct benchmark of equal selection generally depends on the unknown dgp. Equation \eqref{eq:deltaresid_limit} gives a simple example of how this limit depends on various features of the unknown dgp when the covariates are exchangeable. Hence the correct benchmark is currently unknown. Moreover, even if the correct benchmark were known, we observed that $\delta_\text{resid}$ can exhibit a counterintuitive reverse monotonicity property. In contrast, we have shown that alternative sensitivity parameters can achieve a limiting value that does \emph{not} depend on the dgp under equal selection (e.g., Theorem  \ref{thm:rx_conv}), and thus provides a feasible benchmark. This is the property we called consistency. In particular, the sensitivity parameters in both \cite{CinelliHazlett2020} and our companion paper \cite{DMP2023v5} are consistent, and also satisfy the monotonicity in selection property we introduced. Based on this criterion, one immediate practical implication of our results is that researchers should consider using sensitivity analyses beyond those involving $\delta_\text{resid}$. Hence we recommend that applied researchers use either of those methods, or any other methods that satisfy those two properties.

We showed that consistency alone is sufficient to rule out some sensitivity parameters. We also showed that some parameters are consistent, but do not satisfy monotonicity in selection, such as $\delta_\text{orig}$, so that this second property is also useful for distinguishing between sensitivity parameters. Nonetheless, there are multiple parameters that satisfy both requirements. We conjecture that further refinements can be obtained by examining the distributional features of covariate sampling distributions, rather than just their probability limits, although it is unlikely that future work will ever recommend a single unique robustness check. Finally, as discussed in section \ref{sec:intro}, our framework can be straightforwardly extended to help researchers narrow down the set of robustness checks in other settings as well, including assessing the parallel trends assumption in difference-in-differences analyses or the exogeneity assumption in instrumental variable models.

\singlespacing
\bibliographystyle{econometrica}
\bibliography{BadControls_paper}

\appendix
\section{Proofs for Section \ref{sec:designBasedAsymptotics}}\label{sec:proofForDesignBasedStuff}

We use $\Lin(A \mid 1, B)$ to denote the linear projection of a random vector $A$ on a random vector $B$ and a constant. Let $[K] \coloneqq \{1,\ldots,K\}$. We often use the following lemma, which is Theorem B(ii) on page 99 of \cite{ScottWu1981}.

\begin{lemma}[Finite Population LLN]\label{lemma:LLN}
Consider a sequence of non-random vectors $\xi^K \coloneqq (\xi_1^K,\ldots,\xi_K^K)$. Suppose $S$ satisfies Assumption \ref{assn:sampling_of_S}. Then
\begin{enumerate}
	\item ${\var}_S(\sum_{i=1}^K S_i \xi_i^K) = \frac{d_1 d_2}{K(K-1)}\sum_{i=1}^K \left(\xi_i^K - \frac{1}{K} \sum_{j=1}^K \xi_j^K\right)^2$.
	\item If ${\var}_S(\frac{1}{d_1} \sum_{i=1}^K S_i \xi_i^K) \rightarrow 0$ as $d_1,K \to \infty$, then
		\[ \frac{1}{d_1} \sum_{i=1}^K S_i \xi_i^K - \frac{1}{K} \sum_{i=1}^K \xi_i^K \pconv 0\]
as $d_1,K \rightarrow\infty$. 
\end{enumerate}
\end{lemma}
We obtain as corollaries the following two results: (1) ${\var}_S(\sum_{i=1}^K S_i \xi_i^K) \leq \frac{d_1 d_2}{K(K-1)}\sum_{i=1}^K \left(\xi_i^K\right)^2$ and (2) ${\var}_S(\frac{1}{d_1} \sum_{i=1}^K S_i \xi_i^K) \rightarrow 0$ as $d_1,K \to \infty$ when $\sup_{K: K \geq 1} \sup_{i \in [K]}|\xi_i^K| \leq C$ for some $C \in \R$.

In the following proofs, we leave the dependence of the covariates and their associated coefficients on $K$ implicit to lighten the notation.

\begin{proof}[Proof of Theorem \ref{thm:rx_conv}]
Note that $\Exp_S[S_i] = d_1/K$ and that $\Exp_S[S_i S_j] = \Prob_S(S_i = 1|S_j = 1)\Prob_S(S_j = 1) = d_1(d_1-1)/(K(K-1))$ whenever $i \neq j$ by Assumption \ref{assn:sampling_of_S}. First, observe that
\begin{align*}
	&\Exp_S\left[\pi_1(S)'\var(W_1(S))\pi_1(S)\right] = \Exp_S\left[\sum_{i=1}^K \sum_{j=1}^K S_i S_j \cov(\pi_i W_i,\pi_j W_j)\right]\\
	&= \Exp_S\left[\sum_{i=1}^K S_i \var(\pi_i W_i)\right] + \Exp_S\left[\sum_{i,j:i\neq j} S_i S_j \cov(\pi_i W_i,\pi_j W_j)\right]\\
	&= \frac{d_1}{K} \sum_{i=1}^K \var(\pi_i W_i) + \frac{d_1(d_1-1)}{K(K-1)}\sum_{i,j:i\neq j}\cov(\pi_i W_i,\pi_j W_j)\\
	&= \frac{d_1(d_1-1)}{K(K-1)}\left(\sum_{i=1}^K \var(\pi_i W_i) + \sum_{i,j:i\neq j}^K\cov(\pi_i W_i,\pi_j W_j)\right) \\
	&\qquad + \left(\frac{d_1}{K} - \frac{d_1(d_1-1)}{K(K-1)}\right)\sum_{i=1}^K \var(\pi_i W_i)\\
	&= \var\left(\pi'W\right)\left(\frac{d_1(d_1-1)}{K(K-1)} + \frac{d_1}{K}\frac{d_2}{K-1}\var(\pi'W)^{-1}\sum_{i=1}^K \var(\pi_i W_i)\right)\\
	&= \var\left(\pi'W\right)\left(\frac{1}{(1+r)^2} + \frac{r}{(1+r)^2}c_\pi + o(1)\right)\\
	&= \var\left(\pi'W\right)\frac{1 + rc_\pi}{(1+r)^2} + o(1).
\end{align*}
The sixth equality follows from assumptions \ref{assn:limit_d2d1}, \ref{assn:pi_and_VarW}.1, and \ref{assn:pi_and_VarW}.3. Using a symmetric argument, we have
\begin{align*}
	&\Exp_S\left[\pi_2(S)'\var(W_2(S))\pi_2(S)\right] \\
	&= \Exp_S\left[\sum_{i=1}^K \sum_{j=1}^K (1-S_i)(1- S_j) \cov(\pi_i W_i,\pi_j W_j)\right]\\
	&= \var\left(\pi'W\right)\left(\frac{d_2(d_2-1)}{K(K-1)} + \frac{d_2}{K}\frac{d_1}{K-1}\var(\pi'W)^{-1}\sum_{i=1}^K \var(\pi_i W_i)\right)\\
	&= \var\left(\pi'W\right)\left(\frac{r^2}{(1+r)^2} + \frac{r}{(1+r)^2}c_\pi + o(1)\right)\\
	&= \var\left(\pi'W\right)\frac{r(r+c_\pi)}{(1+r)^2} + o(1).
\end{align*}

Second, we show that 
\begin{align*}
	\pi_l(S)'\var(W_l(S))\pi_l(S) - \Exp_S[\pi_l(S)'\var(W_l(S))\pi_l(S)] &= o_p(1)
\end{align*}
for $l = 1,2$. This follows from Chebyshev's inequality and from
\begin{align*}
	&\Exp_S\left[|\pi_l(S)'\var(W_l(S))\pi_l(S) - \Exp_S[\pi_l(S)'\var(W_l(S))\pi_l(S)]|^2\right] \\
	&= {\var}_S(\pi_l(S)'\var(W_l(S))\pi_l(S))\\
	&={\var}_S\left(\sum_{i=1}^K \sum_{j=1}^K S_i^{2-l}(1-S_i)^{l-1} S_j^{2-l}(1-S_j)^{l-1} \cov(\pi_i W_i,\pi_j W_j)\right) \rightarrow 0,
\end{align*}
which is implied by Assumption \ref{assn:pi_and_VarW}.2 and by examining the $l = 1,2$ cases separately. 

Third, we combine the convergence of the numerators and denominators to obtain
\begin{align*}
	r_X(S)^2 &= \frac{\pi_2(S)'\var(W_2(S))\pi_2(S)}{\pi_1(S)'\var(W_1(S))\pi_1(S)}\\
	&= \frac{\Exp_S[\pi_2(S)'\var(W_2(S))\pi_2(S)] + o_p(1)}{\Exp_S[\pi_1(S)'\var(W_1(S))\pi_1(S)] + o_p(1)}\\
	&= \frac{\var(\pi'W)r(r+c_\pi)/(1+r)^2 + o(1) + o_p(1)}{\var(\pi'W)(1 + rc_\pi)/(1+r)^2 + o(1) + o_p(1)}\\
	&= \frac{r(r+c_\pi) + (1+r)^2\var(\pi'W)^{-1} \cdot o_p(1)}{(1 + rc_\pi) + (1 + r)^2 \var(\pi'W)^{-1} \cdot o_p(1)}\\
	&\pconv \frac{r(r+c_\pi)}{1+rc_\pi}.
\end{align*}
The result follows from the continuous mapping theorem with mapping $x \mapsto \sqrt{x}$.
\end{proof}

\begin{proof}[Proof of Theorem \ref{thm:delta_orig_conv}]
First, we study the denominators of the two fractions in $\delta_\text{orig}(S)$. Similarly to the proof of Theorem \ref{thm:ry_conv}, we have that
\begin{align*}
	\gamma_1(S)'\var(W_1(S))\gamma_1(S) &= \var\left(\gamma'W\right)(1 + r c_\gamma)/(1+r)^2 + o_p(1)\\
	\gamma_2(S)'\var(W_2(S))\gamma_2(S) &= \var\left(\gamma'W\right)r(r+c_\gamma)/(1+r)^2 + o_p(1)
\end{align*}
under the theorem assumptions.

Second, we show that 
\begin{align*}
	\Exp_S\left[\cov(X,\gamma_1(S)'W_1(S))\right] &=  \Exp_S[\cov(X,\sum_{i=1}^K S_i \gamma_i W_i)]\\
	&= \cov(X,\sum_{i=1}^K\Exp_S[S_i] \gamma_i W_i)\\
	&= \cov(X,\gamma'W)d_1/K\\
	&= \cov(X,\gamma'W) \left(1/(1+r) + o(1)\right)\\
	&= \cov(X,\gamma'W)/(1+r) + o(1),
\end{align*}
where the fourth equality follows from Assumption \ref{assn:limit_d2d1}, and the fifth from $\cov(X,\gamma'W)$ being bounded.
Similarly
\begin{align*}
	\Exp_S\left[\cov(X,\gamma_2(S)'W_2(S))\right] &= \cov(X,\sum_{i=1}^K(1 - \Exp_S[S_i]) \gamma_i W_i)\\
	&= \cov(X,\gamma'W)d_2/K\\
	&= \cov(X,\gamma'W) r/(1+r) + o(1).
\end{align*}

We can also see that
\begin{align*}
	{\var}_S(\cov(X,\gamma_1(S)'W_1(S))) &= {\var}_S( \sum_{i=1}^K S_i \cov(X,\gamma_i W_i))\\
	&= \frac{d_1 d_2}{K(K-1)} \sum_{i=1}^K \left(\cov(X,\gamma_i W_i) - \frac{1}{K}\sum_{j=1}^K \cov(X,\gamma_j W_j)\right)^2\\
	&\leq \frac{d_1 d_2}{K(K-1)} \sum_{i=1}^K \cov(X,\gamma_i W_i)^2\\
	&\rightarrow 0
\end{align*}
by Lemma \ref{lemma:LLN}, Assumption \ref{assn:limit_d2d1}, and Assumption \ref{assn:limited_dep}. Similarly, 
\begin{align*}
	{\var}_{S}(\cov(X,\gamma_2(S)'W_2(S))) &= {\var}_{S}(\cov(X,\gamma'W) - \cov(X,\gamma_1(S)'W_1(S)))\\
	&= {\var}_{S}(\cov(X,\gamma_1(S)'W_1(S)))\\
	&\rightarrow 0
\end{align*}
by the previous derivation. Therefore, by Chebyshev's inequality,
\begin{align*}
	\cov(X,\gamma_l(S)'W_l(S)) - \Exp_S\left[\cov(X,\gamma_l(S)'W_l(S))\right] = o_p(1),
\end{align*}
for $l = 1,2$.

Third, we combine the convergence of the numerators and denominators to obtain
\begin{align*}
	\delta_\text{orig}(S) &= \frac{\cov(X,\gamma_2(S)'W_2(S))}{\var(\gamma_2(S)'W_2(S))}\bigg/\frac{\cov(X,\gamma_1(S)'W_1(S))}{\var(\gamma_1(S)'W_1(S))}\\
	&= \frac{\Exp_S\left[\cov(X,\gamma_2(S)'W_2(S))\right] + o_p(1)}{\Exp_S\left[\gamma_2(S)'\var(W_2(S))\gamma_2(S)\right] + o_p(1)}\bigg/\frac{\Exp_S\left[\cov(X,\gamma_1(S)'W_1(S))\right] + o_p(1)}{\Exp_S\left[\gamma_1(S)'\var(W_1(S))\gamma_1(S)\right] + o_p(1)}\\
	&= \frac{\cov(X,\gamma'W)r/(1+r)  + o_p(1)}{\var\left(\gamma'W\right)r(r + c_\gamma)/(1+r)^2 + o_p(1)}\bigg/\frac{ \cov(X,\gamma'W)/(1+r) + o_p(1)}{\var\left(\gamma'W\right)(1 + rc_\gamma)/(1+r)^2 + o_p(1)}\\
	&= \frac{r}{r(r + c_\gamma)}\bigg/\frac{1}{1 + r c_\gamma} + o_p(1)\\
	&\pconv \frac{1 + rc_\gamma}{r + c_\gamma},
\end{align*}
where the fourth equality follows from $\var(\gamma'W)$ and $\cov(X,\gamma'W)$ being bounded and bounded away from 0.
\end{proof}

\begin{proof}[Proof of Theorem \ref{thm:delta_resid}]
Before computing the limit of $\delta_\text{resid}(S)$, we first rewrite its four components: 
\begin{align*}
	\cov(X,\gamma_2(s)'W_2(s)^{\perp W_1(s)}) &= \cov(\pi_1(s)'W_1(s) + \pi_2(s)'W_2(s) + X^{\perp W},\gamma_2(s)'W_2(s)^{\perp W_1(s)})\\
	&= \gamma_2(s)'\var(W_2(s)^{\perp W_1(s)})\pi_2(s)\\[0.5em]
	\var(\gamma_2(s)'W_2(s)^{\perp W_1(s)}) &= \gamma_2(s)'\var(W_2(s)^{\perp W_1(s)})\gamma_2(s)
\end{align*}
and using $\phi(s)' W_1(s) = \Lin(\gamma_2(s)'W_2(s) \mid 1,W_1(s)) = \gamma_2(s)'(W_2(s) - W_2(s)^{\perp W_1(s)}) + \text{const.}$ yields
\begin{align*}
	\cov(X,(\gamma_1(s) + \phi(s))'W_1(s)) 
	&= \cov(X,\gamma_1(s)'W_1(s) + \gamma_2(s)'W_2(s) - \gamma_2(s)'W_2(s)^{\perp W_1(s)})\\
	&= \gamma'\var(W)\pi - \gamma_2(s)'\var(W_2(s)^{\perp W_1(s)})\pi_2(s)\\[0.5em]
	\var((\gamma_1(s) + \phi(s))'W_1(s)) 
	&= \var(\gamma'W - \gamma_2(s)'W_2(s)^{\perp W_1(s)})\\
	&= \gamma'\var(W)\gamma - \gamma_2(s)'\var(W_2(s)^{\perp W_1(s)})\gamma_2(s).
\end{align*}
These components depend on the matrix $\var(W_2(s)^{\perp W_1(s)})$, which we now derive. From the theorem assumptions, $\var(W) = \rho \iota_K \iota_K'  + (1-\rho)\textbf{I}_K$ for some $\rho \in (0,1)$. First observe that for all $s$ such that $\sum_{i=1}^K s_i = d_1$,
\begin{align*}
	\Lin(W_2(s) \mid 1,W_1(s)) &= \text{const.} + \cov(W_2(s),W_1(s))\var(W_1(s))^{-1}W_1(s)\\
	&= \text{const.} + \rho\iota_{d_2}\iota_{d_1}'\left(\rho \iota_{d_1}\iota_{d_1}' + (1-\rho)\textbf{I}_{d_1}\right)^{-1}W_1(s)\\
	&= \text{const.} + \frac{\rho}{(d_1-1)\rho + 1} \iota_{d_2} \sum_{i=1}^K s_i W_i.
\end{align*}
We then have that
\begin{align*}
	\var(\Lin(W_2(s) \mid 1,W_1(s))) &= \var\left(\frac{\rho}{(d_1-1)\rho + 1} \iota_{d_2} \iota_{d_1}' W_1(s)\right)\\
	&= \frac{\rho^2}{((d_1-1)\rho + 1)^2} \iota_{d_2} \iota_{d_1}' \left(\rho \iota_{d_1}\iota_{d_1}' + (1-\rho)\textbf{I}_{d_1}\right)	\iota_{d_1} \iota_{d_2}'\\
	&= \frac{d_1\rho^2}{(d_1-1)\rho + 1} \iota_{d_2} \iota_{d_2}'.
\end{align*}
Thus, by properties of linear projections,
\begin{align*}
	\var(W_2(s)^{\perp W_1(s)}) &= \var(W_2(s)) - \var(\Lin(W_2(s) \mid 1,W_1(s)))\\
	&= \left(\rho \iota_{d_2} \iota_{d_2}' + (1-\rho)\textbf{I}_{d_2}\right) -  \frac{d_1\rho^2}{(d_1-1)\rho + 1} \iota_{d_2} \iota_{d_2}'\\
	&= \frac{\rho(1-\rho)}{(d_1 - 1)\rho + 1}\iota_{d_2} \iota_{d_2}' + (1-\rho)\textbf{I}_{d_2}
\end{align*}
for all $s$ such that $\sum_{i=1}^K s_i = d_1$. Therefore,
\begin{align}
	\var(\gamma_2(s)'W_2(s)^{\perp W_1(s)})
	&= \frac{\rho(1-\rho)}{(d_1-1)\rho + 1}\left(\sum_{i=1}^K \gamma_i(1-s_i)\right)^2  +  (1-\rho)\left(\sum_{i=1}^K \gamma_i^2 (1-s_i)\right)\label{eq:varw2perp}
\end{align}
and
\begin{align*}
	\gamma_2(s)'\var(W_2(s)^{\perp W_1(s)})\pi_2(s) &= \frac{\rho(1-\rho)}{(d_1-1)\rho + 1}\left(\sum_{i=1}^K \gamma_i(1-s_i)\right)\left(\sum_{i=1}^K \pi_i(1-s_i)\right)\\
	&\qquad  +  (1-\rho)\left(\sum_{i=1}^K \gamma_i \pi_i (1-s_i)\right).
\end{align*}
Moreover,
\begin{align*}
	\gamma'\var(W)\gamma &= \rho \left(\sum_{i=1}^K \gamma_i\right)^2 + (1-\rho) \left(\sum_{i=1}^K \gamma_i^2\right)\\
	\gamma'\var(W)\pi &= \rho \left(\sum_{i=1}^K \gamma_i\right)\left(\sum_{i=1}^K \pi_i\right) + (1-\rho) \left(\sum_{i=1}^K \gamma_i \pi_i\right).
\end{align*}
Combining these expressions, we can write $\delta_\text{resid}(S) = \frac{\text{Num}_1(S)/\text{Num}_2(S)}{\text{Den}_1(S)/ \text{Den}_2(S)}$ where 
\begin{align*}
	\text{Den}_1(s)	&\coloneqq \gamma'\var(W)\pi - \gamma_2(s)'\var(W_2(s)^{\perp W_1(s)})\pi_2(s)\\
	&= \rho \left(\sum_{i=1}^K \gamma_i\right)\left(\sum_{i=1}^K \pi_i\right) + (1-\rho) \left(\sum_{i=1}^K \gamma_i \pi_i s_i\right)\\
	&\qquad - \frac{\rho(1-\rho)}{(d_1-1)\rho + 1}\left(\sum_{i=1}^K \gamma_i(1-s_i)\right)\left(\sum_{i=1}^K \pi_i(1-s_i)\right)\\
	\text{Den}_2(s) &\coloneqq \gamma'\var(W)\gamma - \gamma_2(s)'\var(W_2(s)^{\perp W_1(s)})\gamma_2(s)\\
	&= \rho \left(\sum_{i=1}^K \gamma_i\right)^2 + (1-\rho) \left(\sum_{i=1}^K \gamma_i^2 s_i\right) - \frac{\rho(1-\rho)}{(d_1-1)\rho + 1}\left(\sum_{i=1}^K \gamma_i(1-s_i)\right)^2\\
	\text{Num}_1(s)	&\coloneqq \gamma_2(s)'\var(W_2(s)^{\perp W_1(s)})\pi_2(s)\\
	&= \frac{\rho(1-\rho)}{(d_1-1)\rho + 1}\left(\sum_{i=1}^K \pi_i(1-s_i)\right)\left( \sum_{i=1}^K \gamma_i(1-s_i)\right) + (1-\rho)\sum_{i=1}^K \pi_i\gamma_i (1-s_i)\\
	\text{Num}_2(s) &\coloneqq \gamma_2(s)'\var(W_2(s)^{\perp W_1(s)})\gamma_2(s)\\
	&= \frac{\rho(1-\rho)}{(d-1)\rho + 1}\left(\sum_{i=1}^K \gamma_i(1-s_i)\right)^2  +  (1-\rho)\left(\sum_{i=1}^K \gamma_i^2 (1-s_i)\right).
\end{align*}
Next, we can write
\begin{align*}
	\sum_{i=1}^K \pi_i S_i^{2-l}(1-S_i)^{l-1} &= \frac{d_l}{K}\sum_{i=1}^K \pi_i + o_p(1) = O_p(1)
\end{align*}
for $l = 1,2$.
This holds for $l=1$ because
\begin{align*}
	\sum_{i=1}^K \pi_i S_i &= \frac{d_1}{K}\left(\frac{1}{d_1}\sum_{i=1}^K (K \cdot \pi_i) S_i \right)\\
	&= \frac{d_1}{K}\left(\avgk (K \cdot \pi_i)  + o_p(1) \right)\\
	&= \frac{d_1}{K}\sum_{i=1}^K \pi_i + o_p(1) = O_p(1),
\end{align*}
where the second line follows from $|K \cdot \pi_i| \leq C$ and Lemma \ref{lemma:LLN}, and the third line from $d_1/K = O(1)$. The result holds for $l=2$ following a similar derivation. Similarly, we can show that $\sum_{i=1}^K \gamma_i S_i^{2-l}(1-S_i)^{l-1} = \frac{d_l}{K}\sum_{i=1}^K \gamma_i + o_p(1)$ for $l= 1,2$.

Also, for $l=1,2$,
\begin{align*}
	K\sum_{i=1}^K \pi_i \gamma_i S_i^{2-l}(1-S_i)^{l-1} &= \frac{d_l}{K} \frac{1}{d_l}\sum_{i=1}^K (K^2\pi_i \gamma_i)  S_i^{2-l}(1-S_i)^{l-1}\\
	&= \frac{d_l}{K} \left(\frac{1}{K}\sum_{i=1}^K (K^2\pi_i \gamma_i)  + o_p(1)\right)\\
	&= d_l \sum_{i=1}^K \pi_i \gamma_i + o_p(1) = O_p(1).
\end{align*}
where the second line follows from $|K^2\pi_i \gamma_i| \leq C^2$ and Lemma \ref{lemma:LLN}, and the third line from $d_l/K = O(1)$. Similarly, we can show that
\begin{align*}
	K\sum_{i=1}^K \gamma_i^2 S_i^{2-l}(1-S_i)^{l-1} &= d_l\sum_{i=1}^K \gamma_i^2 + o_p(1) = O_p(1).
\end{align*}

Therefore,
\begin{align*}
	\text{Den}_1(S)	&= \rho \left(\sum_{i=1}^K \gamma_i\right)\left(\sum_{i=1}^K \pi_i\right) + (1-\rho) \left(\sum_{i=1}^K \gamma_i \pi_i S_i\right)\\
	&\qquad - \frac{\rho(1-\rho)}{(d_1-1)\rho + 1}\left(\sum_{i=1}^K \gamma_i(1-S_i)\right)\left(\sum_{i=1}^K \pi_i(1-S_i)\right)\\
	&= \rho \left(\sum_{i=1}^K \gamma_i\right)\left(\sum_{i=1}^K \pi_i\right) + (1-\rho) O_p(K^{-1})\\
	&\qquad - O(K^{-1}) \left(\frac{r}{1+r}\sum_{i=1}^K \gamma_i + o_p(1)\right)\left(\frac{r}{1+r}\sum_{i=1}^K \pi_i + o_p(1)\right)\\
	&= \rho \left(\sum_{i=1}^K \gamma_i\right)\left(\sum_{i=1}^K \pi_i\right) + o_p(1)\\
	\text{Den}_2(S) &= \rho \left(\sum_{i=1}^K \gamma_i\right)^2 + (1-\rho) \left(\sum_{i=1}^K \gamma_i^2 S_i\right) - \frac{\rho(1-\rho)}{(d_1-1)\rho + 1}\left(\sum_{i=1}^K \gamma_i(1-S_i)\right)^2\\
	&=\rho \left(\sum_{i=1}^K \gamma_i\right)^2 + (1-\rho) O_p(K^{-1}) - O(K^{-1})\left(\frac{r}{1+r}\sum_{i=1}^K \gamma_i + o_p(1)\right)^2\\
	&= \rho \left(\sum_{i=1}^K \gamma_i\right)^2 + o_p(1)
\end{align*}
where we used the stochastic orders of the terms derived above. For the numerator terms, multiplying each by $K$ reveals that
\begin{align*}
	K \cdot \text{Num}_1(S)	&= \frac{K\rho(1-\rho)}{(d_1-1)\rho + 1}\left(\sum_{i=1}^K \pi_i(1-S_i)\right)\left( \sum_{i=1}^K \gamma_i(1-S_i)\right)\\
	&\qquad + (1-\rho)\sum_{i=1}^K K\pi_i\gamma_i (1-S_i)\\
	&= \frac{K\rho(1-\rho)}{(d_1-1)\rho + 1}\left(\frac{r}{1+r}\sum_{i=1}^K \pi_i  + o_p(1)\right)\left(\frac{r}{1+r}\sum_{i=1}^K \gamma_i  + o_p(1)\right)\\
	&\qquad + (1-\rho)\left(\frac{r}{1+r}\sum_{i=1}^K K \pi_i \gamma_i + o_p(1)\right)\\
	&= \frac{(1-\rho)r^2}{1+r}\left(\sum_{i=1}^K \pi_i\right)\left(\sum_{i=1}^K \gamma_i \right) + \frac{K(1-\rho)r}{1+r}\sum_{i=1}^K \pi_i \gamma_i + o_p(1)\\
	K \cdot \text{Num}_2(S)	&=\frac{K\rho(1-\rho)}{(d_1-1)\rho + 1}\left(\sum_{i=1}^K \gamma_i(1-S_i)\right)^2  +  (1-\rho)\left(\sum_{i=1}^K K\gamma_i^2 (1-S_i)\right)\\
	&= \frac{(1-\rho)r^2}{1+r}\left(\sum_{i=1}^K \gamma_i\right)^2 + \frac{K(1-\rho)r}{1+r}\sum_{i=1}^K \gamma_i^2 + o_p(1).
\end{align*}
Thus the numerator terms' ratio is
\begin{align*}
	\frac{\text{Num}_1(S)}{\text{Num}_2(S)} &= \frac{\frac{(1-\rho)r^2}{1+r}\left(\sum_{i=1}^K \pi_i\right)\left(\sum_{i=1}^K \gamma_i \right) + \frac{K(1-\rho)r}{1+r}\sum_{i=1}^K \pi_i \gamma_i + o_p(1)}{\frac{(1-\rho)r^2}{1+r}\left(\sum_{i=1}^K \gamma_i\right)^2 + \frac{K(1-\rho)r}{1+r}\sum_{i=1}^K \gamma_i^2 + o_p(1)}\\
	&= \frac{r\left(\sum_{i=1}^K \pi_i\right)\left(\sum_{i=1}^K \gamma_i \right) + K\sum_{i=1}^K \pi_i \gamma_i + o_p(1)}{r\left(\sum_{i=1}^K \gamma_i\right)^2 + K\sum_{i=1}^K \gamma_i^2 + o_p(1)}\\
	&= \frac{r\left(\sum_{i=1}^K \pi_i\right)\left(\sum_{i=1}^K \gamma_i \right) + K\sum_{i=1}^K \pi_i \gamma_i}{r\left(\sum_{i=1}^K \gamma_i\right)^2 + K\sum_{i=1}^K \gamma_i^2} + o_p(1)
\end{align*}
which follows from $\rho \neq 1$ and $ \sum_{i=1}^K \gamma_i$ being bounded away from zero. The denominator terms' ratio is
\begin{align*}
	\frac{\text{Den}_1(S)}{\text{Den}_2(S)} &= \left(\rho \left(\sum_{i=1}^K \gamma_i\right)\left(\sum_{i=1}^K \pi_i\right) + o_p(1)\right)/\left(\rho \left(\sum_{i=1}^K \gamma_i\right)^2 + o_p(1)\right)\\
	&=\left(\sum_{i=1}^K \pi_i\right) /\left(\sum_{i=1}^K \gamma_i\right)+ o_p(1)
\end{align*}
by the same argument and $\rho \neq 0$. Note that this denominator is bounded and bounded away from zero by the theorem assumptions. Thus, combining the convergence of the numerator and denominator ratios yields that
\begin{align}\label{eq:deltaresid_limit2}
	\delta_\text{resid}(S) &= \frac{r \left(\sum_{i=1}^K \pi_i\right)\left(\sum_{i=1}^K \gamma_i\right)^2 + K\left(\sum_{i=1}^K \pi_i \gamma_i\right) \left(\sum_{i=1}^K \gamma_i\right) }{r\left(\sum_{i=1}^K \pi_i\right)\left(\sum_{i=1}^K \gamma_i \right)^2  +   K\left(\sum_{i=1}^K \pi_i\right)\left(\sum_{i=1}^K \gamma_i^2\right) } + o_p(1),
\end{align}
which follows from $\sum_{i=1}^K \pi_i/\sum_{i=1}^K \gamma_i$ being bounded away from zero. 
\end{proof}

\begin{proof}[Proof of Corollary \ref{corr:delta_r_noconv}]

Fix $C \in \R$ and let $C' = C(r+2) - r$, $\gamma_i = (2/K) \mathbbm{1}(i \in \{2,4,6,\ldots\})$, and 
\begin{align*}
	\pi_i = \frac{2 - C'}{K}\mathbbm{1}(i \in \{1,3,5,\ldots\}) + \frac{C'}{K}\mathbbm{1}(i \in \{2,4,6,\ldots\}).
\end{align*}

We show that Assumption \ref{assn:EXvar} holds with $R = 1$, $\Lambda^K = \iota_K\sqrt{\rho}$, and $\sigma^2_E = 1 - \rho$. Together with assumptions \ref{assn:sampling_of_S} and \ref{assn:limit_d2d1}, this is sufficient to show that assumptions \ref{assn:pi_and_VarW} and \ref{assn:gamma_and_VarW} hold, by Proposition \ref{prop:EXvar}. 

We can directly see that these choices satisfy assumptions \ref{assn:EXvar}.1--\ref{assn:EXvar}.3. Moreover, given $\rho \in(0,1)$ and $\sum_{i=1}^K \pi_i^2 > 0$ for $K > 1$, note that 
\begin{align*}
	\var\left(\sum_{i=1}^K \pi_i W_i\right) &= (1-\rho)\sum_{i=1}^K \pi_i^2 + \rho\left(\sum_{i=1}^K \pi_i \right)^2 > 0
\end{align*}
and that $\var\left(\sum_{i=1}^K \pi_i W_i\right) = O(K^{-1}) + \rho (1 + O(K^{-1}))^2 \rightarrow \rho > 0$ as $K \to \infty$. Therefore, there exists $\epsilon > 0$ such that  $\var\left(\sum_{i=1}^K \pi_i W_i\right)> \epsilon$ for all $K$. We can similarly show that $\var\left(\sum_{i=1}^K \gamma_i W_i\right) > \epsilon$ for sufficiently small $\epsilon > 0$. Therefore Assumption \ref{assn:EXvar}.4 holds, which implies that assumptions 
\ref{assn:sampling_of_S}--\ref{assn:gamma_and_VarW} hold. Finally, we note that $\sum_{i=1}^K \gamma_i$ and $\sum_{i=1}^K \pi_i$ are bounded away from zero, and therefore the assumptions of Theorem \ref{thm:delta_resid} hold.

Next, we substitute these sequences in the expression for $\delta_\text{resid}(S)$ from equation \eqref{eq:deltaresid_limit2}. We note that
\begin{align*}
	\left(\sum_{i=1}^K \pi_i, \sum_{i=1}^K \gamma_i, K\sum_{i=1}^K \gamma_i \pi_i,K\sum_{i=1}^K \gamma_i^2\right) &\rightarrow (1,1,C',2).
\end{align*}
Therefore,
\begin{align*}
	\delta_\text{resid}(S) &= \frac{r  + C'}{r + 2} + o_p(1) \pconv \frac{r + (C(r+2) - r)}{r+2} = C.
\end{align*}
\end{proof}

\begin{proof}[Proof of Corollary \ref{corr:delta_r_noconv_pt2}]

For the second part, we note that
\begin{align*}
	\frac{\partial}{\partial r}\frac{r \left(\sum_{i=1}^K \pi_i\right)\left(\sum_{i=1}^K \gamma_i\right)^2 + K\left(\sum_{i=1}^K \pi_i \gamma_i\right) \left(\sum_{i=1}^K \gamma_i\right) }{r\left(\sum_{i=1}^K \pi_i\right)\left(\sum_{i=1}^K \gamma_i \right)^2  +   K\left(\sum_{i=1}^K \pi_i\right)\left(\sum_{i=1}^K \gamma_i^2\right) }
\end{align*}
has the same sign as
\begin{align}\label{eq:derivative_denom}
	\left(\sum_{i=1}^K \pi_i\right)^2\left(\sum_{i=1}^K \gamma_i\right)^2 K\left(\sum_{i=1}^K \gamma_i^2\right) - K\left(\sum_{i=1}^K \pi_i \gamma_i\right)  \left(\sum_{i=1}^K \pi_i\right)\left(\sum_{i=1}^K \gamma_i \right)^3.
\end{align}
As in the proof of Corollary \ref{corr:delta_r_noconv}, let $\gamma_i = (2/K) \mathbbm{1}(i \in \{2,4,6,\ldots\})$ and	$\pi_i = (2-C')/K\mathbbm{1}(i \in \{1,3,5,\ldots\}) + C'/K\mathbbm{1}(i \in \{2,4,6,\ldots\})$ for an arbitrary $C' > 2$. Using the previous results, the term in \eqref{eq:derivative_denom} converges to 
$2 - C'$, which is negative.
\end{proof}

\begin{proof}[Proof of Theorem \ref{thm:delta_ACET}]
We split this proof into two parts.

\noindent\textbf{Part 1: Rewriting Expressions}

We begin by rewriting the numerators and denominators in the expression for $\delta_\text{ACET}(S)$. 
By properties of linear projections, we have that
\begin{align*}
	\var(\gamma_2(s)'W_2(s)^{\perp \gamma_1(s)'W_1(s)}) &= \var(\gamma_2(s)'W_2(s)) - \frac{\cov( \gamma_2(s)'W_2(s),\gamma_1(s)'W_1(s))^2}{\var(\gamma_1(s)'W_1(s))}\\
	\var(\gamma_2(s)'W_2(s)^{\perp \gamma_1(s)'W_1(s)}) &= \var(\gamma_2(s)'W_2(s)) - \frac{\cov( \gamma_2(s)'W_2(s),\gamma_1(s)'W_1(s))^2}{\var(\gamma_1(s)'W_1(s))}
\end{align*}
where
\begin{align}
	\cov(\gamma_1(s)'W_1(s),\gamma_2(s)'W_2(s)) &= \cov(\gamma_1(s)'W_1(s),\gamma'W) -\var(\gamma_1(s)'W_1(s))\notag\\
	&= \sum_{i=1}^K s_i \cov(\gamma_i W_i,\gamma'W) -\var(\gamma_1(s)'W_1(s)).\label{eq:covar_gamma1gamma_2}
\end{align}
Also,
\begin{align*}
	\cov(X,\gamma_2(s)'W_2(s)^{\perp \gamma_1(s)'W_1(s)}) &= \cov(X,\gamma_2(s)'W_2(s))\\
	&\quad - \frac{\cov( \gamma_1(s)'W_1(s),\gamma_2(s)'W_2(s))\cov(\gamma_1(s)'W_1(s),X)}{\var(\gamma_1(s)'W_1(s))} \\
	\cov(X,\gamma_1(s)'W_1(s)^{\perp \gamma_2(s)'W_2(s)}) &= \cov(X,\gamma_1(s)'W_1(s))\\
	&\quad - \frac{\cov( \gamma_2(s)'W_2(s),\gamma_1(s)'W_1(s))\cov(\gamma_2(s)'W_2(s),X)}{\var(\gamma_2(s)'W_2(s))}.
\end{align*}

\noindent\textbf{Part 2: Convergence of Terms}

Using the proof of Theorem \ref{thm:ry_conv} in the Supplemental Appendix, and similarly to the proof of Theorem \ref{thm:rx_conv}, we have that
\begin{align*}
	\var(\gamma_1(S)'W_1(S)) &= \var(\gamma'W)(1 + rc_\gamma)/(1+r)^2 + o_p(1)\\
	\var(\gamma_2(S)'W_2(S)) &= \var(\gamma'W)r(r + c_\gamma)/(1+r)^2 + o_p(1).
\end{align*}
Using the proof of Theorem \ref{thm:delta_orig_conv}, we have that
\begin{align*}
	\cov(X,\gamma_1(S)'W_1(S)) &= \cov(X,\gamma'W)/(1+r) + o_p(1)\\
	\cov(X,\gamma_2(S)'W_2(S)) &= \cov(X,\gamma'W)r/(1+r) + o_p(1).
\end{align*}
We now investigate the asymptotic behavior of $\cov(\gamma_1(s)'W_1(s),\gamma'W)$. We have that
\begin{align*}
	\Exp_S[\cov(\gamma_1(S)'W_1(S),\gamma'W)] &= \frac{d_1}{K} \sum_{i=1}^K \cov(\gamma_i W_i,\gamma'W) = \frac{d_1}{K}\var(\gamma'W) = \frac{\var(\gamma'W)}{1+r} + o(1)
\end{align*}
by Assumption \ref{assn:limit_d2d1}. Also,
\begin{align*}
	&{\var}_S\left(\cov(\gamma_1(S)'W_1(S),\gamma'W)\right)\\
	&= {\var}_S( \sum_{i=1}^K S_i \cov(\gamma_i W_i,\gamma'W))\\
	&= \frac{d_1 d_2}{K(K-1)} \sum_{i=1}^K \left(\cov(\gamma_i W_i,\gamma'W) - \frac{1}{K}\sum_{j=1}^K \cov(\gamma_j W_j,\gamma'W)\right)^2\\
	&\leq \frac{d_1 d_2}{K(K-1)} \sum_{i=1}^K \cov(\gamma_i W_i,\gamma'W)^2 \rightarrow 0
\end{align*}
where we used Lemma \ref{lemma:LLN}, Assumption \ref{assn:limit_d2d1}, and Assumption \ref{assn:limited_dep}. Therefore,
\begin{align*}
	\cov(\gamma_1(S)'W_1(S),\gamma'W) &= \var(\gamma'W)/(1+r) + o_p(1).
\end{align*}
Combining this with equation \eqref{eq:covar_gamma1gamma_2}, we obtain.
\begin{align*}
	\cov(\gamma_1(S)'W_1(S),\gamma_2(S)'W_2(S)) &= \frac{1}{1+r}\var(\gamma'W)  - \frac{1 + rc_\gamma}{(1+r)^2}\var(\gamma'W) + o_p(1)\\
	&= \frac{r(1-c_\gamma)}{(1+r)^2}\var(\gamma'W) + o_p(1).
\end{align*}
Combining all these expressions, we derive the asymptotic behavior of the numerator terms:
\begin{align*}
	&\cov(X,\gamma_2(S)'W_2(S)^{\perp \gamma_1(S)'W_1(S)})\\
	&= \cov(X,\gamma_2(S)'W_2(S)) - \frac{\cov( \gamma_1(S)'W_1(S),\gamma_2(S)'W_2(S))\cov(\gamma_1(S)'W_1(S),X)}{\var(\gamma_1(S)'W_1(S))}\\
	&= \frac{\cov(X,\gamma'W)r}{1+r} - \left(\frac{r(1-c_\gamma)\cov(X,\gamma'W)\var(\gamma'W)}{(1+r)^3}\right)/\left(\frac{\var(\gamma'W)(1+ r c_\gamma)}{(1+r)^2}\right) + o_p(1)\\
	&= \frac{\cov(X,\gamma'W)r}{1+r}\left(1 - \frac{1-c_\gamma}{1 + r c_\gamma}\right) + o_p(1)\\
	&= \cov(X,\gamma'W)r c_\gamma/(1+rc_\gamma) + o_p(1)
\end{align*}
and
\begin{align*}
	&\var(\gamma_2(S)'W_2(S)^{\perp \gamma_1(S)'W_1(S)})\\
	&= \var(\gamma_2(S)'W_2(S)) -  \cov(\gamma_2(S)'W_2(S),\gamma_1(S)'W_1(S))^2/\var(\gamma_1(S)'W_1(S))\\
	&= \frac{r(r + c_\gamma)\var(\gamma'W)}{(1+r)^2} - \frac{r^2(1-c_\gamma)^2\var(\gamma'W)^2}{(1+r)^4}/\frac{(1 + rc_\gamma)\var(\gamma'W)}{(1+r)^2}  + o_p(1)\\
	&= \frac{\var(\gamma'W)r}{(1+r)^2}\left(r + c_\gamma - \frac{r(1-c_\gamma)^2}{1+rc_\gamma}\right) + o_p(1)\\
	&= \var(\gamma'W) r c_\gamma/(1+r c_\gamma) + o_p(1),
\end{align*}
where we used that $\var(\gamma'W)$ is uniformly bounded away from 0. Similar derivations for the denominator yield
\begin{align*}
	\cov(X,\gamma_1(S)'W_1(S)^{\perp \gamma_2(S)'W_2(S)}) &= \cov(X,\gamma'W) c_\gamma/(r + c_\gamma) + o_p(1)\\
	\var(\gamma_1(S)'W_1(S)^{\perp \gamma_2(S)'W_2(S)}) &= \var(\gamma'W) c_\gamma/(r + c_\gamma) + o_p(1).
\end{align*}
Combining the results for the numerator and denominator, we can write
\begin{align*}
	\delta_\text{ACET}(S) &= \frac{ \cov(X,\gamma'W)r c_\gamma/(1+rc_\gamma) + o_p(1)}{\var(\gamma'W) r c_\gamma/(1+r c_\gamma) + o_p(1)}\bigg/\frac{ \cov(X,\gamma'W) c_\gamma/(r + c_\gamma) + o_p(1)}{\var(\gamma'W) c_\gamma/(r + c_\gamma) + o_p(1)}\\
	&= \frac{ \cov(X,\gamma'W)r c_\gamma/(1+rc_\gamma)}{\var(\gamma'W) r c_\gamma/(1+r c_\gamma) }\bigg/\frac{ \cov(X,\gamma'W) c_\gamma/(r + c_\gamma)}{\var(\gamma'W) c_\gamma/(r + c_\gamma) } + o_p(1)\\
	&\pconv 1
\end{align*}
where the second equality follows from $\cov(X,\gamma'W)$ and $\var(\gamma'W)$ and bounded away from 0, and from $r,c_\gamma > 0$.
\end{proof}

\begin{proof}[Proof of Theorem \ref{thm:k_X_conv}]
First, we use results obtained in the proof of Theorem \ref{thm:delta_resid} to rewrite the expression for $k_X(S)$. By properties of R-squared, we can write
\begin{align*}
	R^2_{X \sim W} - R^2_{X \sim W_1(s)} &= R^2_{X \sim W_2(s)^{\perp W_1(s)}}
\end{align*}
where $R^2_{X \sim W_2(s)^{\perp W_1(s)}}\var(X) = \var(\pi_2(s)'W_2(s)^{\perp W_1(s)})$ and $R^2_{X \sim W}\var(X) = \var(\pi'W)$. Substituting these expressions in the formula for $k_X(S)$ yields
\begin{align*}
	k_X(S) &= \frac{R^2_{X \sim W} - R^2_{X \sim W_1(S)}}{R^2_{X \sim W_1(S)}} 
	 = \frac{\var(\pi_2(s)'W_2(s)^{\perp W_1(s)})}{\var(\pi'W) - \var(\pi_2(s)'W_2(s)^{\perp W_1(s)})}.
\end{align*}
Since the covariates are assumed exchangeable, we have that
\begin{align*}
	\var(\pi'W) &= 
	 \alpha \left(\frac{1}{\sqrt{K}}\sum_{i=1}^K \pi_i\right)^2 + \left(1-\frac{\alpha}{K}\right) \sum_{i=1}^K \pi_i^2.
\end{align*}
By an analog of equation \eqref{eq:varw2perp}, we also have that
\begin{align*}
	\var(\pi_2(s)'W_2(s)^{\perp W_1(s)}) &= \frac{\alpha(1-\alpha/K)}{(d_1-1)\alpha + K}\left(\sum_{i=1}^K \pi_i(1-S_i)\right)^2  +  \frac{K-\alpha}{K}\sum_{i=1}^K \pi_i^2 (1-S_i).
\end{align*}

Second, we analyze the asymptotic properties of the terms in $\var(\pi_2(s)'W_2(s)^{\perp W_1(s)})$ and $\var(\pi'W)$. To begin, we note that
\begin{align*}
	K \cdot \frac{\alpha(1-\alpha/K)}{(d_1-1)\alpha + K} &= \frac{\alpha(1-\alpha/K)}{\alpha(d_1-1)/K + 1} = \frac{\alpha}{\alpha / (1 + r) + 1} + o(1) = \frac{\alpha(1+r)}{1 + \alpha + r} + o(1).
\end{align*}
Next, consider the normalized sum $K^{-1/2}\sum_{i=1}^K \pi_i(1-S_i)$. Under Assumption \ref{assn:EXvar_shrink},
\begin{align*}
	\Exp_S\left[K^{-1/2}\sum_{i=1}^K \pi_i(1-S_i)\right] &= 
	 \left(\frac{r}{1+r} + o(1)\right)\frac{1}{\sqrt{K}}\sum_{i=1}^K \pi_i = \frac{r}{1+r}\frac{1}{\sqrt{K}}\sum_{i=1}^K \pi_i + o(1).
\end{align*}
The last equality follows from
\begin{align*}
	\left|\frac{1}{\sqrt{K}}\sum_{i=1}^K \pi_i\right| &\leq \frac{1}{\sqrt{K}}\sum_{i=1}^K |\pi_i| \leq \frac{1}{\sqrt{K}} \cdot K \cdot \sup_{i \in [K]} |\pi_i| \leq  \frac{1}{\sqrt{K}} \cdot K \cdot \frac{C}{\sqrt{K}} = C = O(1),
\end{align*}
which follows from Assumption \ref{assn:EXvar_shrink}.2. Using this assumption again, we obtain that
\begin{align*}
	{\var}_S\left(\frac{1}{\sqrt{K}}\sum_{i=1}^K \pi_i(1-S_i)\right) 
	&\leq \frac{1}{K}\frac{d_1d_2}{K(K-1)} \sum_{i=i}^K \pi_i^2\\
	&\leq \frac{d_1d_2}{K^2(K-1)} K \cdot \sup_{i\in[K]}\pi_i^2\\
	&=  o(1),
\end{align*}
where we also used Lemma \ref{lemma:LLN} and Assumption \ref{assn:limit_d2d1}. Using Chebyshev's inequality, we conclude that
\begin{align*}
	\frac{1}{\sqrt{K}}\sum_{i=1}^K \pi_i(1-S_i) &= \frac{r}{1+r}\frac{1}{\sqrt{K}}\sum_{i=1}^K \pi_i + o_p(1) = O_p(1).
\end{align*}
We now analyze the limiting behavior of $\sum_{i=1}^K \pi_i^2 (1-S_i)$. Its expectation is
\begin{align*}
	\Exp_S\left[\sum_{i=1}^K \pi_i^2(1-S_i)\right] &= \frac{d_2}{K}\sum_{i=1}^K \pi_i^2 = \left(\frac{r}{1+r} + o(1)\right)\sum_{i=1}^K \pi_i^2 = \frac{r}{1+r}\sum_{i=1}^K \pi_i^2 + o(1).
\end{align*}
The last line follows from
\begin{align*}
	\left|\sum_{i=1}^K \pi_i^2\right| &\leq \sum_{i=1}^K |\pi_i|^2 \leq  K \cdot \sup_{i \in [K]} |\pi_i|^2 \leq   K \cdot \frac{C^2}{K} = C^2 = O(1),
\end{align*}
where Assumption \ref{assn:EXvar_shrink}.2 was used. Similarly, its variance is
\begin{align*}
	{\var}_S\left(\sum_{i=1}^K \pi_i^2(1-S_i)\right) 
	&\leq \frac{d_1d_2}{K(K-1)} \sum_{i=i}^K \pi_i^4\\
	&\leq \frac{d_1d_2}{K(K-1)} K \cdot \sup_{i\in[K]}|\pi_i|^4\\
	&= o(1).
\end{align*}
where we also used Lemma \ref{lemma:LLN} and Assumption \ref{assn:limit_d2d1}. Using Chebyshev's inequality, we conclude that
\begin{align*}
	\sum_{i=1}^K \pi_i^2(1-S_i) &= \frac{r}{1+r}\sum_{i=1}^K \pi_i^2 + o_p(1) = O_p(1).
\end{align*}

Therefore,
\begin{align*}
	\var(\pi_2(s)'W_2(s)^{\perp W_1(s)}) &= \left(\frac{\alpha(1+r)}{1 + \alpha + r} + o(1)\right)\left(\frac{r}{1+r}\frac{1}{\sqrt{K}}\sum_{i=1}^K \pi_i + o_p(1)\right)^2\\
	&\qquad +  \frac{r}{1+r}\sum_{i=1}^K \pi_i^2 + o_p(1)\\
	&= \frac{\alpha r^2}{(1 + \alpha + r)(1+r)}\frac{1}{K}\left(\sum_{i=1}^K \pi_i\right)^2 + \frac{r}{1+r}\sum_{i=1}^K \pi_i^2 + o_p(1)\\
	\var(\pi'W) &= \frac{\alpha}{K}\left(\sum_{i=1}^K \pi_i\right)^2 + \sum_{i=1}^K \pi_i^2 + o(1),
\end{align*}
and combining these two expressions yields
\begin{align*}
	k_X(S) &= \frac{\frac{\alpha r^2}{(1 + \alpha + r)(1+r)}\frac{1}{K}\left(\sum_{i=1}^K \pi_i\right)^2 + \frac{r}{1+r}\sum_{i=1}^K \pi_i^2 + o_p(1)}{ \frac{\alpha}{K}\left(\sum_{i=1}^K \pi_i\right)^2 + \sum_{i=1}^K \pi_i^2 - \frac{\alpha r^2}{(1 + \alpha + r)(1+r)}\frac{1}{K}\left(\sum_{i=1}^K \pi_i\right)^2 - \frac{r}{1+r}\sum_{i=1}^K \pi_i^2 + o_p(1)}\\
	&= \frac{\frac{\alpha r^2}{(1 + \alpha + r)(1+r)} + \frac{r}{1+r}\left(K \sum_{i=1}^K \pi_i^2\right)/\left(\sum_{i=1}^K \pi_i\right)^2 + o_p(1)}{\alpha\left(1 - \frac{r^2}{(1 + \alpha + r)(1+r)}\right)   + \left(K \sum_{i=1}^K \pi_i^2\right)/\left(\sum_{i=1}^K \pi_i\right)^2\left(1 -\frac{r}{1+r}\right)+ o_p(1)}\\
	&=	\frac{\alpha r^2/[(1 + \alpha + r)(1+r)] + d_{\pi,\text{EX}}r/(1+r) + o_p(1)}{\alpha((1+\alpha)(1+r) + r)/[(1 + \alpha + r)(1+r)] + d_{\pi,\text{EX}}/(1+r)+ o_p(1)}\\
	&=	\frac{\alpha r^2 + d_{\pi,\text{EX}}r(1+\alpha+r)}{\alpha((1+\alpha)(1+r) + r) + d_{\pi,\text{EX}}(1 + \alpha + r) }+ o_p(1).
\end{align*}
The last equality follows from $\alpha((1+\alpha)(1+r) + r) + d_{\pi,\text{EX}}(1 + \alpha + r) \geq (1+\alpha)^2(1+r) \geq 0$, which follows from $d_{\pi,\text{EX}} \geq 1$ and $\alpha > -1$.

We finish by computing the derivative of this limit with respect to $r$:
\begin{align*}
	&\frac{\partial }{\partial r}\frac{\alpha r^2 + d_{\pi,\text{EX}} r (1 + r + \alpha) }{\alpha((1+r)(1 + \alpha) + r) + d_{\pi,\text{EX}}  (1 + r + \alpha)}\\
	 &\qquad = \frac{(\alpha + d_{\pi,\text{EX}})(r^2 + (1 + \alpha)^2) + 2(\alpha + d_{\pi,\text{EX}})^2(1 + \alpha)r }{\left(\alpha((1+r)(1 + \alpha) + r) + d_{\pi,\text{EX}} (1 + r + \alpha)\right)^2}.
\end{align*}
The numerator of this expression is positive because $\alpha > -1$, $d_{\pi,\text{EX}} \geq 1$, and $r > 0$. Therefore this derivative is strictly positive and the limit of $k_X(S)$ is strictly increasing $r$.
\end{proof}
\makeatletter\@input{DMP2aux.tex}\makeatother
\end{document}


\maketitle

\singlespacing
\normalsize
\appendix
\allowdisplaybreaks

\setcounter{section}{1}

This document contains appendices to accompany the main text. Appendix \ref{sec:additionalAsymptotics} derives the probability limit of $r_Y(S)$. Appendix \ref{sec:osterredef} explains how and why the $\delta_\text{resid}$ parameter is derived as a variation on $\delta_\text{orig}$. Appendix \ref{sec:empiricalAppendix} provides additional results for our empirical comparison of selection ratios. Finally, Appendix \ref{sec:onlineAppendixProofSection} gives proofs for section \ref{sec:AR-MA-Ex}.

\section{Additional Asymptotic Analysis}\label{sec:additionalAsymptotics}

Let 
\[
	r_Y(s) \coloneqq \frac{\sqrt{\var(\gamma_2(s)'W_2(s))}}{\sqrt{\var(\gamma_1(s)'W_1(s))}}.
\]

\begin{theorem}[Convergence of $r_Y$]\label{thm:ry_conv}
Suppose assumptions \ref{assn:sampling_of_S}, \ref{assn:limit_d2d1}, and \ref{assn:gamma_and_VarW} hold. Then, as $K \rightarrow \infty$,
\[
	r_Y(S) \pconv \sqrt{ \frac{r(r+c_\gamma)}{1+rc_\gamma}}.
\]
\end{theorem}

\begin{proof}[Proof of Theorem \ref{thm:ry_conv}]
This proof is identical to that of Theorem  \ref{thm:rx_conv}, after replacing the vector $\pi$ with $\gamma$ and the constant $c_\pi$ with $c_\gamma$.
\end{proof}

\section{Details on Residualization in \cite{Oster2019}}\label{sec:osterredef}

\cite{Oster2019} derives all of her identification results using the sensitivity parameter $\delta_\text{orig}$ combined with the exogenous controls assumption $\cov(W_1, \gamma_2' W_2) = 0$. To remove this assumption, she suggests replacing $\delta_\text{orig}$ with a different sensitivity parameter, which leads to $\delta_\text{resid}$. Here we explain how this residualized sensitivity parameter is derived.

For simplicity, suppose $W_1$ is scalar. $W_2$ in Oster's notation is equivalent to $\gamma_2' W_2$ in our notation. For the discussion below, we will also assume $W_2$ is a scalar and $\gamma_2 = 1$ for simplicity. This is not required, but it implies that Oster's $W_2$ and ours are the same, which makes the derivations and comparisons clearer. With these simplifying assumptions, equation \eqref{eq:DeltaOrig} becomes
\[
	\delta_\text{orig} = \frac{ \cov(X, W_2) }{ \var(W_2) } \Bigg/ \frac{\cov(X,\gamma_1 W_1) }{\var(\gamma_1 W_1)}.
\]
In her Appendix A.1, Oster notes that if $W_1$ and $W_2$ are correlated, we can consider the linear projection of $W_2$ onto $(1,W_1)$:
\[
	W_2 = \phi W_1 + W_2^{\perp W_1}.
\]
Here $\cov(W_1, W_2^{\perp W_1}) = 0$ by construction, and $\phi = \cov(W_2, W_1) / \var(W_1)$. Hence we could instead define $\delta$ based on the residuals $W_2^{\perp W_1}$ rather than the original covariate $W_2$:
\[
	\tilde{\delta} \coloneqq \frac{ \cov(X, W_2^{\perp W_1}) }{ \var(W_2^{\perp W_1}) } \Bigg/ \frac{\cov(X,\gamma_1 W_1) }{\var(\gamma_1 W_1)}.
\]
Oster defines this $\tilde{\delta}$ in her Appendix A.1 and then suggests that all of the previous analysis can proceed as before, using $\tilde{\delta}$ instead of $\delta_\text{orig}$. This is slightly incorrect, however, because the coefficient on $W_1$ needs to be adjusted. For example, without this adjustment, which we describe below, Oster's Proposition 1 does not hold if we replace $W_2$ with $W_2^{\perp W_1}$ and $\delta_\text{orig}$ with $\tilde{\delta}$. 

To see the issue, we can write the outcome equation as
\begin{align*}
	Y
		&= \beta_\text{long} X + \gamma_1 W_1 + W_2 + Y^{\perp X,W_1,W_2} \\
		&= \beta_\text{long} X + \gamma_1 W_1 + (\phi W_1 + W_2^{\perp W_1})  + Y^{\perp X,W_1,W_2} \\
		&= \beta_\text{long} X + (\gamma_1 + \phi) W_1 + W_2^{\perp W_1} + Y^{\perp X,W_1,W_2}.
\end{align*}
This is the correct version of the outcome equation after residualizing the unobserved covariates. In particular, the $\phi W_1$ term cannot be absorbed into the residual because then the residual would no longer be uncorrelated with $W_1$. Applying equation \eqref{eq:DeltaOrig} to this equation yields
\[
	\delta_\text{resid} = \frac{ \cov(X, W_2^{\perp W_1}) }{ \var(W_2^{\perp W_1}) } \Bigg/ \frac{\cov(X,(\gamma_1 + \phi) W_1) }{\var((\gamma_1 + \phi) W_1)}.
\]
This is a special case of $\delta_\text{resid}$ from equation \eqref{eq:deltaResid}. This version of the sensitivity parameters matches the outcome equation, and thus all of Oster's identification results go through if we replace $W_2$ with $W_2^{\perp W_1}$ and $\delta_\text{orig}$ with $\delta_\text{resid}$. 

Based on this analysis, and shortly after stating the exogenous controls assumption $\cov(W_1, \gamma_2' W_2) = 0$, \cite{Oster2019} says
\begin{quote}
``As in AET [Altonji, Elder, and Taber (2005)], the orthogonality of $W_1$ and $W_2$ is central to deriving the results and may be somewhat at odds with the intuition that the observables and the unobservables are ``related.'' In practice, the weight of this assumption is in \emph{how we think about the proportionality condition}. In Appendix A.1, I show formally that if we begin with a case in which the elements of $W_1$ are correlated with $W_2$ we can always redefine $W_2$ such that the results hold under some value of $\delta$.'' (page 192, emphasis added)
\end{quote}
Here Oster argues that the main implication of using $\delta_\text{resid}$ instead of $\delta_\text{orig}$ is that it changes the interpretation of the sensitivity parameter. However, as we discussed in section \ref{sec:frameworkImplications}, it \emph{also} changes the benchmark of equal selection, which can lead researchers to draw the wrong conclusions about robustness.

\section{Additional Empirical Results}\label{sec:empiricalAppendix}

\begin{figure}[th]
    \centering
    \begin{tabular}{
    >{\centering\arraybackslash} m{0.01\textwidth}
    >{\centering\arraybackslash} m{0.29\textwidth}
    >{\centering\arraybackslash} m{0.29\textwidth}
    >{\centering\arraybackslash} m{0.29\textwidth}
    }
    & {\footnotesize More Observed (19 of 22)} & {\footnotesize Equal Selection (11 of 22)} & {\footnotesize More Unobserved (3 of 22)} \\

    \rotatebox{90}{{\footnotesize $r_Y (S)$}} &
    \includegraphics[width=0.26\textwidth]{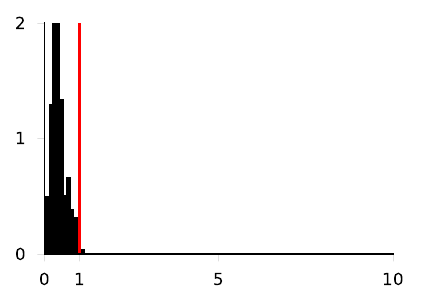} &
    \includegraphics[width=0.26\textwidth]{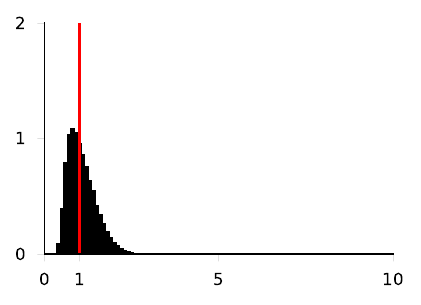} &
    \includegraphics[width=0.26\textwidth]{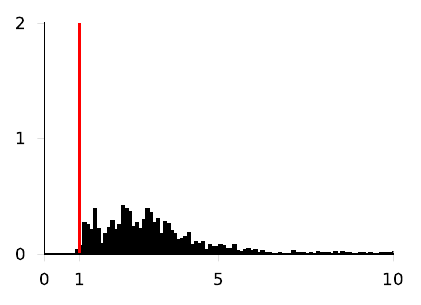} \\
    
    \rotatebox{90}{{\footnotesize $k_X (S)$}}&
    \includegraphics[width=0.26\textwidth]{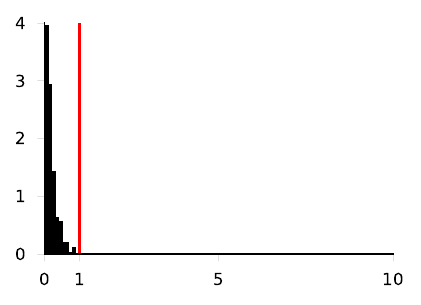} &
    \includegraphics[width=0.26\textwidth]{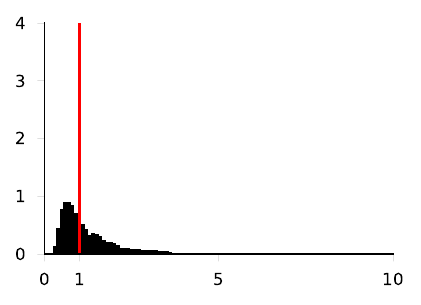} &
    \includegraphics[width=0.26\textwidth]{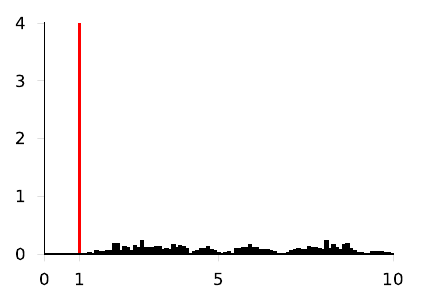} \\

    \rotatebox{90}{{\footnotesize $|\delta_{\text{orig}} (S)|$}}&
    \includegraphics[width=0.26\textwidth]{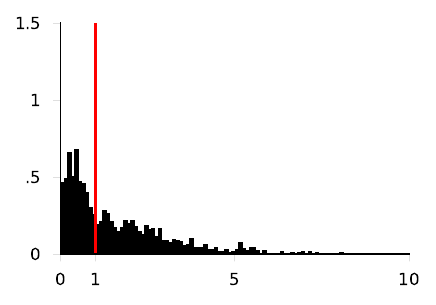} &
    \includegraphics[width=0.26\textwidth]{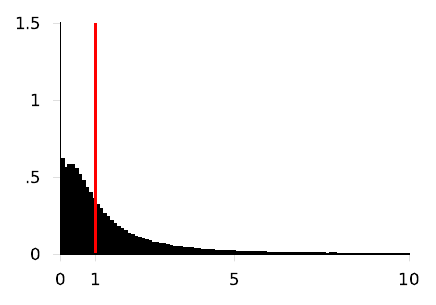} &
    \includegraphics[width=0.26\textwidth]{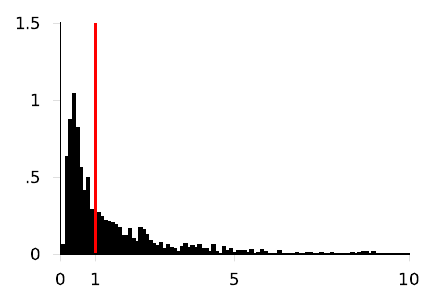} \\

    \rotatebox{90}{{\footnotesize $|\delta_{\text{ACET}} (S)|$}}&
    \includegraphics[width=0.26\textwidth]{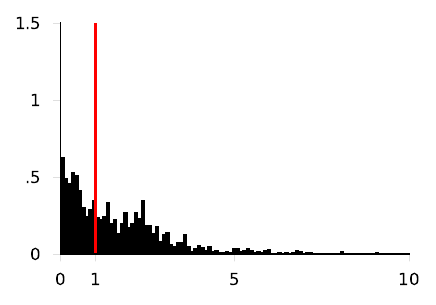} &
    \includegraphics[width=0.26\textwidth]{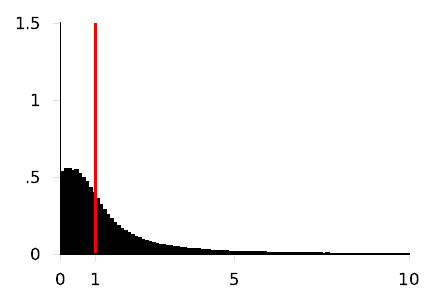} &
    \includegraphics[width=0.26\textwidth]{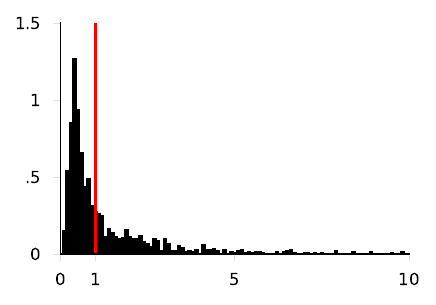} \\
    \end{tabular}
    \\

    \caption{Covariate Sampling Distributions (Additional Parameters).}
    \label{fig:designBasedEmpiricalAppendix}
\end{figure}

Figure \ref{fig:designBasedEmpiricalAppendix} completes figure \ref{fig:designBasedEmpirical} by plotting the covariate sampling distributions for $r_Y(S)$, $k_X(S)$, $| \delta_\text{orig}(S) |$, and $| \delta_\text{ACET}(S) |$. The summary statistics for these distributions are shown in Table \ref{table:designAppendixTable1}, which completes Table \ref{table:designTable2}.

\begin{table}[h]
\centering
\SetTblrInner[talltblr]{rowsep=0pt}
\resizebox{.95\textwidth}{!}{
\begin{talltblr}[
    caption = {Covariate Sampling Distribution Summary Statistics (Additional Parameters).\label{table:designAppendixTable1}},
    remark{Note} = {These summary statistics correspond to the full distributions displayed in Figure \ref{fig:designBasedEmpiricalAppendix}.},
  ]{p{0.35\textwidth} | *{3}{>{\centering \arraybackslash}p{0.17\textwidth}}}

\toprule
\rule{0pt}{1.25\normalbaselineskip}
\rule{0pt}{\normalbaselineskip}

& \SetCell[c=3]{c} Number of Observed Covariates \\
& More Observed & Equal Selection & More Unobserved \\
& 19 of 22& 11 of 22& 3 of 22\\
Number of Support Points, ${K \choose d_1}$ & 1,540 & 705,432 & 1,540 \\
\hline
\SetCell[c=4]{l} Panel A. Distribution of $r_Y(S)$ \\
\hline
\ \ \ \ Minimum &  0.033 &  0.246 &  0.889 \\[2pt]
\ \ \ \ 25th Percentile &  0.263 &  0.762 &  2.086 \\[2pt]
\ \ \ \ Median &  0.346 &  1.000 &  2.890 \\[2pt]
\ \ \ \ 25th Percentile &  0.479 &  1.312 &  3.799 \\[2pt]
\ \ \ \ Maximum & 1.125& 4.063& 30.658\\[2pt]
\ \ \ \ Mean &  0.395 &  1.071 &  3.315 \\[2pt]
\ \ \ \ Standard Deviation &  0.198 &  0.403 &  2.15 \\[5pt]
\hline
\SetCell[c=4]{l} Panel D. Distribution of $k_X (S)$ \\
\hline
\ \ \ \ Minimum &  0.001 &  0.130 &  1.232 \\[2pt]
\ \ \ \ 25th Percentile &  0.075 &  0.654 &  3.945 \\[2pt]
\ \ \ \ Median &  0.131 &  0.971 &  7.713 \\[2pt]
\ \ \ \ 25th Percentile &  0.235 &  1.570 &  17.011 \\[2pt]
\ \ \ \ Maximum & 1.382& 21.760& 2,578\\[2pt]
\ \ \ \ Mean &  0.185 &  1.259 &  21.749 \\[2pt]
\ \ \ \ Standard Deviation &  0.169 &  0.918 &  76.4 \\[5pt]
\hline
\SetCell[c=4]{l} Panel B. Distribution of $|\delta_{\text{orig}}(S)|$ \\
\hline
\ \ \ \ Minimum &  0.001 &  0.000 &  0.029 \\[2pt]
\ \ \ \ 25th Percentile &  0.440 &  0.435 &  0.417 \\[2pt]
\ \ \ \ Median &  1.138 &  1.000 &  0.879 \\[2pt]
\ \ \ \ 25th Percentile &  2.399 &  2.297 &  2.272 \\[2pt]
\ \ \ \ Maximum & 34.615& 267,598& 1,818\\[2pt]
\ \ \ \ Mean &  1.679 &  8.510 &  4.465 \\[2pt]
\ \ \ \ Standard Deviation &  1.99 &  527 &  48.6 \\[5pt]
\hline
\SetCell[c=4]{l} Panel C. Distribution of $|\delta_{\text{ACET}}(S)|$ \\
\hline
\ \ \ \ Minimum &  0.002 &  0.000 &  0.033 \\[2pt]
\ \ \ \ 25th Percentile &  0.475 &  0.462 &  0.419 \\[2pt]
\ \ \ \ Median &  1.320 &  1.000 &  0.758 \\[2pt]
\ \ \ \ 25th Percentile &  2.388 &  2.163 &  2.103 \\[2pt]
\ \ \ \ Maximum & 29.852& 153,935& 475.874\\[2pt]
\ \ \ \ Mean &  1.790 &  6.565 &  3.815 \\[2pt]
\ \ \ \ Standard Deviation &  2.09 &  321 &  20.8 \\[5pt]
\hline
\end{talltblr}
}
\end{table}

\clearpage

\section{Proofs for Section \ref{sec:AR-MA-Ex}}\label{sec:onlineAppendixProofSection}

In this section we prove that the four sets of low-level conditions introduced in section \ref{sec:AR-MA-Ex} of the main text imply our higher-order conditions \ref{assn:pi_and_VarW}--\ref{assn:limited_dep} of section \ref{sec:designBasedAsymptotics}. 

We begin with a preliminary lemma that states that any of these four lower-level assumptions implies Assumption \ref{assn:limited_dep} holds. 

\begin{lemma}\label{lem:lowlevelconds_deltaorig}
Suppose assumptions \ref{assn:sampling_of_S} and \ref{assn:limit_d2d1} hold. Suppose either assumption \ref{assn:MAvar}, \ref{assn:ARvar}, \ref{assn:EXvar}  or \ref{assn:EXvar_shrink} holds. Then $\sum_{i=1}^K \cov(X,\gamma_i^K W_i^K)^2 \rightarrow 0$ and $\sum_{i=1}^K \cov(\gamma^{K\prime}W^K,\gamma_i^K W_i^K)^2 \rightarrow 0$ as $K \rightarrow \infty$.
\end{lemma}

We omit the $K$ superscripts from all proofs for notational simplicity.

\begin{proof}[Proof of Lemma \ref{lem:lowlevelconds_deltaorig}]

 We first prove the result under Assumption  \ref{assn:MAvar}, then prove it for the subsequent assumptions in order. We focus our proof on showing the limit of $\sum_{i=1}^K \cov(X,\gamma_i W_i)^2$ is 0 since derivations involving $\sum_{i=1}^K \cov(\gamma'W,\gamma_i W_i)^2$ are identical up to replacing $\pi$ by $\gamma$. Throughout these derivations, we use as a first step that 
\begin{align*}
	\sum_{i=1}^K \cov(X,\gamma_i W_i)^2 &\leq K \sup_{i\in[K]} |\cov(X,\gamma_i W_i)|^2\\
	&\leq K\sup_{k\in[K]}|\gamma_k|^2 \sup_{i\in[K]}\left|\cov\left(\sum_{j=1}^K \pi_j W_j, W_i\right)\right|^2\\
	&\leq K\sup_{k\in[K]}|\gamma_k|^2 \sup_{i\in[K]}\left(\sup_{l \in [K]} |\pi_l| \sum_{j=1}^K  |\cov(W_j, W_i)|\right)^2\\
	&= K\sup_{k\in[K]}|\gamma_k|^2 \sup_{l\in[K]}|\pi_l|^2 \sup_{i\in[K]}\left(\sum_{j=1}^K |\cov( W_j, W_i)|\right)^2.
\end{align*}

Suppose Assumption \ref{assn:MAvar} holds. Then,
\begin{align*}
		K\sup_{k\in[K]}|\gamma_k|^2 \sup_{l\in[K]}|\pi_l|^2 \sup_{i\in[K]}\left(\sum_{j=1}^K |\cov( W_j, W_i)|\right)^2
		&\leq K \cdot \frac{C^4}{K^2} \cdot \sup_{i\in[K]}\left(\sum_{j=1}^K |\cov( W_j, W_i)|\right)^2 \\
		&\leq \frac{C^4}{K} \cdot \sup_{i\in[K]}\left(\max\{1 + |\rho|,1 + 2|\rho|\}\right)^2 \\
		&= O(K^{-1}),
\end{align*}
where we used Assumption \ref{assn:MAvar}.2 in the first inequality, and Assumption \ref{assn:MAvar}.1 in the second inequality.
Therefore, $\sum_{i=1}^K \cov(X,\gamma_i W_i)^2 \rightarrow 0$ as $K \rightarrow \infty$.
	
If Assumption \ref{assn:ARvar} holds, then
\begin{align*}
	&K\sup_{k\in[K]}|\gamma_k|^2 \sup_{l\in[K]}|\pi_l|^2 \sup_{i\in[K]}\left(\sum_{j=1}^K |\cov( W_j, W_i)|\right)^2\\
	&\leq K \cdot \frac{C^4}{K^2} \cdot \sup_{i\in[K]}\left(\sum_{j=1}^K |\cov( W_j, W_i)|\right)^2 \\
	&\leq \frac{C^4}{K} \cdot \sup_{i\in[K]}\left(\sum_{j=1}^K |\rho|^{|j-i|}\right)^2\\
	&= O(K^{-1}),
\end{align*}
where we used Assumption \ref{assn:ARvar}.2 in the first inequality, and Assumption \ref{assn:ARvar}.1 in the second inequality. The fourth line follows from 
\begin{align}\label{eq:doublesum_rho}
	\sum_{j=1}^K |\rho|^{|j-i|} &\leq \sum_{j = -\infty}^\infty |\rho|^{|j-i|} \leq 2 \sum_{j=0}^{\infty} |\rho|^j = \frac{2}{1-|\rho|} < \infty.
\end{align}
Therefore, $\sum_{i=1}^K \cov(X,\gamma_i W_i)^2 \rightarrow 0$ as $K \rightarrow \infty$.

If Assumption \ref{assn:EXvar} holds, then
\begin{align*}
		&K\sup_{k\in[K]}|\gamma_k|^2 \sup_{l\in[K]}|\pi_l|^2 \sup_{i\in[K]}\left(\sum_{j=1}^K |\cov( W_j, W_i)|\right)^2\\
		&\leq K \cdot \frac{C^4}{K^4} \cdot \sup_{i\in[K]}\left(\sum_{j=1}^K |\cov( W_j, W_i)|\right)^2 \\
		&\leq \frac{C^4}{K^3} \cdot \sup_{i\in[K]}\left(\sum_{j=1}^K \left|\Lambda_i \Lambda_j' + \sigma^2_E \1(i = j)\right|\right)^2 \\
		&= O(K^{-1}),
\end{align*}
where we used Assumption \ref{assn:EXvar}.2 in the first inequality, and Assumption \ref{assn:EXvar}.1 in the second inequality. The fourth line follows from 
\begin{align*}
	\sum_{j=1}^K \left|\Lambda_i \Lambda_j' + \sigma^2_E \1(i = j)\right| &\leq K \sup_{i \in [K]} \|\Lambda_i\|^2  + \sigma^2_E =  K R\overline{\lambda}^2 + \sigma^2_E =O(K).
\end{align*}
Therefore, $\sum_{i=1}^K \cov(X,\gamma_i W_i)^2 \rightarrow 0$ as $K \rightarrow \infty$.

If Assumption \ref{assn:EXvar_shrink} holds, then
\begin{align*}
	&K\sup_{k\in[K]}|\gamma_k|^2 \sup_{l\in[K]}|\pi_l|^2 \sup_{i\in[K]}\left(\sum_{j=1}^K |\cov( W_j, W_i)|\right)^2\\
	&\leq K \cdot \frac{C^4}{K^2} \cdot \sup_{i\in[K]}\left(\sum_{j=1}^K |\cov( W_j, W_i)|\right)^2 \\
	&\leq \frac{C^4}{K} \cdot \sup_{i\in[K]}\left(\sum_{j=1}^K \left|\frac{\alpha}{K} + \1(i = j)\left(1 - \frac{\alpha}{K}\right)\right| \right)^2\\
	&= O(K^{-1}),
\end{align*}
where we used Assumption \ref{assn:EXvar_shrink}.2 in the first inequality, and Assumption \ref{assn:EXvar_shrink}.1 in the second inequality. The third line follows from 
\begin{align*}
	\sum_{j=1}^K \left|\frac{\alpha}{K} + \1(i = j)\left(1 - \frac{\alpha}{K}\right)\right| &\leq 1 + |\alpha| <\infty.
\end{align*}
Therefore, $\sum_{i=1}^K \cov(X,\gamma_i W_i)^2 \rightarrow 0$ as $K \rightarrow \infty$.

As mentioned earlier, derivations for $\sum_{i=1}^K \cov(\gamma'W,\gamma_i W_i)^2$ are identical except that all instances of $\pi_i$ are replaced by $\gamma_i$.
\end{proof}

\begin{proof}[Proof of Proposition \ref{prop:MAvar}]
We first prove that all three parts of Assumption \ref{assn:pi_and_VarW} hold. 

\medskip

\noindent\textbf{Proof that Assumption \ref{assn:pi_and_VarW}.1 holds}. We have
\begin{align*} 
	\var \left( \sum_{i=1}^K \pi_i W_i \right)
	&\leq \sum_{i=1}^K \pi_i^2 + 2 |\rho| \sum_{i=1}^{K-1} |\pi_i \pi_{i+1}|\\
	&\leq K \cdot \left(\frac{C}{\sqrt{K}}\right)^2 + 2|\rho| (K-1) \cdot \left(\frac{C}{\sqrt{K}}\right)^2\\
	&\leq (1 + 2|\rho|)C^2 < \infty.
\end{align*}
The first line follows from assumptions \ref{assn:MAvar}.1 and \ref{assn:MAvar}.2. The second line uses Assumption \ref{assn:MAvar}.3. By \ref{assn:MAvar}.4, this variance is also bounded away from zero uniformly in $K$. Hence Assumption \ref{assn:pi_and_VarW}.1 holds.

\medskip

\noindent \textbf{Proof that Assumption \ref{assn:pi_and_VarW}.2 holds}. We have that
\begin{align*}
	&{\var}_S\left(\sum_{i=1}^K \sum_{j=1}^K S_i S_j \cov(\pi_i W_i,\pi_j W_j)\right)\\
	&= {\var}_S\left(\sum_{i=1}^K S_i \pi_i^2  + 2\rho \sum_{i=1}^{K-1} S_i S_{i+1} \pi_i\pi_{i+1}\right)\\
	&\leq 2{\var}_S\left(\sum_{i=1}^K S_i \pi_i^2\right)  + 8\rho^2 {\var}_S\left(\sum_{i=1}^{K-1} S_i S_{i+1} \pi_i\pi_{i+1}\right).
\end{align*}
We show that both of these terms converge to zero. First, we see that
\begin{align*}
	{\var}_S\left(\sum_{i=1}^K S_i \pi_i^2\right) &= \frac{d_1 d_2}{K(K-1)} \sum_{i=1}^K \left(\pi_i^2 - \frac{1}{K} \sum_{j=1}^K \pi_j^2 \right)^2\\
	&\leq \frac{d_1 d_2}{K^2} \sum_{i=1}^K \pi_i^4 \\
	&\leq \frac{d_1 d_2}{K^2} K \cdot \left(\frac{C}{\sqrt{K}}\right)^4 \rightarrow 0
\end{align*}
as $K\to \infty$. The first line follows by Lemma \ref{lemma:LLN}.1. The second since the sample average minimizes the sum of squares. The third by Assumption \ref{assn:MAvar}.2, and the convergence to zero follows since $d_1/K$ and $d_2/K$ are bounded by Assumption \ref{assn:limit_d2d1}. 

Second, by Assumption \ref{assn:MAvar}.2,
\begin{align*}
	&{\var}_S\left(\sum_{i=1}^{K-1} S_i S_{i+1} \pi_i\pi_{i+1}\right)
	\leq \sum_{i=1}^{K-1} {\var}_S(S_i S_{i+1} \pi_i\pi_{i+1})\\
	&\hspace{60mm} + \sum_{i,j:|i-j| = 1}\left|{\cov}_S(S_i S_{i+1} \pi_i \pi_{i+1}, S_j S_{j+1} \pi_j\pi_{j+1})\right|\\
	&\hspace{60mm} + \sum_{i,j:|i-j| \geq 2}\left|{\cov}_S(S_i S_{i+1} \pi_i\pi_{i+1}, S_j S_{j+1} \pi_j\pi_{j+1})\right|\\
	&\leq (K-1) {\var}_S(S_iS_{i+1}) \left(\frac{C}{\sqrt{K}}\right)^4 + 2(K-2) \left|{\cov}_S(S_iS_{i+1},S_{i+1}S_{i+2})\right| \left(\frac{C}{\sqrt{K}}\right)^4\\
	&\qquad + (K-2)(K-3) \left|{\cov}_S(S_iS_{i+1},S_{j}S_{j+1})\right|\mathbbm{1}(|i-j| \geq 2) \left(\frac{C}{\sqrt{K}}\right)^4. 
\end{align*}
To compute the order of this term, we compute the following quantities. Noting that $S_iS_{i+1}$ is binary and that $\Exp_S[S_iS_{i+1}] = \Prob_S(S_{i+1} = 1|S_i = 1)\Prob(S_i = 1) = d_1(d_1 - 1)/(K(K-1))$,
\begin{align*}
{\var}_S(S_iS_{i+1}) = \Exp_S[S_i S_{i+1}](1 - \Exp_S[S_i S_{i+1}]) = \frac{d_1-1}{K-1}\frac{d_1}{K}\left(1 - \frac{d_1-1}{K-1}\frac{d_1}{K}\right).
\end{align*}
Similarly,
\begin{align*}
	{\cov}_S(S_iS_{i+1},S_{i+1}S_{i+2}) &= \Exp_S[S_iS_{i+1}S_{i+2}] - \Exp_S[S_i S_{i+1}]\Exp_S[S_{i+1}S_{i+2}]\\
	&= \frac{(d_1 - 2)(d_1 - 1)d_1}{(K-2)(K-1)K} - \frac{(d_1 - 1)^2d_1^2}{(K-1)^2 K^2}.
\end{align*}
Both ${\var}_S(S_iS_{i+1})$ and ${\cov}_S(S_iS_{i+1},S_{i+1}S_{i+2})$ are $O(1)$ since $d_1/K \to 1/(1+r) \in (0,1)$. We also have that, for $|i - j|\geq 2$,
\begin{align*}
	{\cov}_S(S_iS_{i+1},S_{j}S_{j+1}) &= \Exp[S_iS_{i+1}S_jS_{j+1}] - \Exp[S_i S_{i+1}]^2\\
	&= \frac{(d_1-3)(d_1 - 2)(d_1 - 1)d_1}{(K-3)(K-2)(K-1)K} - \frac{(d_1 - 1)^2 d_1^2}{(K-1)^2 K^2}\\
	&= \frac{(d_1 - 1)d_1}{(K-1)K}\frac{(K-d_1)(-4d_1 K + 6 d_1 + 6K - 6)}{(K-3)(K-2)(K-1)K}.
\end{align*}
By Assumption \ref{assn:limit_d2d1}, this term is of order $O(K^{-1})$. Therefore,
\begin{align*}
	&{\var}_S\left(\sum_{i=1}^{K-1} S_i S_{i+1} \pi_i\pi_{i+1}\right)\\
	&\leq \left(\frac{C}{\sqrt{K}}\right)^4\left((K-1) O(1)  + 2(K-2) O(1)  + (K-2)(K-3) O(K^{-1}) \right) = o(1)
\end{align*}
since $K\to\infty$. We can similarly show ${\var}_S\left(\sum_{i=1}^K \sum_{j=1}^K (1-S_i) (1-S_j) \cov(\pi_i W_i,\pi_j W_j)\right) = o(1)$. Hence Assumption \ref{assn:pi_and_VarW}.2 holds.

\medskip

\noindent\textbf{Proof that Assumption \ref{assn:pi_and_VarW}.3 holds}. Consider the ratio in Assumption \ref{assn:pi_and_VarW}.3:
\begin{align*}
	\frac{\sum_{i=1}^K  \var(\pi_i W_i)}{\var\left(\sum_{i=1}^K \pi_i W_i\right)} &= \frac{\sum_{i=1}^K \pi_i^2}{\sum_{i=1}^K \pi_i^2 + 2\rho \sum_{i=1}^{K-1}\pi_i \pi_{i+1}} = \left(1 + 2\rho \frac{\sum_{i=1}^{K-1}\pi_i \pi_{i+1}}{\sum_{i=1}^K \pi_i^2}\right)^{-1}.
\end{align*}
Its denominator $\var(\sum_{i=1}^K \pi_i W_i)$ is bounded away from zero by Assumption \ref{assn:MAvar}.4, and its numerator $\sum_{i=1}^K \pi_i^2$ is bounded by Assumption \ref{assn:MAvar}.2. Therefore this nonnegative ratio does not diverge to $+\infty$. By Assumption \ref{assn:MAvar}.3, this ratio's limit as $K\rightarrow\infty$ is $(1 + 2 \rho d_{\pi,\text{MA}})^{-1}$, which is nonnegative by $|\rho| < 1/2$ and $|d_{\pi,\text{MA}}| \leq 1$.

That Assumption \ref{assn:gamma_and_VarW} holds follows from similar derivations where $\pi$ is replaced by $\gamma$. Assumption \ref{assn:limited_dep} holds by Lemma \ref{lem:lowlevelconds_deltaorig}.
\end{proof}

\begin{proof}[Proof of Proposition \ref{prop:ARvar}]
We again break this proof into three parts.

\medskip

\noindent\textbf{Proof that Assumption \ref{assn:pi_and_VarW}.1 holds}. We have
\begin{align*}
	\var \left( \sum_{i=1}^K \pi_i W_i \right)
	&\leq \sum_{i=1}^K \sum_{j=1}^K |\pi_i| |\pi_j| |\rho|^{|i-j|}\\
	&\leq  \left(\frac{C}{\sqrt{K}}\right)^2 \sum_{i=1}^K \sum_{j=1}^K |\rho|^{|i-j|}\\
	&\leq 2C^2 (1-|\rho|)^{-1} < \infty,
\end{align*}
where we used equation \eqref{eq:doublesum_rho} to show the third inequality. By Assumption \ref{assn:ARvar}.4, this variance is also bounded away from zero uniformly in $K$. Therefore Assumption \ref{assn:pi_and_VarW}.1 holds.

\medskip

\noindent\textbf{Proof that Assumption \ref{assn:pi_and_VarW}.2 holds}. We have
\begin{align*}
	&{\var}_S\left(\sum_{i=1}^K \sum_{j=1}^K S_i S_j \cov(\pi_i W_i,\pi_j W_j)\right)\\
	&= \sum_{i=1}^K \sum_{j=1}^K \sum_{k=1}^K \sum_{l=1}^K \pi_i \pi_j \pi_k \pi_l \cov(W_i, W_j)\cov( W_k, W_l) {\cov}_S(S_iS_j, S_kS_l)\\
	&\leq \frac{C^4}{K^2} \sum_{i=1}^K \sum_{j=1}^K \sum_{k=1}^K \sum_{l=1}^K  |\rho|^{|i-j| + |k-l|} |{\cov}_S(S_iS_j, S_kS_l)|.
\end{align*}
We can decompose this quadruple sum into various partial sums:
\begin{align}
	&\sum_{i=1}^K \sum_{j=1}^K \sum_{k=1}^K \sum_{l=1}^K  |\rho|^{|i-j| + |k-l|} |{\cov}_S(S_iS_j, S_kS_l)| = \sum_{i,j,k,l:i=j=k=l} {\cov}_S(S_i,S_i) \label{eq:AR1}\\
	&+ \sum_{i,j,k,l:i=j, k=l, \{i,j\} \cap \{k,l\} = \emptyset} |{\cov}_S(S_i,S_k)| \label{eq:AR2}\\
	&+ \sum_{i,j,k,l:i=j, k\neq l, i \in \{k,l\}} |\rho|^{|k-l|} |{\cov}_S(S_i,S_iS_l)| + \sum_{i,j,k,l:i\neq j, k = l, k \in \{i,j\}} |\rho|^{|i-j|} |{\cov}_S(S_iS_j,S_i)| \label{eq:AR3}\\
	&+ \sum_{i,j,k,l:i=j, k\neq l, i \notin \{k,l\}} |\rho|^{|k-l|} |{\cov}_S(S_i,S_kS_l)| + \sum_{i,j,k,l:i\neq j, k = l, k \notin \{i,j\}} |\rho|^{|i-j|} |{\cov}_S(S_iS_j,S_k)| \label{eq:AR4}\\
	&+ \sum_{i,j,k,l:i \neq j, k\neq l, |\{i,j\}\cap \{k,l\}| = 2} |\rho|^{2|i-j|} |{\cov}_S(S_iS_j,S_iS_j)| \label{eq:AR5}\\
	&+ \sum_{i,j,k,l:i \neq j, k\neq l, |\{i,j\}\cap \{k,l\}| = 1} |\rho|^{|i-j| + |k-l|} |{\cov}_S(S_iS_j,S_iS_l)| \label{eq:AR6}\\
	&+ \sum_{i,j,k,l:i \neq j, k\neq l, |\{i,j\}\cap \{k,l\}| = 0} |\rho|^{|i-j| + |k-l|} |{\cov}_S(S_iS_j,S_kS_l)|.\label{eq:AR7}
\end{align}
We analyze terms \eqref{eq:AR1}--\eqref{eq:AR7} separately. Term \eqref{eq:AR1} is a sum of $K$ terms and equals
\begin{align*}
	\sum_{i,j,k,l:i=j=k=l} {\cov}_S(S_i,S_i) &= K {\var}_S(S_i) = K \frac{d_1}{K}\left(1 - \frac{d_1}{K}\right) = \frac{d_1d_2}{K} = O(K).	
\end{align*}
The last equality follows from Assumption \ref{assn:limit_d2d1}. Term \eqref{eq:AR2} is a sum of $K(K-1)$ terms and equals
\begin{align*}
	\sum_{i,j,k,l:i=j, k=l, \{i,j\} \cap \{k,l\} = \emptyset} |{\cov}_S(S_i,S_k)| &= K(K-1)|\Exp_S[S_iS_k] - \Exp_S[S_i]\Exp_S[S_k]|\\
	&= K(K-1) \left|\frac{d_1(d_1-1)}{K(K-1)} - \frac{d_1^2}{K^2}\right|\\
	&= K(K-1) \left|\frac{Kd_1(d_1-1) - d_1^2(K-1)}{K^2(K-1)}\right|\\
	&= \frac{d_1d_2}{K} = O(K).
\end{align*}

Term \eqref{eq:AR3} is a sum of two identical terms. Hence we can write it as
\begin{align*}
	&\sum_{i,j,k,l:i=j, k\neq l, i \in \{k,l\}} 2|\rho|^{|k-l|} |{\cov}_S(S_i,S_iS_l)|\\
	&= \sum_{i,j,k,l:i=j, k\neq l, i \in \{k,l\}} 2|\rho|^{|k-l|} |\Exp_S[S_iS_l] - \Exp_S[S_i]\Exp_S[S_iS_l]|\\
	&= \sum_{i,j,k,l:i=j, k\neq l, i \in \{k,l\}} 2|\rho|^{|k-l|} \frac{d_1(d_1-1)}{K(K-1)}\left(1 - \frac{d_1}{K}\right)\\
	&= \frac{4d_1d_2(d_1-1)}{K^2(K-1)} \sum_{k,l:k\neq l} |\rho|^{|k-l|} \\
	&\leq \frac{4d_1d_2(d_1-1)}{K^2(K-1)} \sum_{k = 1}^K \frac{2}{1-|\rho|}\\
	&= \frac{8d_1d_2(d_1-1)}{K(K-1)(1-|\rho|)} = O(K),
\end{align*}
where the inequality follows from equation \eqref{eq:doublesum_rho}. Term \eqref{eq:AR4} is similarly a sum of two identical terms. Hence we can write it as
\begin{align*}
	&\sum_{i,j,k,l:i=j, k\neq l, i \notin \{k,l\}} 2|\rho|^{|k-l|} |{\cov}_S(S_i,S_kS_l)|\\
	&= \sum_{i,j,k,l:i=j, k\neq l, i \notin \{k,l\}} 2|\rho|^{|k-l|} |\Exp_S[S_iS_kS_l] - \Exp_S[S_i]\Exp[S_kS_l]|\\
	&= \sum_{i,j,k,l:i=j, k\neq l, i \notin \{k,l\}} 2|\rho|^{|k-l|} \left|\frac{d_1(d_1-1)(d_1-2)}{K(K-1)(K-2)} - \frac{d_1^2(d_1-1)}{K^2(K-1)} \right|\\
	&= 2\frac{d_1(d_1-1)}{K(K-1)}\left|\frac{d_1-2}{K-2} - \frac{d_1}{K}\right|\sum_{i,j,k,l:i=j, k\neq l, i \notin \{k,l\}} |\rho|^{|k-l|}\\
	&= 4\frac{d_1(d_1-1)}{K(K-1)}\frac{d_2}{K(K-2)}(K-2) \sum_{k,l:k\neq l} |\rho|^{|k-l|} = O(K).
\end{align*}
The last line follows from $\sum_{k,l:k\neq l} |\rho|^{|k-l|} = O(K)$, which we showed in our analysis of term \eqref{eq:AR3}. Term \eqref{eq:AR5} is a sum of $2K(K-1)$ terms and equals
\begin{align*}
	&\sum_{i,j,k,l:i \neq j, k\neq l, |\{i,j\}\cap \{k,l\}| = 2} |\rho|^{2|i-j|} |{\cov}_S(S_iS_j,S_iS_j)|\\
	&= 2\sum_{i,j: i \neq j}|\rho|^{2|i-j|} \Exp_S[S_iS_j](1 - \Exp_S[S_iS_j])\\
	&= 2 \frac{d_1(d_1-1)}{K(K-1)}\left(1 - \frac{d_1(d_1-1)}{K(K-1)}\right) \sum_{i,j: i \neq j}|\rho|^{2|i-j|} = O(K).
\end{align*}
since $\sum_{i,j: i \neq j}|\rho|^{2|i-j|} = O(K)$. Term \eqref{eq:AR6} is a sum of $4K(K-1)(K-2)$ terms and equals
\begin{align*}
	&\sum_{i,j,k,l:i \neq j, k\neq l, |\{i,j\}\cap \{k,l\}| = 1} |\rho|^{|i-j| + |k-l|} |{\cov}_S(S_iS_j,S_iS_l)|\\
	&= \sum_{i,j,k,l:i \neq j, k\neq l, |\{i,j\}\cap \{k,l\}| = 1} |\rho|^{|i-j| + |k-l|} \left(\frac{d_1(d_1-1)(d_1-2)}{K(K-1)(K-2)} - \frac{d_1^2(d_1-1)^2}{K^2(K-1)^2}\right)\\
	&= \frac{d_1(d_1-1)}{K(K-1)} \frac{d_2(d_1K - 2K - 2d_1 + 2)}{K(K-1)(K-2)} \sum_{i,j,k,l:i \neq j, k\neq l, |\{i,j\}\cap \{k,l\}| = 1} |\rho|^{|i-j| + |k-l|} \\
	&= O(1) \cdot \sum_{i,j: i \neq j}  |\rho|^{|i-j|} \left(\sum_{k,l:k \neq l, k=i, l \neq j} + \sum_{k,l:k \neq l, k=j, l \neq i} + \sum_{k,l:k \neq l, l=i, k \neq j} + \sum_{k,l:k \neq l, l=j, k \neq i}\right)|\rho|^{|k-l|}\\
	&= O(1) \cdot 4 \sum_{i,j: i \neq j} |\rho|^{|i-j|} \sum_{k,l:k \neq l, k=i, l \neq j}  |\rho|^{|k-l|}\\
	&= O(1) \cdot 4 \sum_{i,j: i \neq j} |\rho|^{|i-j|} \sum_{l: l \notin \{i,j\}}  |\rho|^{|i-l|}\\
	&= O(1)\cdot \sum_{i,j: i \neq j} |\rho|^{|i-j|}= O(K).
\end{align*}
We used $\sum_{l: l \notin \{i,j\}}  |\rho|^{|i-l|}=O(1)$ in the second to last equality. Finally, term \eqref{eq:AR7} is a sum of $K(K-1)(K-2)(K-3)$ terms and equals
\begin{align*}
	&\sum_{i,j,k,l:i \neq j, k\neq l, |\{i,j\}\cap \{k,l\}| = 0} |\rho|^{|i-j| + |k-l|} |{\cov}_S(S_iS_j,S_kS_l)|\\
	&= \sum_{i,j,k,l:i \neq j, k\neq l, |\{i,j\}\cap \{k,l\}| = 0} |\rho|^{|i-j| + |k-l|} \left|\Exp_S[S_iS_jS_kS_l] - \Exp_S[S_iS_j]^2\right|\\
	&= \underbrace{\left|\frac{(d_1 - 1)d_1}{(K-1)K}\frac{(K-d_1)(-4d_1 K + 6 d_1 + 6K - 6)}{(K-3)(K-2)(K-1)K}\right|}_{=O(K^{-1})} \sum_{i,j,k,l:i \neq j, k\neq l, |\{i,j\}\cap \{k,l\}| = 0} |\rho|^{|i-j| + |k-l|}\\
	&\leq O(K^{-1}) \sum_{i,j:i \neq j} \sum_{k,l:k\neq l}|\rho|^{|i-j| + |k-l|}\\
	&= O(K^{-1}) \sum_{i,j:i \neq j} |\rho|^{|i-j|} \sum_{k,l:k\neq l}|\rho|^{|k-l|} = O(K^{-1}) \cdot O(K) \cdot O(K) = O(K).
\end{align*}
Since terms \eqref{eq:AR1}--\eqref{eq:AR7} are all of order $O(K)$, we conclude that\\ ${\var}_S\left(\sum_{i=1}^K \sum_{j=1}^K S_i S_j \cov(\pi_i W_i,\pi_j W_j)\right)$ is of order $O(K^{-1})$, and hence converges to 0 as $K \rightarrow \infty$. The second part of Assumption \ref{assn:pi_and_VarW}.2 holds by a similar argument.

\medskip

\noindent\textbf{Proof that Assumption \ref{assn:pi_and_VarW}.3 holds}. We can write the ratio in Assumption \ref{assn:pi_and_VarW}.3 as
\begin{align*}
	\frac{\sum_{i=1}^K  \var(\pi_i W_i)}{\var\left(\sum_{i=1}^K \pi_i W_i\right)}
	&= \frac{\sum_{i=1}^K \pi_i^2}{\sum_{i=1}^K \pi_i^2 + \sum_{i,j:i \neq j} \pi_i \pi_j \rho^{|i-j|}} 
	= \left(1 +  \frac{\sum_{i,j:i \neq j} \pi_i \pi_j \rho^{|i-j|}}{\sum_{i=1}^K \pi_i^2}\right)^{-1}.
\end{align*}
Its denominator $\var(\sum_{i=1}^K \pi_i W_i)$ is bounded away from zero by Assumption \ref{assn:ARvar}.4, and its numerator $\sum_{i=1}^K \pi_i^2$ is bounded by Assumption \ref{assn:ARvar}.2. Therefore this nonnegative ratio does not diverge to $+\infty$. By Assumption  \ref{assn:ARvar}.3, this ratio's limit as $K\rightarrow\infty$ is $(1 + d_{\pi,\text{AR}})^{-1}$, which is nonnegative since $d_{\pi,\text{AR}} > -1$.

That Assumption  \ref{assn:gamma_and_VarW} holds follows from similar derivations where $\pi$ is replaced by $\gamma$. Assumption \ref{assn:limited_dep} holds by Lemma \ref{lem:lowlevelconds_deltaorig}.
\end{proof}

\begin{proof}[Proof of Proposition \ref{prop:EXvar}]
We again break this proof into three parts.

\medskip

\noindent\textbf{Proof that Assumption \ref{assn:pi_and_VarW}.1 holds}. We have
\begin{align*}
	\var\left(\sum_{i=1}^K \pi_i W_i\right) &= \var(\pi'\Lambda F + \pi' E)\\
	&= \pi'\Lambda \Lambda'\pi + \sigma^2_E \sum_{i=1}^K \pi_i^2\\
	&= \sum_{r=1}^R (\pi'\lambda_r)^2 + \sigma^2_E \sum_{i=1}^K \pi_i^2\\
	&\leq R \sup_{r \in [R]} \left(\sum_{i=1}^K \pi_i \lambda_{i,r}\right)^2 + \sigma^2_E K \sup_{i \in [K]} |\pi_i|^2\\
	&\leq R \sup_{r \in [R]} K \sum_{i=1}^K (\pi_i \lambda_{i,r})^2 + \sigma^2_E K \frac{C^2}{K^2}\\
	&\leq R K^2 \sup_{r \in [R]} \sup_{i\in [K]} |\pi_i|^2 |\lambda_{i,r}|^2 + \sigma^2_E \frac{C^2}{K}\\
	&\leq R K^2 \frac{C^2}{K^2} \overline{\lambda} + \sigma^2_E \frac{C^2}{K} < \infty.
\end{align*}
The first line follows by Assumption \ref{assn:EXvar}.1. The fifth by Assumption \ref{assn:EXvar}.2, and the seventh by Assumption \ref{assn:EXvar}.3. By \ref{assn:EXvar}.4, this variance is also uniformly bounded away from zero. Hence Assumption \ref{assn:pi_and_VarW}.1 holds.

\medskip

\noindent\textbf{Proof that Assumption \ref{assn:pi_and_VarW}.2 holds}. We have
\begin{align}
	&{\var}_S\left(\sum_{i=1}^K \sum_{j=1}^K S_i S_j \cov(\pi_i W_i,\pi_j W_j)\right)\notag\\
	&= {\var}_S\left(\sum_{i=1}^K S_i \var(\pi_i W_i) + \sum_{i,j:i \neq j} S_i S_j \cov(\pi_i W_i,\pi_j W_j)\right)\notag\\
	&\leq 2{\var}_S\left(\sum_{i=1}^K S_i \var(\pi_i W_i)\right) + 2 {\var}_S \left(\sum_{i,j:i \neq j} S_i S_j \cov(\pi_i W_i,\pi_j W_j)\right).\label{eq:EXvar_prop_equation}
\end{align}
The first term in equation \eqref{eq:EXvar_prop_equation} converges to zero since
\begin{align*}
	{\var}_S\left(\sum_{i=1}^K S_i \var(\pi_i W_i)\right) &= \frac{d_1 d_2}{K(K-1)} \sum_{i=1}^K \var(\pi_i W_i)^2\\
	&= \frac{d_1 d_2}{K(K-1)} \sum_{i=1}^K \pi_i^4 \left(\Lambda_i \Lambda_i' + \sigma^2_E\right)^2\\
	&\leq \frac{d_1 d_2}{K(K-1)} \cdot K \cdot \sup_{i \in [K]} |\pi_i|^4\left(R \overline{\lambda}^2 + \sigma^2_E\right)^2\\
	&\leq \frac{d_1 d_2}{K-1} \cdot \frac{C^4}{K^4}\left(R \overline{\lambda}^2 + \sigma^2_E\right)^2\\
	&= o(1).
\end{align*}
Here we used Lemma \ref{lemma:LLN} for the first line, Assumption \ref{assn:EXvar}.3 for the third line, and Assumption \ref{assn:EXvar}.2 for the fourth line. The last line follows from the previous derivations and from Assumption \ref{assn:limit_d2d1}.

We now investigate the convergence of the second term in equation \eqref{eq:EXvar_prop_equation}. We can write
\begin{align*}
	&{\var}_S \left(\sum_{i,j:i \neq j} S_i S_j \cov(\pi_i W_i,\pi_j W_j)\right)\\
	&= \sum_{i,j:i \neq j}\sum_{k,l:k \neq l}{\cov}_S \left( S_i S_j \cov(\pi_i W_i,\pi_j W_j), S_k S_l \cov(\pi_k W_k,\pi_l W_l)\right)\\
	&\leq \sum_{i,j:i \neq j}\sum_{k,l:k \neq l}|{\cov}_S \left( S_i S_j , S_k S_l \right)|\left|\cov(\pi_i W_i,\pi_j W_j)\right|\left|\cov(\pi_k W_k,\pi_l W_l)\right|\\
	&\leq \left(\sup_{i,j:i \neq j}\left|\cov(\pi_i W_i,\pi_j W_j)\right|^2\right) \cdot \sum_{i,j:i \neq j}\sum_{k,l:k \neq l}|{\cov}_S \left( S_i S_j , S_k S_l \right)|.
\end{align*}
The first supremum is bounded above by
\begin{align*}
	\sup_{i,j:i \neq j}\left|\cov(\pi_i W_i,\pi_j W_j)\right|^2 &= \sup_{i,j:i \neq j}|\pi_i|^2|\pi_j|^2\left|\Lambda_i\Lambda_j'\right|^2 \leq \frac{C^4}{K^4} \cdot R^2 \overline{\lambda}^4 = O(K^{-4})
\end{align*}
The summation is bounded above by
\begin{align*}
	&\sum_{i,j:i \neq j}\sum_{k,l:k \neq l}|{\cov}_S \left( S_i S_j , S_k S_l \right)|\\
	&= \sum_{i,j:i \neq j} \left( 2|{\cov}_S(S_iS_j, S_iS_j)| + \sum_{k,l:k \neq l, |\{i,j\}\cap\{k,l\}| = 1} \left|{\cov}_S(S_iS_j,S_i S_l)\right|\right.\\
	&\left. \hspace{70mm} + \sum_{k,l:k \neq l, |\{i,j\}\cap\{k,l\}| = 0} \left|{\cov}_S(S_iS_j,S_k S_l)\right|\right)\\
	&= \sum_{i,j:i \neq j} \left( 2O(1) + \sum_{k,l:k \neq l, |\{i,j\}\cap\{k,l\}| = 1} O(1) + \sum_{k,l:k \neq l, |\{i,j\}\cap\{k,l\}| = 0} O(K^{-1})\right)\\
	&= \left(2K(K-1) O(1) + 4K(K-1)(K-2)O(1) + K(K-1)(K-2)(K-3) O(K^{-1})\right)\\
	&= O(K^3).
\end{align*}
The second equality follows from derivations in the proofs of Propositions \ref{prop:MAvar} and \ref{prop:ARvar}. Therefore \allowbreak ${\var}_S\left(\sum_{i=1}^K \sum_{j=1}^K S_i S_j \cov(\pi_i W_i,\pi_j W_j)\right) \to 0$ as $K \to \infty$. The second part of Assumption \ref{assn:pi_and_VarW}.2 holds by a similar argument.

\medskip

\noindent\textbf{Proof that Assumption \ref{assn:pi_and_VarW}.3 holds}. We can write the ratio in Assumption \ref{assn:pi_and_VarW}.3 as
\begin{align*}
	\frac{\sum_{i=1}^K  \var(\pi_i W_i)}{\var\left(\sum_{i=1}^K \pi_i W_i\right)} &=  \frac{\sum_{i=1}^K \pi_i^2 \left(\Lambda_i \Lambda_i' + \sigma^2_E\right)}{ \sum_{r=1}^R (\pi'\lambda_r)^2 + \sigma^2_E \sum_{i=1}^K \pi_i^2}.
\end{align*}
Its denominator $\var(\sum_{i=1}^K \pi_i W_i)$ is bounded away from zero by Assumption \ref{assn:EXvar}.3. Its numerator converges to zero because
\begin{align*}
	\sum_{i=1}^K \pi_i^2 \left(\Lambda_i \Lambda_i' + \sigma^2_E\right) &\leq K \sup_{i \in [K]} |\pi_i|^2 \left(R \overline{\lambda}^2 + \sigma^2_E\right) = O(K^{-1}) = o(1).
\end{align*} 
Therefore this ratio's limit as $K\rightarrow\infty$ is zero.

That Assumption \ref{assn:gamma_and_VarW} holds follows from similar derivations where $\pi$ is replaced by $\gamma$. Assumption \ref{assn:limited_dep} holds by Lemma \ref{lem:lowlevelconds_deltaorig}.
\end{proof}

\begin{proof}[Proof of Proposition \ref{prop:EXvar_shrink}]
We again break this proof into three parts.

\medskip

\noindent\textbf{Proof that Assumption \ref{assn:pi_and_VarW}.1 holds}. We have
\begin{align*}
	\var\left(\sum_{i=1}^K \pi_i W_i\right) 
	&\leq \sum_{i=1}^K \pi_i^2 + \frac{| \alpha |}{K} \sum_{i,j:i \neq j} |\pi_i| |\pi_j|\\
	&\leq K \cdot \left(\frac{C}{\sqrt{K}}\right)^2 + \frac{| \alpha |}{K} K(K-1) \left(\frac{C}{\sqrt{K}}\right)^2\\
	&\leq C^2(1 + |\alpha|) < \infty.
\end{align*}
The first line follows by Assumption \ref{assn:EXvar_shrink}.1. The second by Assumption \ref{assn:EXvar_shrink}.2. By Assumption \ref{assn:EXvar_shrink}.4, this variance is also uniformly bounded away from zero. Hence Assumption \ref{assn:pi_and_VarW}.1 holds.

\medskip

\noindent\textbf{Proof that Assumption \ref{assn:pi_and_VarW}.2 holds}. We have
\begin{align}
	{\var}_S\left(\sum_{i=1}^K \sum_{j=1}^K S_i S_j \cov(\pi_i W_i,\pi_j W_j)\right) &= {\var}_S\left(\sum_{i=1}^K S_i \pi_i^2  + \frac{\alpha}{K} \sum_{i,j:i\neq j} \pi_i \pi_j S_i S_j\right)\notag\\
	\leq 2{\var}_S&\left(\sum_{i=1}^K S_i \pi_i^2\right) + 2\frac{\alpha^2}{K^2} {\var}_S\left(\sum_{i,j:i\neq j} \pi_i \pi_j S_i S_j\right).\label{eq:EXvar_shrink_prop_equation}
\end{align}
The first line followed from Assumption \ref{assn:EXvar_shrink}.1. The first term in equation \eqref{eq:EXvar_shrink_prop_equation} converges to zero since
\begin{align*}
	{\var}_S\left(\sum_{i=1}^K S_i \pi_i^2\right) &\leq \frac{d_1d_2}{K(K-1)}\sum_{i=1}^K \pi_i^4 \leq \frac{d_1d_2}{K(K-1)} K \cdot \left(\frac{C}{\sqrt{K}}\right)^4 = o(1)
\end{align*}
as $K \rightarrow \infty$. Here we used Lemma \ref{lemma:LLN}.1 in the first inequality and Assumption \ref{assn:EXvar_shrink}.2 in the second. For the second term in equation \eqref{eq:EXvar_shrink_prop_equation},
\begin{align*}
	&\frac{1}{K^2} {\var}_S\left(\sum_{i,j:i\neq j} \pi_i \pi_j S_i S_j\right) = \frac{1}{K^2}\sum_{i,j:i \neq j} \sum_{k,l:k\neq l} {\cov}_S(S_iS_j, S_kS_l) \pi_i \pi_j \pi_k \pi_l\\
	&\leq \frac{1}{K^2}\cdot\frac{C^4}{K^2} \sum_{i,j:i \neq j} \sum_{k,l:k\neq l} |{\cov}_S(S_iS_j, S_kS_l)|\\
	&= \frac{C^4}{K^4} \sum_{i,j:i \neq j} \left( 2|{\cov}_S(S_iS_j, S_iS_j)| + \sum_{k,l:k \neq l, |\{i,j\}\cap\{k,l\}| = 1} \left|{\cov}_S(S_iS_j,S_i S_l)\right|\right.\\
	&\left. \hspace{70mm} + \sum_{k,l:k \neq l, |\{i,j\}\cap\{k,l\}| = 0} \left|{\cov}_S(S_iS_j,S_k S_l)\right|\right)\\
	&= \frac{C^4}{K^4} \sum_{i,j:i \neq j} \left( 2O(1) + \sum_{k,l:k \neq l, |\{i,j\}\cap\{k,l\}| = 1} O(1) + \sum_{k,l:k \neq l, |\{i,j\}\cap\{k,l\}| = 0} O(K^{-1})\right)\\
	&= \frac{C^4}{K^4} \left(2K(K-1) O(1) + 4K(K-1)(K-2)O(1) + K(K-1)(K-2)(K-3) O(K^{-1})\right)\\
	&= o(1).
\end{align*}
We used Assumption \ref{assn:EXvar_shrink}.1 in line two. The fourth line follows from derivations in the proofs of Propositions \ref{prop:MAvar} and \ref{prop:ARvar}. Therefore ${\var}_S\left(\sum_{i=1}^K \sum_{j=1}^K S_i S_j \cov(\pi_i W_i,\pi_j W_j)\right) \to 0$ as $K \to \infty$. The second part of Assumption \ref{assn:pi_and_VarW}.2 holds by a similar argument.

\medskip

\noindent\textbf{Proof that Assumption \ref{assn:pi_and_VarW}.3 holds}. We can write the ratio in Assumption \ref{assn:pi_and_VarW}.3 as
\begin{align*}
	\frac{\sum_{i=1}^K  \var(\pi_i W_i)}{\var\left(\sum_{i=1}^K \pi_i W_i\right)} 
	&=  \frac{\sum_{i=1}^K \pi_i^2}{\left( 1 - \frac{\alpha}{K} \right) \sum_{i=1}^K \pi_i^2 + \frac{\alpha}{K} \left(\sum_{i=1}^K \pi_i \right)^2 }\\
	&= \left( 1- \frac{\alpha}{K} + \alpha\left(\sum_{i=1}^K \pi_i \right)^2/\left(K\sum_{i=1}^K \pi_i^2\right) \right)^{-1}.
\end{align*}
Its denominator $\var(\sum_{i=1}^K \pi_i W_i)$ is bounded away from zero by Assumption \ref{assn:EXvar_shrink}.4, and its numerator $\sum_{i=1}^K \pi_i^2$ is bounded by Assumption \ref{assn:EXvar_shrink}.2. Therefore this nonnegative ratio does not diverge to $+\infty$. By Assumption  \ref{assn:EXvar_shrink}.3, this ratio's limit as $K\rightarrow\infty$ is $d_{\pi,\text{EX}}/(d_{\pi,\text{EX}} + \alpha)$, which is nonnegative since $d_{\pi,\text{EX}} \geq 1$ and $\alpha > -1$.

That Assumption \ref{assn:gamma_and_VarW} holds follows from similar derivations where $\pi$ is replaced by $\gamma$. Assumption \ref{assn:limited_dep} holds by Lemma \ref{lem:lowlevelconds_deltaorig}.
\end{proof}

\bibliographystyle{econometrica}
\bibliography{BadControls_paper}
\makeatletter\@input{DMP1aux.tex}\makeatother